\documentclass[aps,prd,twocolumn,floatfix,10pt,amssymb,amsfont,amsmath,superscriptaddress,nofootinbib,float,floatfix,longbibliography]{revtex4-1}

\usepackage[utf8]{inputenc}
\usepackage{lmodern}
\usepackage[american]{babel}
\usepackage[autostyle=try,english=american]{csquotes}
\usepackage{xfrac}
\usepackage{microtype}
\usepackage{epstopdf}
\usepackage{verbatim}
\usepackage[usenames,dvipsnames]{xcolor}
\usepackage{graphicx}
\usepackage[pdftex, plainpages=false]{hyperref}
\usepackage{todonotes}
\usepackage{mathtools}
\usepackage{tikz-cd}
\usepackage{tikz}
\usetikzlibrary{shapes.geometric}
\usetikzlibrary{arrows,matrix,calc,scopes,decorations.markings}
\usetikzlibrary{arrows.meta}
\usetikzlibrary{intersections}
\usetikzlibrary{decorations.pathreplacing,angles,quotes}
\usetikzlibrary{external}
\tikzsetexternalprefix{fig/}
\usepackage{array}
\usepackage{tabularx}
\usepackage{booktabs}
\usepackage[export]{adjustbox}
\usepackage{bbm, yfonts}
\usepackage{enumitem}
\usepackage{gensymb}
\usepackage{lipsum}
\usepackage{calc}
\usepackage{mdframed, framed}
\usepackage{bm}
\mathtoolsset{showonlyrefs=true}
\usepackage{footnote}

\makeatletter
\def\l@subsubsection#1#2{}
\makeatother

\renewenvironment{leftbar}[1][\hsize]
{%
	\MakeFramed{\hsize#1\advance\hsize-\width\FrameRestore}%
}
{\endMakeFramed}
\hypersetup{%
	bookmarks=true,         
	bookmarksopen=true,
	unicode=true,           
	pdftoolbar=true,        
	pdfmenubar=true,        
	pdffitwindow=false,     
	pdfstartview={FitH},    
	pdftitle={},
	pdfauthor={},
	pdfsubject={},          
	pdfcreator={},          
	pdfproducer={pdfLaTeX}, 
	pdfkeywords={string net} {topological phases} {category theory},
	pdfnewwindow=true,      
	colorlinks,
	citecolor=BrickRed,
	filecolor=BrickRed,
	linkcolor=BlueViolet,
	urlcolor=BrickRed,
	linktocpage=true
}

\newcommand{\unit}{\mathbbm{1}}
\newcommand{\Z}{\,\mathbb{Z}}

\newcommand{\tr}{\triangleright}

\newcommand{\nn}{\nonumber}

\newcommand{\cN}{\mathcal{N}}
\newcommand{\cM}{\mathcal{M}}

\newcommand{\la}{\langle}
\newcommand{\ra}{\rangle}
\newcommand{\q}{\quad}
\newcommand{\bul}{\bullet}
\newcommand{\sss}{\scriptstyle}
\newcommand{\uni}{{\scriptstyle \mathbbm{1}}}
\newcommand{\smilo}{\! \smile \!}
\newcommand{\bfa}{{ a}}
\newcommand{\bfb}{{ b}}
\newcommand{\bfc}{{ c}}
\newcommand{\bfg}{{ g}}
\newcommand{\bfh}{{ h}}
\newcommand{\bfla}{{ \lambda}}

\tikzset{->1-/.style={decoration={
			markings,
			mark=at position #1 with {\arrow{Triangle[fill=black, width=4pt, length=4pt]}}},postaction={decorate}}}

\newcommand{\convONE}[4]{
	\begin{tikzpicture}[scale=#1,baseline=0.4em]
	\coordinate (0) at (0,{sin(60)});
	\coordinate (1) at ({cos(60)},0);
	\coordinate (2) at (-{cos(60)},0);
	\coordinate (3) at (0,0.3);
	
	\draw[] (0) to (1);
	\draw[] (1) to (2);
	\draw[] (2) to (0);
	
	\node[] at (3) {{$\otimes$}};
	
	\node[above=-0.2em] at (0) {{$\scriptstyle{#2}$}};
	\node[right=-0.2em] at (1) {{$\scriptstyle{#3}$}};	
	\node[left=-0.2em] at (2) {{$\scriptstyle{#4}$}};
	\end{tikzpicture}
}

\newcommand{\locONE}[2]{
	\begin{tikzpicture}[scale=#1,baseline=-0.3em]
	\coordinate (0) at (-1,0);
	\coordinate (1) at (0,0);
	\coordinate (2) at (1,0);
	
	\ifthenelse{#2 = 1}{
		\draw[bend right=40] (0) to (1);
		\draw[bend left =40] (1) to (2);}
	{	
		\draw[] (0) to (1);
		\draw[] (1) to (2);
	}
	
	\node[draw,circle,scale=0.3,fill] at (0) {};
	\node[draw,circle,scale=0.3,fill] at (2) {};
	
	\node[above] at (0) {{$\scriptstyle{0}$}};	
	\node[above] at (2) {{$\scriptstyle{2}$}};
	\end{tikzpicture}
}

\newcommand{\locTHREE}[2]{
	\begin{tikzpicture}[scale=#1,baseline=-0.3em]
	\coordinate (0) at (0,0);
	\coordinate (1) at (0,0.6);
	\coordinate (2) at (0,-0.6);

	\ifthenelse{#2 = 1}{
		\draw[] (1) to (2);
		\node[draw,circle,scale=0.3,fill] at (1) {};
		\node[draw,circle,scale=0.3,fill] at (2) {};
		\node[above] at (1) {{$\scriptstyle{1}$}};
		\node[below] at (2) {{$\scriptstyle{2}$}};
	}
	{
		\draw (0,0) circle (0.6);
		\draw[] (0) to (1);
		\draw[] (1) to (2);
		\node[draw,circle,scale=0.3,fill] at (0) {};
		\node[left] at (0) {{$\scriptstyle{0}$}};
		\node[above] at (1) {{$\scriptstyle{1}$}};
		\node[below] at (2) {{$\scriptstyle{2}$}};	
	}
	\end{tikzpicture}
}

\newcommand{\locFIVE}[2]{
	\begin{tikzpicture}[scale=#1,baseline=-0.3em]
	\coordinate (0) at (-1,0);
	\coordinate (1) at (0,0);
	\coordinate (2) at (1,0);
	
	\ifthenelse{#2 = 0}{
		\draw[] (0) to (1);
		\draw[densely dashed] (1) to (2);}
	{	
		\draw[densely dashed] (0) to (1);
		\draw[] (1) to (2);
	}
	
	\node[draw,circle,scale=0.3,fill] at (0) {};
	\node[draw,circle,scale=0.3,fill] at (1) {};
	\node[draw,circle,scale=0.3,fill] at (2) {};
	
	\node[above] at (0) {{$\scriptstyle{0}$}};
	\node[above] at (1) {{$\scriptstyle{1}$}};	
	\node[above] at (2) {{$\scriptstyle{2}$}};
	\end{tikzpicture}
}

\newcommand{\squ}[3]{
	\begin{tikzpicture}[scale=#2,baseline=-0.3em]
	\coordinate (0) at (0,0.5);
	\coordinate (1) at (1,0);
	\coordinate (2) at (0,-0.5);
	\coordinate (3) at (-1,0);
	
	\draw[] (0) to (1);
	\draw[] (1) to (2);
	\draw[] (2) to (3);
	\draw[] (3) to (0);

	\ifthenelse{#1 = 02 \OR #1 = 20}{\draw[] (0) to (2);
		\node[draw,circle,scale=0.3,fill] at (1) {};
		\node[draw,circle,scale=0.3,fill] at (3) {};}{}
	\ifthenelse{#1 = 13 \OR #1 = 31}{\draw[] (1) to (3);
		\node[draw,circle,scale=0.3,fill] at (0) {};
		\node[draw,circle,scale=0.3,fill] at (2) {};}{}

	\ifthenelse{#3 = 1}{
		\node[above] at (0) {{$\scriptstyle{0}$}};	
		\node[right] at (1) {{$\scriptstyle{1}$}};
		\node[below] at (2) {{$\scriptstyle{2}$}};
		\node[left] at (3) {{$\scriptstyle{3}$}};}{}
	\end{tikzpicture}
}

\newcommand{\pent}[4]{
	\begin{tikzpicture}[scale=#3,baseline=-0.3em]
	\coordinate (0) at(0,1);
	\coordinate (1) at ({sin(72)} ,{cos(72)} );
	\coordinate (2) at ({sin(144)} ,{-cos(36)} );
	\coordinate (3) at ({-sin(144)} ,{-cos(36)} );
	\coordinate (4) at ({-sin(72)} ,{cos(72)} );
	
	\draw[] (0) to (1);
	\draw[] (1) to (2);
	\draw[] (2) to (3);
	\draw[] (3) to (4);
	\draw[] (4) to (0);
	
	\foreach \x in {#1, #2}
	{
		\ifthenelse{\x = 20 \OR \x = 02}{\draw[] (0) to (2);}{}
		\ifthenelse{\x = 30 \OR \x = 03}{\draw[] (0) to (3);}{}
		\ifthenelse{\x = 13 \OR \x = 31}{\draw[] (1) to (3);}{}
		\ifthenelse{\x = 14 \OR \x = 41}{\draw[] (1) to (4);}{}
		\ifthenelse{\x = 24 \OR \x = 42}{\draw[] (2) to (4);}{}
	}
	
	\ifthenelse{#4 = 1}{
	\node[above] at (0) {{$\scriptstyle{0}$}};	
	\node[right] at (1) {{$\scriptstyle{1}$}};
	\node[below] at (2) {{$\scriptstyle{2}$}};
	\node[below] at (3) {{$\scriptstyle{3}$}};
	\node[left] at (4) {{$\scriptstyle{4}$}};}{}	
	\end{tikzpicture}
}

\newcommand{\hex}[6]{
	\begin{tikzpicture}[scale=#4,baseline=-0.3em]
	\coordinate (0) at (0,1);
	\coordinate (1) at ({sin(60)} ,{cos(60)} );
	\coordinate (2) at ({sin(60)} ,{-cos(60)} );
	\coordinate (3) at (0,-1);
	\coordinate (4) at ({-sin(60)} ,{-cos(60)} );
	\coordinate (5) at ({-sin(60)} ,{cos(60)} );
	
	\draw[opacity = #6] (0) to (1);
	\draw[opacity = #6] (1) to (2);
	\draw[opacity = #6] (2) to (3);
	\draw[opacity = #6] (3) to (4);
	\draw[opacity = #6] (4) to (5);
	\draw[opacity = #6] (5) to (0);
	
	\foreach \x in {#1, #2, #3}
	{
		\ifthenelse{\x = 20 \OR \x = 02}{\draw[opacity = #6] (0) to (2);}{}
		\ifthenelse{\x = 30 \OR \x = 03}{\draw[opacity = #6] (0) to (3);}{}
		\ifthenelse{\x = 40 \OR \x = 04}{\draw[opacity = #6] (0) to (4);}{}
		\ifthenelse{\x = 13 \OR \x = 31}{\draw[opacity = #6] (1) to (3);}{}
		\ifthenelse{\x = 14 \OR \x = 41}{\draw[opacity = #6] (1) to (4);}{}
		\ifthenelse{\x = 15 \OR \x = 51}{\draw[opacity = #6] (1) to (5);}{}
		\ifthenelse{\x = 24 \OR \x = 42}{\draw[opacity = #6] (2) to (4);}{}
		\ifthenelse{\x = 25 \OR \x = 52}{\draw[opacity = #6] (2) to (5);}{}
		\ifthenelse{\x = 35 \OR \x = 53}{\draw[opacity = #6] (3) to (5);}{}
	}
	
	\ifthenelse{#5 = 1}{
	\node[above] at (0) {{$\scriptstyle{0}$}};	
	\node[right] at (1) {{$\scriptstyle{1}$}};
	\node[right] at (2) {{$\scriptstyle{2}$}};
	\node[below] at (3) {{$\scriptstyle{3}$}};
	\node[left] at (4) {{$\scriptstyle{4}$}};
	\node[left] at (5) {{$\scriptstyle{5}$}};}{}	
	\end{tikzpicture}
}

\newcommand{\PPONE}[2]{
	\begin{tikzpicture}[scale=#1,baseline=-0.3em]
	\coordinate (0) at(0,1);
	\coordinate (1) at ({sin(72)} ,{cos(72)} );
	\coordinate (2) at ({sin(144)} ,{-cos(36)} );
	\coordinate (3) at ({-sin(144)} ,{-cos(36)} );
	\coordinate (4) at ({-sin(72)} ,{cos(72)} );
	
	\draw[] (0) to (1);
	\draw[] (1) to (2);
	\draw[] (2) to (3);
	\draw[] (3) to (4);
	\draw[] (4) to (0);
	
	\draw[] (0) to (3);
	\draw[] (1) to (3);
	\coordinate (a) at (intersection of 3--1 and 0--2);
	\draw[shorten >= 0.3em] (0) to (a);
	\draw[shorten <= 0.3em] (a) to (2);
	\coordinate (b) at (intersection of 0--3 and 4--2);
	\coordinate (c) at (intersection of 3--1 and 4--2);
	\draw[shorten >= 0.3em] (4) to (b);
	\draw[shorten <= 0.3em, shorten >= 0.3em] (b) to (c);
	\draw[shorten <= 0.3em] (c) to (2);
	
	\ifthenelse{#2 = 1}{
	\node[above] at (0) {{$\scriptstyle{0}$}};	
	\node[right] at (1) {{$\scriptstyle{1}$}};
	\node[below] at (2) {{$\scriptstyle{2}$}};
	\node[below] at (3) {{$\scriptstyle{3}$}};
	\node[left] at (4) {{$\scriptstyle{4}$}};}{}	
	\end{tikzpicture}
}

\newcommand{\PPTWO}[3]{
	\begin{tikzpicture}[scale=#1,baseline=#3 em]
	\coordinate (0) at(0,1);
	\coordinate (1) at ({sin(72)} ,{cos(72)} );
	\coordinate (2) at ({sin(144)} ,{-cos(36)} );
	\coordinate (3) at ({-sin(144)} ,{-cos(36)} );
	\coordinate (4) at ({-sin(72)} ,{cos(72)} );
	
	\draw[] (0) to (1);
	\draw[] (1) to (2);
	\draw[] (2) to (3);
	\draw[] (3) to (4);
	\draw[] (4) to (0);
	
	\draw[] (0) to (3);
	\draw[] (1) to (3);
	\coordinate (b) at (intersection of 3--1 and 0--2);
	\coordinate (a) at (intersection of 4--1 and 0--2);
	\draw[shorten >= 0.3em] (0) to (a);
	\draw[shorten <= 0.3em, shorten >= 0.3em] (a) to (b);
	\draw[shorten <= 0.3em] (b) to (2);
	\coordinate (c) at (intersection of 0--3 and 4--2);
	\coordinate (d) at (intersection of 3--1 and 4--2);
	\draw[shorten >= 0.3em] (4) to (c);
	\draw[shorten <= 0.3em, shorten >= 0.3em] (c) to (d);
	\draw[shorten <= 0.3em] (d) to (2);
	\coordinate (e) at (intersection of 0--3 and 4--1);
	\draw[shorten >= 0.3em] (4) to (e);
	\draw[shorten <= 0.3em] (e) to (1);
	
	\ifthenelse{#2 = 1}{	
	\node[above] at (0) {{$\scriptstyle{0}$}};	
	\node[right] at (1) {{$\scriptstyle{1}$}};
	\node[below] at (2) {{$\scriptstyle{2}$}};
	\node[below] at (3) {{$\scriptstyle{3}$}};
	\node[left] at (4) {{$\scriptstyle{4}$}};}{}	
	\end{tikzpicture}
}

\newcommand{\PPPONE}[1]{
	\begin{tikzpicture}[scale=1,baseline=-0.3em]
	\coordinate (0) at (0,1);
	\coordinate (1) at ({sin(60)} ,{cos(60)} );
	\coordinate (2) at ({sin(60)} ,{-cos(60)} );
	\coordinate (3) at (0,-1);
	\coordinate (4) at ({-sin(60)} ,{-cos(60)} );
	\coordinate (5) at ({-sin(60)} ,{cos(60)} );
	
	\draw[] (0) to (1);
	\draw[] (1) to (2);
	\draw[] (2) to (3);
	\draw[] (3) to (4);
	\draw[] (4) to (5);
	\draw[] (5) to (0);
	
	\draw[] (1) to (3);
	\draw[] (0) to (3);
	\draw[] (5) to (3);
	\coordinate (a) at (intersection of 3--1 and 0--2);
	\draw[shorten >= 0.3em] (0) to (a);
	\draw[shorten <= 0.3em, shorten >= 0em] (a) to (2);
	\coordinate (b) at (intersection of 5--3 and 4--2);
	\coordinate (c) at (intersection of 0--3 and 4--2);
	\coordinate (d) at (intersection of 3--1 and 4--2);
	\draw[shorten >= 0.3em] (4) to (b);
	\draw[shorten <= 0.3em, shorten >= 0.3em] (b) to (c);
	\draw[shorten <= 0.3em, shorten >= 0.3em] (c) to (d);
	\draw[shorten <= 0.3em] (d) to (2);
	\coordinate (e) at (intersection of 0--3 and 5--2);
	\coordinate (f) at (intersection of 3--1 and 5--2);
	\draw[shorten >= 0.3em] (5) to (e);
	\draw[shorten <= 0.3em, shorten >= 0.3em] (e) to (f);
	\draw[shorten <= 0.3em] (f) to (2);

	\ifthenelse{#1 = 1}{
	\node[above] at (0) {{$\scriptstyle{0}$}};	
	\node[right] at (1) {{$\scriptstyle{1}$}};
	\node[right] at (2) {{$\scriptstyle{2}$}};
	\node[below] at (3) {{$\scriptstyle{3}$}};
	\node[left] at (4) {{$\scriptstyle{4}$}};
	\node[left] at (5) {{$\scriptstyle{5}$}};}{}
	\end{tikzpicture}
}

\newcommand{\PPPTWO}[1]{
	\begin{tikzpicture}[scale=1,baseline=-0.3em]
	\coordinate (0) at (0,1);
	\coordinate (1) at ({sin(60)} ,{cos(60)} );
	\coordinate (2) at ({sin(60)} ,{-cos(60)} );
	\coordinate (3) at (0,-1);
	\coordinate (4) at ({-sin(60)} ,{-cos(60)} );
	\coordinate (5) at ({-sin(60)} ,{cos(60)} );
	
	\draw[] (0) to (1);
	\draw[] (1) to (2);
	\draw[] (2) to (3);
	\draw[] (3) to (4);
	\draw[] (4) to (5);
	\draw[] (5) to (0);
	
	\draw[] (1) to (3);
	\draw[] (0) to (3);
	\draw[] (5) to (3);
	\coordinate (a) at (intersection of 3--1 and 0--2);
	\draw[shorten >= 0.3em] (0) to (a);
	\draw[shorten <= 0.3em, shorten >= 0em] (a) to (2);
	\coordinate (b) at (intersection of 5--3 and 4--2);
	\coordinate (c) at (intersection of 0--3 and 4--2);
	\coordinate (d) at (intersection of 3--1 and 4--2);
	\draw[shorten >= 0.3em] (4) to (b);
	\draw[shorten <= 0.3em, shorten >= 0.3em] (b) to (c);
	\draw[shorten <= 0.3em, shorten >= 0.3em] (c) to (d);
	\draw[shorten <= 0.3em] (d) to (2);
	\coordinate (h) at (intersection of 0--4 and 5--2);
	\coordinate (e) at (intersection of 0--3 and 5--2);
	\coordinate (f) at (intersection of 3--1 and 5--2);
	\draw[shorten >= 0.3em] (5) to (h);
	\draw[shorten >= 0.3em, shorten <= 0.3em] (h) to (e);
	\draw[shorten <= 0.3em, shorten >= 0.3em] (e) to (f);
	\draw[shorten <= 0.3em] (f) to (2);
	\coordinate (g) at (intersection of 0--4 and 5--3);
	\draw[shorten >= 0.3em] (0) to (g);
	\draw[shorten <= 0.3em, shorten >= 0.3em] (g) to (4);
	
	\ifthenelse{#1 = 1}{
	\node[above] at (0) {{$\scriptstyle{0}$}};	
	\node[right] at (1) {{$\scriptstyle{1}$}};
	\node[right] at (2) {{$\scriptstyle{2}$}};
	\node[below] at (3) {{$\scriptstyle{3}$}};
	\node[left] at (4) {{$\scriptstyle{4}$}};
	\node[left] at (5) {{$\scriptstyle{5}$}};}{}
	\end{tikzpicture}
}

\newcommand{\PPPTHREE}[1]{
	\begin{tikzpicture}[scale=1,baseline=-0.3em]
	\coordinate (0) at (0,1);
	\coordinate (1) at ({sin(60)} ,{cos(60)} );
	\coordinate (2) at ({sin(60)} ,{-cos(60)} );
	\coordinate (3) at (0,-1);
	\coordinate (4) at ({-sin(60)} ,{-cos(60)} );
	\coordinate (5) at ({-sin(60)} ,{cos(60)} );
	
	\draw[] (0) to (1);
	\draw[] (1) to (2);
	\draw[] (2) to (3);
	\draw[] (3) to (4);
	\draw[] (4) to (5);
	\draw[] (5) to (0);
	
	\draw[] (1) to (3);
	\draw[] (0) to (3);
	\draw[] (5) to (3);
	\coordinate (a) at (intersection of 3--1 and 0--2);
	\coordinate (k) at (intersection of 1--4 and 0--2);
	\draw[shorten >= 0.3em] (0) to (k);
	\draw[shorten <= 0.3em, shorten >= 0.3em] (k) to (a);
	\draw[shorten <= 0.3em] (a) to (2);
	\coordinate (b) at (intersection of 5--3 and 4--2);
	\coordinate (c) at (intersection of 0--3 and 4--2);
	\coordinate (d) at (intersection of 3--1 and 4--2);
	\draw[shorten >= 0.3em] (4) to (b);
	\draw[shorten <= 0.3em, shorten >= 0.3em] (b) to (c);
	\draw[shorten <= 0.3em, shorten >= 0.3em] (c) to (d);
	\draw[shorten <= 0.3em] (d) to (2);
	\coordinate (h) at (intersection of 0--4 and 5--2);
	\coordinate (e) at (intersection of 0--3 and 5--2);
	\coordinate (f) at (intersection of 3--1 and 5--2);
	\draw[shorten >= 0.3em] (5) to (h);
	\draw[shorten >= 0.3em, shorten <= 0.3em] (h) to (e);
	\draw[shorten <= 0.3em, shorten >= 0.3em] (e) to (f);
	\draw[shorten <= 0.3em] (f) to (2);
	\coordinate (g) at (intersection of 0--4 and 5--3);
	\draw[shorten >= 0.3em] (0) to (g);
	\draw[shorten <= 0.3em, shorten >= 0.3em] (g) to (4);
	\coordinate (i) at (intersection of 0--3 and 1--4);
	\coordinate (j) at (intersection of 5--3 and 1--4);
	\draw[shorten >= 0.3em] (1) to (i);
	\draw[shorten <= 0.3em, shorten >= 0.3em] (i) to (j);
	\draw[shorten <= 0.3em] (j) to (4);

	\ifthenelse{#1 = 1}{	
	\node[above] at (0) {{$\scriptstyle{0}$}};	
	\node[right] at (1) {{$\scriptstyle{1}$}};
	\node[right] at (2) {{$\scriptstyle{2}$}};
	\node[below] at (3) {{$\scriptstyle{3}$}};
	\node[left] at (4) {{$\scriptstyle{4}$}};
	\node[left] at (5) {{$\scriptstyle{5}$}};}{}
	\end{tikzpicture}
}

\newcommand{\PPPFOUR}[1]{
	\begin{tikzpicture}[scale=1,baseline=-0.3em]
	\coordinate (0) at (0,1);
	\coordinate (1) at ({sin(60)} ,{cos(60)} );
	\coordinate (2) at ({sin(60)} ,{-cos(60)} );
	\coordinate (3) at (0,-1);
	\coordinate (4) at ({-sin(60)} ,{-cos(60)} );
	\coordinate (5) at ({-sin(60)} ,{cos(60)} );
	
	\draw[] (0) to (1);
	\draw[] (1) to (2);
	\draw[] (2) to (3);
	\draw[] (3) to (4);
	\draw[] (4) to (5);
	\draw[] (5) to (0);
	
	\draw[] (1) to (3);
	\draw[] (0) to (3);
	\draw[] (5) to (3);
	\coordinate (a) at (intersection of 3--1 and 0--2);
	\coordinate (k) at (intersection of 1--4 and 0--2);
	\coordinate (m) at (intersection of 1--5 and 0--2);
	\draw[shorten >= 0.3em] (0) to (m);
	 \draw[shorten <= 0.3em, shorten >= 0.3em] (m) to (k);
	\draw[shorten <= 0.3em, shorten >= 0.3em] (k) to (a);
	\draw[shorten <= 0.3em] (a) to (2);
	\coordinate (b) at (intersection of 5--3 and 4--2);
	\coordinate (c) at (intersection of 0--3 and 4--2);
	\coordinate (d) at (intersection of 3--1 and 4--2);
	\draw[shorten >= 0.3em] (4) to (b);
	\draw[shorten <= 0.3em, shorten >= 0.3em] (b) to (c);
	\draw[shorten <= 0.3em, shorten >= 0.3em] (c) to (d);
	\draw[shorten <= 0.3em] (d) to (2);
	\coordinate (h) at (intersection of 0--4 and 5--2);
	\coordinate (e) at (intersection of 0--3 and 5--2);
	\coordinate (f) at (intersection of 3--1 and 5--2);
	\draw[shorten >= 0.3em] (5) to (h);
	\draw[shorten >= 0.3em, shorten <= 0.3em] (h) to (e);
	\draw[shorten <= 0.3em, shorten >= 0.3em] (e) to (f);
	\draw[shorten <= 0.3em] (f) to (2);
	\coordinate (g) at (intersection of 0--4 and 5--3);
	\draw[shorten >= 0.3em] (0) to (g);
	\draw[shorten <= 0.3em, shorten >= 0.3em] (g) to (4);
	\coordinate (i) at (intersection of 0--3 and 1--4);
	\coordinate (j) at (intersection of 5--3 and 1--4);
	\draw[shorten >= 0.3em] (1) to (i);
	\draw[shorten <= 0.3em, shorten >= 0.3em] (i) to (j);
	\draw[shorten <= 0.3em] (j) to (4);
	\coordinate (l) at (intersection of 0--4 and 5--1);
	\coordinate (m) at (intersection of 0--3 and 5--1);
	\draw[shorten >= 0.3em] (5) to (l);
	\draw[shorten <= 0.3em, shorten >= 0.3em] (l) to (m);
	\draw[shorten <= 0.3em] (m) to (1);
	
	\ifthenelse{#1 = 1}{	
	\node[above] at (0) {{$\scriptstyle{0}$}};	
	\node[right] at (1) {{$\scriptstyle{1}$}};
	\node[right] at (2) {{$\scriptstyle{2}$}};
	\node[below] at (3) {{$\scriptstyle{3}$}};
	\node[left] at (4) {{$\scriptstyle{4}$}};
	\node[left] at (5) {{$\scriptstyle{5}$}};}{}
	\end{tikzpicture}
}

\newcommand{\PPPFIVE}[1]{
	\begin{tikzpicture}[scale=1,baseline=-0.3em]
	\coordinate (0) at (0,1);
	\coordinate (1) at ({sin(60)} ,{cos(60)} );
	\coordinate (2) at ({sin(60)} ,{-cos(60)} );
	\coordinate (3) at (0,-1);
	\coordinate (4) at ({-sin(60)} ,{-cos(60)} );
	\coordinate (5) at ({-sin(60)} ,{cos(60)} );
	
	\draw[] (0) to (1);
	\draw[] (1) to (2);
	\draw[] (2) to (3);
	\draw[] (3) to (4);
	\draw[] (4) to (5);
	\draw[] (5) to (0);
	
	\draw[] (1) to (3);
	\draw[] (0) to (3);
	\draw[] (5) to (3);
	\coordinate (a) at (intersection of 3--1 and 0--2);
	\coordinate (k) at (intersection of 1--4 and 0--2);
	\coordinate (m) at (intersection of 1--5 and 0--2);
	\draw[shorten >= 0.3em] (0) to (m);
	\draw[shorten <= 0.3em, shorten >= 0.3em] (m) to (k);
	\draw[shorten <= 0.3em, shorten >= 0.3em] (k) to (a);
	\draw[shorten <= 0.3em] (a) to (2);
	\coordinate (b) at (intersection of 5--3 and 4--2);
	\coordinate (c) at (intersection of 0--3 and 4--2);
	\coordinate (d) at (intersection of 3--1 and 4--2);
	\draw[shorten >= 0.3em] (4) to (b);
	\draw[shorten <= 0.3em, shorten >= 0.3em] (b) to (c);
	\draw[shorten <= 0.3em, shorten >= 0.3em] (c) to (d);
	\draw[shorten <= 0.3em] (d) to (2);
	\coordinate (e) at (intersection of 0--3 and 5--2);
	\coordinate (f) at (intersection of 3--1 and 5--2);
	\draw[shorten >= 0.3em] (5) to (e);
	\draw[shorten <= 0.3em, shorten >= 0.3em] (e) to (f);
	\draw[shorten <= 0.3em] (f) to (2);
	\coordinate (m) at (intersection of 0--3 and 5--1);
	\draw[shorten >= 0.3em] (5) to (m);
	\draw[shorten <= 0.3em] (m) to (1);
	\coordinate (i) at (intersection of 0--3 and 1--4);
	\coordinate (j) at (intersection of 5--3 and 1--4);
	\draw[shorten >= 0.3em] (1) to (i);
	\draw[shorten <= 0.3em, shorten >= 0.3em] (i) to (j);
	\draw[shorten <= 0.3em] (j) to (4);
	
	\ifthenelse{#1 = 1}{	
	\node[above] at (0) {{$\scriptstyle{0}$}};	
	\node[right] at (1) {{$\scriptstyle{1}$}};
	\node[right] at (2) {{$\scriptstyle{2}$}};
	\node[below] at (3) {{$\scriptstyle{3}$}};
	\node[left] at (4) {{$\scriptstyle{4}$}};
	\node[left] at (5) {{$\scriptstyle{5}$}};}{}
	\end{tikzpicture}
}

\newcommand{\PPPSIX}[1]{
	\begin{tikzpicture}[scale=1,baseline=-0.3em]
	\coordinate (0) at (0,1);
	\coordinate (1) at ({sin(60)} ,{cos(60)} );
	\coordinate (2) at ({sin(60)} ,{-cos(60)} );
	\coordinate (3) at (0,-1);
	\coordinate (4) at ({-sin(60)} ,{-cos(60)} );
	\coordinate (5) at ({-sin(60)} ,{cos(60)} );
	
	\draw[] (0) to (1);
	\draw[] (1) to (2);
	\draw[] (2) to (3);
	\draw[] (3) to (4);
	\draw[] (4) to (5);
	\draw[] (5) to (0);
	
	\draw[] (1) to (3);
	\draw[] (0) to (3);
	\draw[] (5) to (3);
	\coordinate (a) at (intersection of 3--1 and 0--2);
	\coordinate (m) at (intersection of 1--5 and 0--2);
	\draw[shorten >= 0.3em] (0) to (m);
	\draw[shorten <= 0.3em, shorten >= 0.3em] (m) to (a);
	\draw[shorten <= 0.3em] (a) to (2);
	\coordinate (b) at (intersection of 5--3 and 4--2);
	\coordinate (c) at (intersection of 0--3 and 4--2);
	\coordinate (d) at (intersection of 3--1 and 4--2);
	\draw[shorten >= 0.3em] (4) to (b);
	\draw[shorten <= 0.3em, shorten >= 0.3em] (b) to (c);
	\draw[shorten <= 0.3em, shorten >= 0.3em] (c) to (d);
	\draw[shorten <= 0.3em] (d) to (2);
	\coordinate (e) at (intersection of 0--3 and 5--2);
	\coordinate (f) at (intersection of 3--1 and 5--2);
	\draw[shorten >= 0.3em] (5) to (e);
	\draw[shorten <= 0.3em, shorten >= 0.3em] (e) to (f);
	\draw[shorten <= 0.3em] (f) to (2);
	\coordinate (m) at (intersection of 0--3 and 5--1);
	\draw[shorten >= 0.3em] (5) to (m);
	\draw[shorten <= 0.3em] (m) to (1);
	
	\ifthenelse{#1 = 1}{	
	\node[above] at (0) {{$\scriptstyle{0}$}};	
	\node[right] at (1) {{$\scriptstyle{1}$}};
	\node[right] at (2) {{$\scriptstyle{2}$}};
	\node[below] at (3) {{$\scriptstyle{3}$}};
	\node[left] at (4) {{$\scriptstyle{4}$}};
	\node[left] at (5) {{$\scriptstyle{5}$}};}{}
	\end{tikzpicture}
}

\newcommand{\assocTWO}[0]{
	\begin{tikzpicture}[scale=1,baseline=0em]
	\matrix[matrix of math nodes,row sep =-1.3em,column sep=0em, ampersand replacement=\&] (m) {
	{} \& {} \& {} \& {} \& {} \& {} \& {}
	\& \hex{13}{35}{51}{0.4}{0}{1}
	{} \& {} \& {} \& {} \& {} \& {} \& {}
	\\
	{} \& {} 
	\& \hex{15}{14}{13}{0.4}{0}{1}
	\& {} \& {} \& {} \& {} \& {} \& {} \& {} \& {} \& {}
	\& \hex{51}{52}{53}{0.4}{0}{1}
	\& {} \& {}
	\\
	{} \& {} \& {} \& {} \& {} \& {} \& {}
	\& \hex{35}{30}{31}{0.4}{0}{1}
	{} \& {} \& {} \& {} \& {} \& {} \& {}
	\\
	\hex{04}{41}{13}{0.4}{0}{1}
	\& {} \& {} \& {} 
	\& \hex{51}{14}{42}{0.4}{0}{0.4}
	\& {} \& {} \& {} \& {} \& {}
	\& \hex{15}{52}{24}{0.4}{0}{0.4}
	\& {} \& {} \& {} 
	\& \hex{02}{25}{53}{0.4}{0}{1}
	\\
	{} \& {} \& {} \& {} \& {}
	\& \hex{40}{03}{31}{0.4}{0}{1}
	\& {} \& {} \& {}
	\& \hex{53}{30}{02}{0.4}{0}{1}
	{} \& {} \& {} \& {} \& {}
	\\
	{} \& {} 
	\& \hex{40}{41}{42}{0.4}{0}{1}
	\& {} \& {} \& {} \& {} \& {} \& {} \& {} \& {} \& {}
	\& \hex{20}{25}{24}{0.4}{0}{1}
	\& {} \& {}
	\\
	{} \& {} \& {} \& {} \& {} \& {} \& {}
	\& \hex{04}{03}{02}{0.4}{0}{1}
	{} \& {} \& {} \& {} \& {} \& {} \& {}
	\\
	{} \& {} \& {} \& {} \& {} \& {} \& {}
	\& \phantom{\hex{13}{35}{51}{0.4}{0}{1}}
	{} \& {} \& {} \& {} \& {} \& {} \& {}
	\\
	{} \& {} \& {} \& {} \& {} \& {} \& {}
	\& \hex{40}{02}{24}{0.4}{0}{1}
	{} \& {} \& {} \& {} \& {} \& {} \& {} \\
	};

	\path
		(m-1-8) edge[shorten <= -0.40em, shorten >= -0.40em] (m-2-3)
			edge[shorten <= -0.40em, shorten >= -0.40em] (m-2-13)
			edge[shorten <= -0.40em, shorten >= -0.40em] (m-3-8)
		(m-3-8) edge[shorten <= -0.99em, shorten >= -0.99em] (m-5-6)
			edge[shorten <= -0.99em, shorten >= -0.99em] (m-5-10)
		(m-2-3) edge[shorten <= -0.99em, shorten >= -0.99em] (m-4-1)
			edge[shorten <= -0.99em, shorten >= -0.99em, opacity = 0.4] (m-4-5)
		(m-2-13) edge[shorten <= -0.99em, shorten >= -0.99em, opacity = 0.4] (m-4-11)
			edge[shorten <= -0.99em, shorten >= -0.99em] (m-4-15)
		(m-5-6) edge[shorten <= -0.99em, shorten >= -0.99em] (m-7-8)
			edge[shorten <= -0.40em, shorten >= -0.40em] (m-4-1)
		(m-5-10) edge[shorten <= -0.99em, shorten >= -0.99em] (m-7-8)
			edge[shorten <= -0.40em, shorten >= -0.40em] (m-4-15)
		(m-9-8) edge[shorten <= -0.40em, shorten >= -0.40em] (m-7-8)
			edge[shorten <= -0.44em, shorten >= -0.44em] (m-6-3)
			edge[shorten <= -0.44em, shorten >= -0.44em] (m-6-13)
		(m-6-3) edge[shorten <= -0.99em, shorten >= -0.99em] (m-4-1)
			edge[shorten <= -0.99em, shorten >= -0.99em, opacity = 0.4] (m-4-5)
		(m-6-13) edge[shorten <= -0.99em, shorten >= -0.99em] (m-4-15)
			edge[shorten <= -0.99em, shorten >= -0.99em, opacity = 0.4] (m-4-11)
		(m-4-5) edge[opacity = 0.4, shorten <= -0.41em, shorten >= -0.41em] (m-4-11);
	\end{tikzpicture}
}

\newcommand{\triONE}[2]{
	\begin{tikzpicture}[scale=#1,baseline=0.4em]
	\coordinate (0) at (0,{sin(60)});
	\coordinate (1) at ({cos(60)},0);
	\coordinate (2) at (-{cos(60)},0);
	
	\draw[] (0) to (1);
	\draw[] (1) to (2);
	\draw[] (2) to (0);

	\node[above=-0.2em] at (0) {{$\scriptstyle{1}$}};
	\node[right=-0.2em] at (1) {{$\scriptstyle{2}$}};
	\node[left=-0.2em] at (2) {{$\scriptstyle{0}$}};	
	
	\end{tikzpicture}
}

\newcommand{\triTHREE}[2]{
	\begin{tikzpicture}[scale=#1,baseline=1em]
	\coordinate (0) at (0,{sin(60)});
	\coordinate (1) at ({cos(60)},0);
	\coordinate (2) at (-{cos(60)},0);
	\coordinate (3) at (-{cos(30)+cos(60)}, {sin(30)});
	\coordinate (4) at (intersection of 0--2 and 1--3);
	
	\draw[densely dashed] (0) to (4);
	\draw[] (2) to (4);
	\draw[] (1) to (2);
	\draw[] (1) to (4);
	\draw[] (0) to (1);

	\node[above=-0.2em] at (0) {{$\scriptstyle{1}$}};
	\node[right=-0.2em] at (1) {{$\scriptstyle{2}$}};
	\node[left=-0.2em] at (2) {{$\scriptstyle{0}$}};
	\node[left=-0.4em] at (3) {{$\scriptstyle{1'}$}};	
	\end{tikzpicture}
}

\newcommand{\triTWO}[2]{
	\begin{tikzpicture}[scale=#1,baseline=1em]
	\coordinate (0) at (0,{sin(60)});
	\coordinate (1) at ({cos(60)},0);
	\coordinate (2) at (-{cos(60)},0);
	\coordinate (3) at ({cos(30)-cos(60)}, {sin(30)});
	\coordinate (4) at (intersection of 0--1 and 2--3);
	
	\draw[densely dashed] (0) to (4);
	\draw[] (1) to (4);
	\draw[] (1) to (2);
	\draw[] (2) to (4);
	\draw[] (0) to (2);

	\node[above=-0.2em] at (0) {{$\scriptstyle{1'}$}};
	\node[right=-0.2em] at (1) {{$\scriptstyle{2}$}};
	\node[left=-0.2em] at (2) {{$\scriptstyle{0}$}};
	\node[right=-0.3em] at (3) {{$\scriptstyle{1}$}};
	\end{tikzpicture}
}

\newcommand{\triPachner}[2]{
	\begin{tikzpicture}[scale=#1,baseline=1em]
	\coordinate (0) at (0,{sin(60)});
	\coordinate (1) at ({cos(60)},0);
	\coordinate (2) at (-{cos(60)},0);
	\coordinate (4) at (0,0.35);
	
	\draw[] (0) to (4);
	\draw[] (1) to (4);
	\draw[] (1) to (2);
	\draw[] (2) to (4);
	\draw[] (0) to (2);
	\draw[] (0) to (1);

	\node[above=-0.2em] at (0) {{$\scriptstyle{1}$}};
	\node[right=-0.2em] at (1) {{$\scriptstyle{2}$}};
	\node[left=-0.2em] at (2) {{$\scriptstyle{3}$}};
	
	\ifthenelse{#2 = 0}{
		\node[below] at (4) {{$\scriptstyle{0}$}};}
		{\node[below] at (4) {{$\scriptstyle{0'}$}};}
	\end{tikzpicture}
}

\newcommand{\SPTtetraONE}[2]{
	\begin{tikzpicture}[scale=#1,baseline=2em]
	\coordinate (0) at (0,0);
	\coordinate (1) at (1,1);
	\coordinate (2) at (-2,0.8);
	\coordinate (3) at (-1, 3);
	\coordinate (4) at (intersection of 1--2 and 0--3);	
	\coordinate (5) at (0.2,1);
	\coordinate (6) at (-1,0.8);	
	\coordinate (7) at (-0.9,1.5);

	\ifthenelse{#2 = 1}{
		\draw[] (0)--(1) node[pos=0.35, right]{{$\scriptstyle{g_{12}}$}};
		\draw[shorten >= 0.3em] (1)--(4);
		\draw[shorten <= 0.3em] (4)--(2) node[pos=0.5, above=-0.2em]{};
		\draw[] (2)--(3) node[pos=0.5, left]{};
		\draw[] (0)--(3) node[pos=0.5, right]{};
		\draw[] (1)--(3) node[pos=0.5, right]{{$\scriptstyle{g_{01}}$}};
		\draw[] (0)--(2) node[pos=0.6, below]{{$\scriptstyle{g_{23}}$}};
		\node[above] at (5) {{$\scriptstyle{h_{012}}$}};
		\node[above] at (6) {{$\scriptstyle{h_{023}}$}};
		\node[above] at (7) {{$\scriptstyle{h_{013}}$}};
		\node[below] at (0) {{$\scriptstyle{2}$}};
		\node[right] at (1) {{$\scriptstyle{1}$}};
		\node[left] at (2) {{$\scriptstyle{3}$}};
		\node[above] at (3) {{$\scriptstyle{0}$}};	
	}{}

	\ifthenelse{#2 = 2}{
		\draw[] (0)--(1) node[pos=0.35, right]{{$\scriptstyle{\lambda_{12}}$}};
		\draw[shorten >= 0.3em] (1)--(4);
		\draw[shorten <= 0.3em] (4)--(2) node[pos=0.5, above=-0.2em]{{$\scriptstyle{\lambda_{13}}$}};
		\draw[] (2)--(3) node[pos=0.5, left]{{$\scriptstyle{\lambda_{03}}$}};
		\draw[] (0)--(3) node[pos=0.5, right]{{$\scriptstyle{\lambda_{02}}$}};
		\draw[] (1)--(3) node[pos=0.5, right]{{$\scriptstyle{\lambda_{01}}$}};
		\draw[] (0)--(2) node[pos=0.6, below]{{$\scriptstyle{\lambda_{23}}$}};	
		\node[below] at (0) {{$\scriptstyle{k_2}$}};
		\node[right] at (1) {{$\scriptstyle{k_1}$}};
		\node[left] at (2) {{$\scriptstyle{k_3}$}};
		\node[above] at (3) {{$\scriptstyle{k_0}$}};
	}{}

	\ifthenelse{#2 = 3}{
		\draw[] (0)--(1) node[pos=0.35, right]{{$\scriptstyle{g_{2}}$}};
		\draw[shorten >= 0.3em] (1)--(4);
		\draw[shorten <= 0.3em] (4)--(2) node[pos=0.5, above=-0.2em]{};
		\draw[] (2)--(3) node[pos=0.5, left]{};
		\draw[] (0)--(3) node[pos=0.5, right]{};
		\draw[] (1)--(3) node[pos=0.5, right]{{$\scriptstyle{g_{1}}$}};
		\draw[] (0)--(2) node[pos=0.6, below]{{$\scriptstyle{g_{3}}$}};
		\node[above] at (5) {{$\scriptstyle{h_{1}}$}};
		\node[above] at (6) {{$\scriptstyle{h_{2}}$}};
		\node[above] at (7) {{$\scriptstyle{h_{3}}$}};
		\node[below] at (0) {{$\scriptstyle{2}$}};
		\node[right] at (1) {{$\scriptstyle{1}$}};
		\node[left] at (2) {{$\scriptstyle{3}$}};
		\node[above] at (3) {{$\scriptstyle{0}$}};	
	}{}

	\end{tikzpicture}
}

\newcommand{\SPTtetraTWO}[2]{
	\begin{tikzpicture}[scale=#1,baseline=2em]
	\coordinate (0) at (0,0);
	\coordinate (1) at (1,1);
	\coordinate (2) at (-2,0.8);
	\coordinate (3) at (-1, 3);
	\coordinate (4) at (intersection of 1--2 and 0--3);	
	\coordinate (5) at (0.2,1);
	\coordinate (6) at (-1,1);	
	\coordinate (7) at (-0.9,1.5);
	\coordinate (8) at (-0.1,-1.7);
	\coordinate (8a) at (intersection of 3--8 and 0--2);	
	\coordinate (9) at (intersection of 3--8 and 4--2);
	\coordinate (10) at (-0.9,1.5);	
	\coordinate (11) at (0.3,1.5);
	\coordinate (12) at (-0.31,0.16);
	\coordinate (13) at (-0.7, 0.35);
	\coordinate (14) at (0.24, -0.6);
	\coordinate (15) at (-0.7, -0.6);		
	
	\draw[shorten >= 0.3em] (1)--(4);
	\draw[] (2)--(3) node[pos=0.5, left]{};
	\draw[] (0)--(3) node[pos=0.5, right]{};
	\node[right=-0.3em, below] at (0) {{$\scriptstyle{1}$}};
	\node[right] at (1) {{$\scriptstyle{2}$}};
	\node[left] at (2) {{$\scriptstyle{0}$}};
	\node[above] at (3) {{$\scriptstyle{3}$}};
	\node[below] at (8) {{$\scriptstyle{4}$}};		
	\draw[] (0)--(8);
	\draw[] (1)--(8);
	\draw[] (2)--(8);
	\draw[] (1)--(3) node[pos=0.5, right]{{$\scriptstyle{g_{3}}$}};
	\draw[] (0)--(2) node[pos=0.6, below]{{$\scriptstyle{g_{1}}$}};
	\draw[] (0)--(1) node[pos=0.17, right=-0.1em]{{$\scriptstyle{g_{2}}$}};
	
	\ifthenelse{#2 = 1}{
		\draw[shorten <= 0.3em] (4)--(2) node[pos=0.5, above=-0.2em]{};
		\node[above, rotate=0] at (5) {{$\scriptstyle{h_{1}}$}};
		\node[above, rotate=0] at (6) {{$\scriptstyle{h_{5}}$}};
		\node[above, rotate=0] at (13) {{$\scriptstyle{h_{4}}$}};
		\node[above, rotate=0] at (14) {{$\scriptstyle{h_{3}}$}};
		\node[above, rotate=0] at (15) {{$\scriptstyle{h_{6}}$}};	
	}{}

	\ifthenelse{#2 = 2}{
		\node[above, rotate=0] at (12) {{$\scriptstyle{h_{2}}$}};
		\node[below, rotate=0] at (10) {{$\scriptstyle{g_4}$}};
		\draw[shorten >= 0.3em] (3)--(8a);
		\draw[shorten <= 0.3em] (8a)--(8);	
		\draw[shorten <= 0.3em, shorten >= 0.3em] (4)--(9) node[pos=0.5, above=-0.2em]{};
		\draw[shorten >= 0.3em] (2)--(9) node[pos=0.5, above=-0.2em]{};
	}{}
		\end{tikzpicture}
}

\newcommand{\actionA}[3]{
	\begin{tikzpicture}[scale=#1,baseline=#3 em]
	\coordinate (0) at (0,0);
	\coordinate (1) at (1,1);
	\coordinate (2) at (-2,0.8);
	\coordinate (3) at (-1, 3);
	\coordinate (4) at (intersection of 1--2 and 0--3);	
	\coordinate (5) at (-0.6,0.6);
	\coordinate (a) at (intersection of 3--5 and 1--2);
	\coordinate (b) at (intersection of 1--5 and 0--3);

	\ifthenelse{#2 = 1}{
		\draw[] (0)--(1); 
		\draw[] (1)--(2);
		\draw[] (0)--(2); 
		\draw[] (5)--(0);
		\draw[] (5)--(1);
		\draw[] (5)--(2);
		\node[below] at (0) {{$\scriptstyle{2}$}};
		\node[right] at (1) {{$\scriptstyle{1}$}};
		\node[left] at (2) {{$\scriptstyle{3}$}};
		\node[below=-0.1em] at (5) {{$\scriptstyle{0}$}};	
	}{}	
	
	\ifthenelse{#2 = 2}{
		\draw[] (0)--(1);
		\draw[shorten >= 0.3em] (1)--(4);
		\draw[shorten >= 0.3em, shorten <= 0.3em] (4)--(a);
		\draw[shorten <= 0.3em] (a)--(2) node[pos=0.5, above=-0.2em]{};
		\draw[] (2)--(3) node[pos=0.5, left]{};
		\draw[] (0)--(3) node[pos=0.5, right]{};
		\draw[] (1)--(3);
		\draw[] (0)--(2);
		\draw[] (5)--(0);
		\draw[shorten >= 0.3em] (1)--(b);
		\draw[shorten <= 0.3em] (b)--(5);
		\draw[] (5)--(2);
		\draw[] (3)--(5);
		\node[below] at (0) {{$\scriptstyle{2}$}};
		\node[right] at (1) {{$\scriptstyle{1}$}};
		\node[left] at (2) {{$\scriptstyle{3}$}};
		\node[below=-0.1em] at (5) {{$\scriptstyle{0}$}};
		\node[above] at (3) {{$\scriptstyle{0'}$}};	
	}{}
	
	\end{tikzpicture}
}

\begin{document}

\title{From gauge to higher gauge models of topological phases}

\author{Clement Delcamp}
\email{cdelcamp@perimeterinstitute.ca}
\affiliation{Perimeter Institute for Theoretical Physics, Waterloo, Ontario N2L 2Y5, Canada}
\affiliation{Department of Physics and Astronomy, University of Waterloo, Waterloo, Ontario N2L 3G1, Canada}

\author{Apoorv Tiwari}
\email{t.apoorv@gmail.com}
\affiliation{Department of Physics and Institute for Condensed Matter Theory,
University of Illinois, 1110 W. Green Street, Urbana, IL 61801, USA}
\affiliation{Perimeter Institute for Theoretical Physics, Waterloo, Ontario N2L 2Y5, Canada}

\date{\today}

\begin{abstract}
    We consider exactly solvable models in (3+1)d whose ground states are described by topological lattice gauge theories. Using simplicial arguments, we emphasize how the consistency condition of the unitary map performing a local change of triangulation is equivalent to the coherence relation of the pentagonator 2-morphism of a monoidal 2-category. By weakening some axioms of such 2-category, we obtain a cohomological model whose underlying 1-category is a 2-group. Topological models from 2-groups together with their lattice realization are then studied from a higher gauge theory point of view. Symmetry protected topological phases protected by higher symmetry structures are explicitly constructed, and the gauging procedure which yields the corresponding topological gauge theories is discussed in detail. We finally study the correspondence between symmetry protected topological phases and 't Hooft anomalies in the context of these higher group symmetries.\\~\\~\\~\\
\end{abstract}

\maketitle

\tableofcontents

\newpage
\section{Introduction}
\label{sec:introduction}
\noindent 
Over the past few years, there has been a lot of progress in our understanding of \emph{quantum phases of matter} \cite{fradkin2013field, wen2004quantum}. A quantum phase may be defined as a path connected component in the space of \emph{models}. Since the language that currently most accurately describes quantum many body phases of matter is \emph{quantum field theory}, one may say that a quantum phase of matter is a path connected component in the space of quantum field theories. 

Such a space is very difficult to study in its full generality but one may make some progress by restricting to smaller and perhaps more manageable subspaces. This is often done by introducing some adjectives which specify what kind of models or phases we are interested in. These adjectives may refer to the spacetime dimension, the kind of matter involved such as fermionic or bosonic, the symmetry structures the theory is endowed with, broad descriptions of entanglement patterns such as \emph{short-range} or \emph{long-range entanglement}, and broad properties about the spectrum of the theory such as \emph{gapped} or \emph{gapless}.   

In this work we will always be interested in gapped phases of matter. Gapped phases are those that have a spectral gap, above the groundstate of the many-body Hamiltonian, that persists in the thermodynamics limit. Focusing on gapped phases greatly simplifies the tasks of classification and characterization due to the expectation that these phases are described by \emph{topological quantum field theories} (TQFTs) in the thermodynamic limit. In other words, all geometric or non-topological correlation functions are exponentially suppressed in some characteristic correlation length scale that depends on the microscopics of the model. Thus, if we consider a setup where the system size is much larger than any microscopic length scale of the system, we would expect that the only correlation functions that survive are topological in nature and can be captured by a topological theory. 
Describing a gapped phase is thus easier because TQFTs are much simpler than QFTs in many ways, e.g. their configuration space usually reduces to a finite sum from an integral over an (often divergent) infinite dimensional space.\footnote{For instance: Functional spaces for scalar theories, space of $q$-forms valued in some space $X$ for form theories, differential cohomology groups $\check{H}^{q+1}(M)$ for $q$-form ${\rm U}(1)$ gauge theories \cite{Freed:2006yc}, etc.} 
This being said, it is not completely clear that there is a bijection between TQFTs and physically realizable phases of matter, i.e whether all such theories can be realized by physically sensible Hamiltonian lattice models for example. In a recent beautiful work \cite{Gaiotto:2017zba} the relation between TQFTs and gapped phases of matter was carefully studied for theories with global symmetries.

TQFTs were originally mathematically formalized by Michael Atiyah \cite{Atiyah:1989vu} and were later shown to have a strong category theoretical flavour \cite{Baez:1995xq, lurie2009higher}. In (1+1)d and (2+1)d the categorical content of TQFTs have been studied in great detail. The situation is however much more complex in (3+1)d. In the present work, we are partly motivated by the study of the categorical structure underlying higher-dimensional models. We provide some evidence that the natural structures are based on categorical groups and their higher generalizations. This supports and extends the proposal by Lan et al. \onlinecite{lan2017classification}. In the above mentioned taxonomy of quantum phases of matter, such TQFTs will describe bosonic, gapped long-range entangled phases of matter without any symmetries. It is possible to discuss enrichment by symmetry \cite{Mesaros:2012yd, Barkeshli:2014cna, Chen:2017zzx, 2016PhRvB..94w5136H, Williamson:2017uzx}, however, we do not cover this topic in this manuscript. 

\bigskip \noindent 
There is a particularly tractable class of TQFTs which have a topological gauge theory interpretation. Given a ($d$+1)-manifold, the data that goes into defining them is simply a pair $(G, [\omega])$ where $G$ is a finite group and $[\omega]\in H^{d+1}(G,{\rm U}(1))$ a cohomology class \cite{dijkgraaf1990topological, Freed:1991bn}. Such cohomological models are typically defined as space-time state-sum models as in the original paper by Dijkgraaf and Witten \cite{dijkgraaf1990topological}. Considering a triangulation $\triangle$ with a $G$-coloring of the 1-simplices, the path integral is just a sum over the moduli space of principal $G$-bundles and the topological action is provided by a cocycle $\omega \in [\omega]$ evaluated on each ($d$+1)-simplices. Because of their gauge theory interpretation, it is particularly easy to define the corresponding Hamiltonian models \cite{Kitaev:2006lla, Levin:2004mi, Hu:2012wx, Wan:2014woa, Mesaros:2012yd}. 

In (2+1)d, such topological gauge theories based on finite groups have been extensively studied. They turn out to have strong connections to \emph{(quasi)-Hopf algebras} \cite{Drinfeld:1989st} and orbifold models in rational conformal field theories \cite{Dijkgraaf1991, Dijkgraaf:1990ne}.  Given a finite group $G$ and an element $[\omega]\in H^{3}(G,{\rm U}(1))$, one can indeed construct a non-trivial quasi-Hopf algebra whose irreducible representations label the anyons of the corresponding topological gauge theory. This quasi-Hopf algebra is the so-called \emph{twisted Drinfel'd double} $\mathcal{D}^\omega(G)$ of the group $G$. The corresponding Hamiltonian model is referred to as the twisted quantum double model \cite{Kitaev:2006lla, Levin:2004mi, Hu:2012wx} which reduces to Kitaev model \cite{Kitaev:2006lla} when $\omega$ is chosen to be trivial. Three-dimensional generalizations of such statements were explored in \onlinecite{Wan:2014woa, Delcamp:2017pcw, Delcamp:2018efi}.

Besides topological gauge theories, there are other TQFTs in (2+1)d which have been actively studied. Among these are theories based on \emph{modular tensor categories} (MTCs). A topological state-sum can be built from an MTC using the Turaev-Viro construction \cite{Turaev:1994xb, Turaev:1992hq, Barrett:1993ab} while the corresponding Hamiltonian realization is provided by the Levin-Wen models. More recently fermionic versions of MTCs known as super-MTCs were studied \cite{Bruillard:2016yio, bruillard2017classification, Aasen:2017ubm}. Objects in such categories have a natural $\mathbb Z_{2}$-grading provided by fermion number parity. The corresponding fermionic or spin-TFT state-sum models were studied in \cite{Bhardwaj:2016clt} and their Hamiltonian versions in \cite{Aasen:2017ubm}. Furthermore, there exists another class of models in (2+1)d that are \emph{almost}-TQFTs. Canonical examples of these are Chern-Simons theories also known as Witten-Reshetekhin-Turaev theories \cite{Witten:1988hf, Reshetikhin:1990pr, Reshetikhin:1991tc}. These models are not TQFTs in a strict sense since they can only be well-defined by either providing a framing of the three-manifold or by thinking of the three-manifold as the boundary of a four manifold which houses an almost trivial TQFT. 

In comparison much less is known in (3+1)d. The only known classes of TQFTs with intrinsic topological order\footnote{We define theories with intrinsic topological order as those that have $(i)$ non-trivial groundstate degeneracy that depends on the topology of the manifold $(ii)$ fractionalized excitations i.e dynamical operators of the theory that have topological correlation functions and $(iii)$ the theory has long range entanglement.} are topological gauge theories built from groups or group-like structures. There has been significant recent progress in the study of such theories \cite{Wang:2014xba, Tiwari:2016zru, Jiang:2014ksa, kapustin2014coupling,Ye:2016czw, putrov2016braiding, Wan:2014woa, Wang:2014oya, Tiwari:2017wqf, Delcamp:2017pcw, Delcamp:2016lux, Wen:2017xwk}. Another class of (3+1)d theories uses \emph{unitary braided fusion categories} as the input. State-sum models for these theories are then provided by Crane-Yetter models \cite{Crane:1993if, Crane:1993cm, Crane:1994ji} and the corresponding Hamiltonian models are the so-called Walker-Wang models \cite{Walker:2011mda}. However, in the case where the fusion category is modular, such models are trivial in the sense mentioned above (groundstate degeneracy and fractionalized excitations). In this work, we focus on the former class, namely topological gauge theories. 

\bigskip \noindent 
Our study follows two parallel approaches. The first one involves Levin-Wen models obtained by coloring the one-skeleton of a triangulated $d$-manifold with objects of a group-like categories. Such models take as an input a map which performs a specific change of triangulation, namely a $d$-dimensional \emph{Pachner move}. We then study the algebraic constraints that emerge from the consistency relations that such maps must satisfy, or equivalently, from requiring topological invariance of the corresponding ($d$+1) state-sum. To do so, we begin by analyzing the above question in the simpler setting of (2+1)d. Starting from the category  $\mathbb{C}$--$\text{Vec}_G$ of $G$-graded vector spaces, which is relevant for topological $G$-gauge theories, we can show that the freedom to weaken the associativity condition is captured by cohomological classes $[\alpha]\in H^{3}(G,{\rm U}(1))$. But the consistency condition that the map performing a two-dimensional Pachner move must satisfy is precisely the 3-cocycle condition. Similarly, using simplicial arguments \cite{Carter:1998qs}, we can show how the consistency condition of the map performing a three-dimensional Pachner move corresponds to the coherence relation of a structural 2-morphism of a given monoidal 2-category. Starting with this 2-category, it is very natural to relax certain identities by introducing additional structural 2-morphisms while preserving topological invariance. In doing so, we obtain a (weak) monoidal 2-category whose underlying (weak) 1-category is given by a 2-group.\footnote{A 2-group is an algebraic structure one obtains by categorifying a group. In its weak version, it can succinctly be defined as a (weak) monoidal category whose 2-morphisms are all invertible and 1-morphism are weakly invertible \cite{baez2004higher}.} This takes us to our second approach.

Our second approach involves directly constructing state-sum models from 2-groups\cite{Baez:2003fs, Baez:2004in, Baez:2005qu, baez2011invitation} as higher gauge theories. This is done in two equivalent ways: $(i)$ By endowing a triangulated four-manifold with a $\mathbb G$-coloring with a given 2-group $\mathbb G$, $(ii)$ building topological gauge theories as sigma models of the \emph{classifying space} $B \mathbb G$ of the 2-group. In $(i)$, the $\mathbb{G}$-coloring together with some constraints define a flat $\mathbb G$-bundle. The Dijkgraaf-Witten approach can then be generalized straightforwardly. By allowing the topological action to be a non-trivial element of the cohomology group $H^{4}(\mathbb G,{\rm U}(1))$, we define distinct topological gauge theories built from 2-groups. Given a topological group $G$, the classifying space $BG$ is a space which satisfies in particular the important property that its fundamental group is $G$ and all other homotopy groups vanish. Furthermore, for a discrete group $G$, homotopy classes of maps from $\mathcal M$ to $BG$ are equivalent to isomorphism classes of flat $G$-bundles that are locally 1-cochains. Similarly, we can define a space $B^{q+1}H$, where $H$ is an abelian group, such that $\pi_{q+1}(B^{q+1}H)=H$ and all other homotopy groups of $B^{q+1}H$ vanish. Homotopy classes of maps from $\mathcal M$ to $B^{q+1}H$ are equivalent to isomorphism classes of flat ($q$+1)-form $H$-bundles denoted by $H_{[q]}$ bundles, i.e. $BH_{[q]}\equiv B^{q+1}H$. Finally the classifying space of a 2-group in $(ii)$ may be understood as a fibration of classifying spaces of groups $H_{[1]}$ over $G$. With such an understanding of the classifying space of a 2-group, we can define topological higher gauge theories quite straightforwardly. The partition function reduces to a sum over homotopy classes of maps and with the topological action provided by the pullback of a representative of a cohomology class on the classifying space. Throughout this manuscript, we contrast and compare the original Dijkgraaf-Witten approach for groups with higher form groups, 2-groups and other generalizations. 

Note that the subject of gauge theories built on categorified groups is not exactly a new one. In \onlinecite{Yetter:1993dh}, D. Yetter defined a TQFT for three-manifolds based on 2-groups in a manner analogous to Dijkgraaf-Witten theories but without any cohomological twist. In \onlinecite{porter1998topological}, the construction of Yetter was extended to homotopy $n$-types such that the classifying space of a 2-group is a homotopy 2-type and that of an ordinary group is a homotopy 1-type. Furthermore, in \onlinecite{mackaay2000finite}, Mackaay provided an explicit construction of TQFTs based on homotopy 3-type whilst including a cohomological twist. In \onlinecite{martins20052, martins2007categorical}, Martins studied invariants of knotted surfaces embedded in $S^4$ based on Yetter's TQFT. More recently, in the condensed matter literature, there have been several papers related to topological constructions based on 2-groups \cite{kapustin2017higher, Cui:2017vek, Williamson:2016evv, Cheng:2017ftw, Bullivant:2016clk, Bullivant:2017sjz}. See also the recent exhaustive study of abelian continuous 2-group global symmetries arising in quantum field theories \cite{cordova2018exploring}.

\bigskip \noindent 
Apart from intrinsic topological orders, it is possible to define \emph{symmetry protected} topological phases of matter (SPTs) with a 2-group as input data. In general, SPTs are gapped short-range entangled symmetric phases of matter that have unique groundstates. A classification of bosonic SPTs based on group cohomology was proposed in \onlinecite{Chen:2011pg}. The authors showed that in ($d$+1) dimensions distinct bosonic SPTs with symmetry $G$ are labeled by classes $[\omega] \in H^{d+1}(G,{\rm U}(1))$. Furthermore, it was shown that given a representative cocycle $\omega$, one can construct an exactly solvable fixed point model based on a state-sum construction. Later, it was shown in \onlinecite{Levin:2012yb} that it is no coincidence if the classification of such SPTs coincides with the classification of $G$-topological gauge theories. In fact gauging the global symmetry $G$ in an SPT labeled by $[\omega]$  yields the Dijkgraaf-Witten model $(G,[\omega])$. Furthermore, one notices that SPT topological invariants are the ${\rm U}(1)$-valued response functions that an SPT furnishes in the presence of a background $G$-bundle \cite{Wen:2013ue, kapustin2014anomalies, Wen:2014zga, Wang:2014pma, Ye:2015eba, Tiwari:2017wqf}. 

Besides containing point-like operators, SPTs may also contain operators that are localized on $q$-dimensional spacetime submanifolds. Global symmetries of $q$-dimensional operators dubbed \emph{generalized global symmetries} were studied by Gaiotto et al. in \onlinecite{gaiotto2015generalized}. Symmetry operators can then be constructed as topological wall operators localized on ($d$$-$$q$)-submanifolds of a ($d$+1)-manifold. Gauging a $q$-form symmetry group $H_{[q]}$ requires the introduction of a ($q$+1)-form gauge field valued in the gauge group. As discussed above this may be understood as a map from $\mathcal M$ to $B^{q+1}H$ in a particular homotopy class. Similarly, one may want to consider theories with $0$-dimensional and $q$-dimensional matter fields so that the the global symmetry group is $\mathbb G_{[q]}$ which is $(G,H_{[q]})$ as a set. In order to gauge such a symmetry we first need to understand what a $\mathbb G_{[q]}$ flat connection looks like. This may be answered by addressing the related question: What does the classifying space $B\mathbb G_{[q]}$ look like? This can be either a product space $B\mathbb G_{[q]}=BG \times B^{q+1}H$ or a non-trivial fibration $B^{q+1}H \to B \mathbb G_{[q]}\to BG$ classified by the extension class $[\alpha]\in H^{q+2}(BG,H)$. A flat $\mathbb G_{[q]}$ connection is then be captured by a system of fields $({\bfg}, {\bfh})\in Z^{1}(\mathcal M,G)\times C^{q+1}( \mathcal M,H)$ that satisfy the conditions $d{\bfg}=0$ and $d{\bfh}=\alpha({\bfg})$. In the special case where $q=1$ we recover a 2-group $\mathbb G$. 

Following the above discussion, we are able construct exactly solvable state-sum models for four scenarios of matter field distribution: $(i)$ Point-like matter with global symmetry $G$, $(ii)$ $q$-dimensional matter with global symmetry $H_{[q]}$, $(iii)$ point-like as well as string-like matter with global symmetry $\mathbb G_{[1]}\equiv \mathbb G$ and $(iv)$ point-like as well as $q$-dimensional matter with global symmetry $\mathbb G_{[q]}$. For each one of these scenarios, gauging the global symmetry requires coupling the theory to appropriate background symmetry bundles and showing that, upon summing over isomorphism classes of such bundles, one obtains the usual or higher-categorical generalizations of Dijkgraaf-Witten theory.

\bigskip \noindent 
A particularly interesting aspect of SPTs is their boundary theories. Consider a ($d$+1)-dimensional SPT with a global symmetry structure $X= G, H_{[q]}, \mathbb G_{[q]}$ labeled by a cohomology class representative   $[\omega]\in H^{d+1}(BX,{\rm U}(1))$, where $BX$ is the classifying space of $X$. Since the theory is symmetric under $X$, it is possible to probe the symmetry by coupling the theory to a flat background $X$-bundle. Such a bundle is provided via a map from $\mathcal M$ to $BX$. This can be done on closed manifolds as well as open manifolds. In the case of an open manifold $\mathcal M$ such that $\partial \mathcal M=\mathcal N$, one can show that the response theory is invariant under reparametrizations of the background gauge field only up to boundary terms. For the theory to be well-defined, there must exist a condition on the $d$-dimensional boundary which precisely cancels the lack of invariance of the ($d$+1)-dimensional topological response action on $\mathcal M$. In other words, there must exist a boundary $X$-symmetric theory on $\mathcal N$ which can be coupled to a background $X$-bundle but so that its partition function is not invariant under gauge variation of $X$. This is precisely what is referred to as \emph{'t Hooft anomaly}. Such scnearios are particularly interesting to investigate in the context of SPTs \cite{Sule:2013qla, kapustin2014anomalies, Hsieh:2014lba, Hsieh:2015xaa, Witten:2015aba} protected by 2-group symmetry \cite{Thorngren:2015gtw, tachikawa2017gauging} and its generalizations. 

\newpage
\subsubsection*{Organization of the paper}
\noindent
In sec.~\ref{sec:2dstring}, we review the construction of (2+1)d string net models focusing on Hamiltonian realizations of the 3d Dijkgraaf-Witten model. We emphasize how the unitary map performing a 2-2 Pachner move is related to the associator of the category of $G$-graded vector spaces. We present the (3+1)d generalization of this construction in sec.~\ref{sec:3dstring} where we now emphasize how the unitary map performing a 3-2 Pacher move is related to the pentagonator of the 2-category which yields the 4d Dijkgraaf-Witten model. We then consider a generalization of this 2-category which naturally leads to a model built on a weak 2-category whose underlying 1-category is a 2-group. In sec.~\ref{sec:higherGauge}, we present the notions of weak 2-group and flat 2-connection allowing us to construct topological models from groups, higher-from groups, 2-groups and further generalizations having a higher gauge theory interpretation. We are then able to recover the model obtained from a category-theoretical point of view. In sec.~\ref{sec:spts}, we provide state-sum models protected by 2-group (and generalizations of 2-groups) symmetry together with their lattice realization. The gauging procedure of such models is also exposed. Finally, we discuss in sec.~\ref{sec:hooft} the relations between SPTs in $d$+1 dimensions and 't Hooft anomalies for higher symmetries in $d$-dimensional TQFTs.\\

\noindent\textsc{Notations}\vspace{-0.5em}
\begin{leftbar} \noindent
	\begin{tabular}{cl}
		$\cM$ & ($d$+1)-manifold \\
		$\mathcal N$ & $d$-manifold such that $\partial \cM = \cN$ \\
		$BG$ & classifying space of topological group $G$ \\
		$G$ & finite group \\
		$H$ & finite abelian group \\
		$\,\,\; H_{[q]}$ & global $q$-form symmetry group\protect\footnotemark[4] 
		\\
		$\triangleright$ & group action \\
		$\mathbb G$ & 2-group \\
		$\;\;\; \mathbb G_{[q]}$ & higher-form generalization of a 2-group\protect\footnotemark[5]\\
		$\triangle$ & triangulation of manifold $\cM$ \\
		$\,\, \triangle_i$ & $i$-simplex of triangulation $\triangle$ \\
		$g$ & $G$-coloring of the 1-simplices of $\triangle$ \\
		$h$ & $H$-coloring of the $2$-simplices of $\triangle$ \\
		${\rm Col}(\cM,\mathbb G)$ & set of $\mathbb G$-colorings of $\cM$ \\
		$k$ & 0-form $G$-gauge symmetry parameter \\
		$\lambda$ & 1-form $H$-gauge symmetry parameter \\
		$K$ & 0-form global symmetry parameter \\
		$\Lambda$ & 1-form global symmetry parameter \\
		$\alpha$ & $M$-valued group 3-cocycle in $Z^3(G,M)$ \\
		$\pi$ & $M$-valued group 4-cocycle in $Z^4(G,M)$ \\ 
		$[\omega]$ & cohomological class in $H^{d+1}(X, \mathbb R /\mathbb Z)$ \\
		$\smile$ & cohomological cup product \\
	\end{tabular}
\end{leftbar}
\footnotetext[4]{Note that for $q > 0$, $H_{[q]}$ as a group is always abelian. When $q=0$, it reduces to a group by $G$ which may be non-abelian. When gauging a $q$-form symmetry, one introduces a ($q$+1)-form flat gauge field whose classifying space is denoted by $B^{q+1}H$.}
\footnotetext[5]{It is a mixed symmetry group acting on point objects as well as $q$-dimensional objects. When $q=1$, we recover the 2-group $\mathbb G$.}
\setcounter{footnote}{5}

\newpage
\section{Review of (2+1)d string net models}
\label{sec:2dstring}

\noindent The construction of higher-dimensional topological models proposed in the next section follows closely the definition of (2+1)d string nets. As such, we briefly review here the general string net picture. More details can be found in \onlinecite{Levin:2004mi}. Since we are interested in gauge models, we restrict our attention to models whose input data is a group-like category. This is the setup of the \emph{twisted quantum double models} \cite{Hu:2012wx}. More specifically we will present the Hamiltonian realization of the 3d Dijkgraaf-Witten model.

\subsection{Fixed point wave functions}
\noindent
Gapped quantum phases of matter can be defined in terms of equivalences classes of states (or many-body wave functions) under \emph{local unitary transformations}. These equivalence classes are associated with a given pattern of long-range entanglement which is the defining feature of intrinsic topological orders. Thinking of these local transformations as implementing a wave function renormalization group flow, the task to find equivalence classes of states boils down to defining fixed-point wave functions. The fixed-point wave functions are expected to capture all the universal long-range features of the corresponding phase. 

\emph{String net models}, or Levin-Wen models \cite{Levin:2004mi}, were introduced as a systematic way to construct ground states exhibiting the phenomenon of string net condensation. These models are expressed in terms of graphs. Each graph, together with a given labeling, defines a state. The Hilbert space of the model is then defined as the linear superposition of spatial configurations of string nets. In particular, the fixed-point wave functions we are interested in are obtained as superpositions of such graph-based states. These wave functions are specified uniquely by the local transformations defined on the lattice and in turn define ground states of given Hamiltonians.

At the most basic level, a string net is a network of strings such that each string is decorated by an object $x \in \mathcal{C}$, where $\mathcal{C}$ is a collection of super-selection sectors. We equip $\mathcal{C}$ with a vacuum sector $\mathbbm{1}_{\mathcal{C}}$ and a duality map $x \mapsto \bar{x}$ such that the type $\bar{x}$ is assigned to a type-$x$ string with opposite orientation. 
We choose the underlying network to be the one-skeleton of a triangulation  $\triangle$ of a 2-dimensional hypersurface $\Sigma$. Furthermore, we introduce some compatibility conditions associated with every 2-simplex $\{\triangle_2\}$ which are referred to as the {\it branching rules}.

The branching rules are constraints between super-selection sectors labeling the 1-simplices surrounding a given 2-simplex. Consider a 2-simplex and let $\{x_1,x_2,x_3\}$ be the super-selection sectors labeling the corresponding 1-simplices. The branching rules are such that if we assign a vector space $V_{x_3}^{x_1,x_2}$ to the 2-simplex, its dimension is non-vanishing only if the branching rules between $x_1$, $x_2$ and $x_3$ are satisfied.

The input data for a given string net model is ({\it a priori}) given by a set $\{\mathcal{C},N,\alpha,\ell,r\}$ where
\begin{enumerate}[itemsep=0.4em,parsep=1pt,leftmargin=*]
	\item[$\circ$] $\ell_{(\triangle_1)}$ and $r_{(\triangle_1)}$ are maps associated to each super-selection sector labeling a 1-simplex.
	\item[$\circ$] $N_{(\triangle_2)}$ is a three-valent tensor which vanish if the triplet of super-selection sectors $\{x_1,x_2,x_3\}$ labeling the 1-simplices bounding the 2-simplex $\triangle_2$ does not satisfy the branching rules. 
	\item[$\circ$] $\alpha_{(\triangle_3)}$ is a map which depends on the super-selection sectors labeling the edges of a 3-simplex (or alternatively a 4-gon).
\end{enumerate}
In the following we will not write the super-selection sectors explicitly. Instead we will label the vertices and refer to the simplices on which the super-selection sectors live in. For instance $(ab)$ refers to the edge going from the vertex $b$ to the vertex $a$ while $x_{ab}$ is the super-selection sector labeling it.

The local unitary transformations which are required to be satisfied by the fixed-point wave functions can be represented as follows
\begin{align}
	\label{loc1}
	\bigg| \locONE{0.7}{1} \bigg\ra &\, = \, \bigg| \locONE{0.7}{3} \bigg\ra  \\[0.3em]
	\label{loc3}\noeqref{loc3}
	\bigg| \locTHREE{0.7}{0} \bigg\ra &\,\propto\, \bigg| \locTHREE{0.7}{1} \bigg\ra
	\delta_{x_{12}^\text{left},x_{12}^\text{right}} \\[0.3em]
	\label{loc4}
	\ell_{(12)}\bigg| \locFIVE{0.7}{1} \bigg\ra &\,=\, \bigg| \locONE{0.7}{3} \bigg\ra \\[0.3em]
	\label{loc5} 
	r_{(01)}\bigg| \locFIVE{0.7}{0} \bigg\ra & \, = \, \bigg| \locONE{0.7}{3} \bigg\ra\\[0.3em]
	\label{loc6}
	\bigg| \squ{02}{0.7}{1} \bigg\ra \,=\, \displaystyle\sum_{x_{13}}&\alpha_{(0123)} \bigg| \squ{13}{0.7}{1} \! \bigg\ra 
\end{align}
where the dots are used to make manifest the presence of certain vertices while the dashed line represents a 1-simplex labeled by the vacuum sector $\mathbbm{1}_\mathcal{C}$. 

For the conditions \eqref{loc1}--\eqref{loc6} to be self-consistent, some coherence relations need to be satisfied. More precisely, not every set $\{\mathcal{C},N,\alpha,\ell,r\}$ can give rise to a string net condensed phase. In particular the maps $\alpha$, $\ell$ and $r$ need to satisfy the so-called {\it pentagon and triangle relations}. 

\subsection{Pentagon relation}
\noindent
Let us first focus on the pentagon equation. From a simplicial point of view, the local transformation \eqref{loc6} corresponds to the so-called 2--2 Pachner move denoted by $\mathcal{P}_{2 \mapsto 2}$. Considering a given triangulation of a pentagon, it is possible to perform five different $\mathcal{P}_{2 \mapsto 2}$ moves so as to reach back the original triangulation. We represent this cyclic property as follows
\begin{equation}
	\label{coher2D}
	\begin{tikzcd}[column sep=-0.5em, row sep=1.4em]
		{}
		& {}
		& \pent{02}{24}{0.4}{0} 
		\ar[drr,"{(0124)}",dash, shorten <= -1em, shorten >= -1em]
		& {}
		& {}
		\\
		\pent{02}{03}{0.4}{0} \ar[urr,"{(0234)}",dash, shorten <= -1em, shorten >= -1em]
		& {}
		& {}
		& {}
		& \pent{14}{24}{0.4}{0} 	
		\ar[dl,"{ (1234)}",dash, shorten <= -0.48em, shorten >= -0.9em]
		\\
		{}
		& \pent{03}{13}{0.4}{0} 	
		\ar[ul,"{(0123)}",dash, shorten >= -0.48em, shorten <= -0.9em]
		& {}
		& \pent{14}{13}{0.4}{0} 	
		\ar[ll,"{(0134)}", dash, shorten <= -0.59em, shorten >= -0.59em]
		& {}
	\end{tikzcd} =: \mathcal{K}_4(01234)
\end{equation}
where each small pentagon is labeled from $0$ to $5$ starting at the upper vertex and going clockwise, while the edges of the global pentagon, whose vertices are given by the small pentagons, represent a $\mathcal{P}_{2 \mapsto 2}$ local transformation. For instance the top left edge labeled by $(0234)$ is associated with
\begin{equation}
	\bigg| \pent{02}{03}{0.4}{1} \bigg\ra = \sum_{x_{24}} \alpha_{(0234)}
	\bigg| \pent{02}{24}{0.4}{1}  \bigg\ra \; .
\end{equation}
Each edge of the global pentagon are labeled by a 4-gon and are associated with a given $\mathcal{P}_{2 \mapsto 2}$ move which can be performed in both directions. As it turns out, we can also think of the edges of this global pentagon as being associated with a $3$-simplex. The 3-simplices $\{\triangle_3\}$ are identified by the vertices labeling the 4-gon on which the $\mathcal{P}_{2 \mapsto 2}$ move acts. The transformation of the corresponding states is performed by the map $\alpha_{(\triangle_3)} {}^{\epsilon(\triangle_3)}$ where $\epsilon(\triangle_3) \pm 1 $ is a factor which is defined according to the following convention:\\

\noindent\textsc{Convention for the 2--2 Pachner move}\vspace{-0.5em}
\begin{leftbar}\noindent 
	Pick one of the two triangles $\triangle_2$ in the source 4-gon whose vertices define a 3-simplex $\triangle_3$. Let us assume this 2-simplex is label by $(abc)$ such that $a < b < c$. The remaining vertex is labeled by $d$. If it takes an odd number of permutations to bring the list $(d,a,b,c)$ to the ascending ordered one, then $\epsilon(\triangle_3)=+1$, otherwise $\epsilon(\triangle_3)= -1$. For instance, in the example above, we can choose as 2-simplex $\triangle_2 = (023)$. It takes three permutations to bring $(4,0,2,3)$ to $(0,2,3,4)$ and therefore $\alpha_{(0234)}^{\epsilon(0234)} = \alpha_{(0234)}$.
\end{leftbar}
\noindent
Interestingly, the five 3-simplices on which the map $\alpha$ is defined bound the 4-simplex $(01234)$. Furthermore, we can associate a higher-dimensional Pachner move to this cycle of $\mathcal{P}_{2 \mapsto 2}$ moves, namely the following 2--3 Pachner move denoted by $\mathcal{P}_{2 \mapsto 3}$:
\begin{equation}
	\label{pachner23}
	\mathcal{P}_{2 \mapsto 3} : \PPONE{1}{1} \longmapsto \PPTWO{1}{1}{-0.3} \; .
\end{equation} 
\noindent
Again the five 3-simplices appearing in this $\mathcal{P}_{2 \mapsto 3}$ move are the fives simplices $(0124)$, $(1234)$, $(0134)$, $(0123)$ and $(0234)$ appearing above. In addition, the triangulated pentagons labeling the vertices of the global pentagon in \eqref{coher2D} correspond to faces of the union of the 3-simplices appearing in the move \eqref{pachner23}.
 
We explained above that the map $\alpha$ performing the $\mathcal{P}_{2 \mapsto 2}$ move comes with a sign which depends on the ordering of the vertices labeling the corresponding source 4-gon whose vertices also define a $3$-simplex. This suggests that each 3-simplex comes with an orientation which determines the sign of $\epsilon(\triangle_3)$ in the action of $\alpha$. Using the explicit representation of these 3-simplices, there is a canonical way to determine such orientation independently of the corresponding $\mathcal{P}_{2 \mapsto 2}$ move:\\

\noindent\textsc{Orientation convention of the 3-simplices}\vspace{-0.6em}
\begin{leftbar} \noindent
	Pick one of the triangles in the 3-simplex $\triangle_3$ and look at the remaining vertex through this triangle. If the vertices of the triangle are ordered in a clock-wise fashion, then the orientation is positive, otherwise it is negative. For instance, we have
	\begin{equation}
	\epsilon \bigg( \convONE{0.7}{0}{1}{2} \bigg) = +1 \q , \q
	\epsilon \bigg( \convONE{0.7}{0}{2}{1} \bigg) = -1
	\end{equation}
	where $\otimes$ represent the fourth vertex as seen from behind the triangle.
\end{leftbar}
\noindent
Applying this convention to the five 3-simplices appearing in \eqref{pachner23}, or equivalently the convention for the 2--2 Pachner moves to the coherence relation \eqref{coher2D}, and requiring that any sequence of $\mathcal{P}_{2 \mapsto 2}$ moves such that the initial triangulation is the same as the final one must be associated with a trivial map, one finds the following formal condition: 

\begin{equation}
	\label{coher2Deq}
		\alpha_{(0234)}\alpha_{(0124)} =\alpha_{(0123)}\alpha_{(0134)}\alpha_{(1234)}
\end{equation}
\vspace{0.5em}

\noindent which is usually referred to as the pentagon relation. More generally, the following result holds: Any sequence of $\mathcal{P}_{2 \mapsto 2}$ moves between two given triangulations of an $m$--gon leads to the same isomorphism between the corresponding vector spaces.\footnote{As we will see later, this is a consequence of Maclane's coherence theorem of category theory.}

Before moving on to the triangle equation, let us make a final remark.
The object denoted by $\mathcal{K}_4(01234)$ introduced in \eqref{coher2D} corresponds to the so-called {\it fourth Stasheff polytope} \cite{stasheff1963homotopy} associated with the permutation of vertices $(01234)$. It is a combinatorial object which appears in many areas of mathematics defined as follows:\\
\newpage
\noindent\textsc{Stasheff polytopes}\vspace{-0.7em}
\begin{leftbar} \noindent
	The $\mathcal{K}_n$ Stasheff polytope, or \emph{associahedron} , is an $(n-2)$-dimensional convex polytope whose vertices correspond to all the correct bracketings of a word containing $n$ letters. Or equivalently, the vertices are associated to all possible triangulations of an $n$-gon. The first five Stasheff polytopes are represented below:
	\begin{equation*}
		\mathcal{K}_{1}  \, = \, \;
		\begin{tikzpicture}[scale=0.4,baseline=-0.3em]
			\coordinate (0) at(0,0);
			\node[draw,circle,scale=0.3,fill] at (0) {};
		\end{tikzpicture} 
		\q , \q
		\mathcal{K}_{2} \, = \, \;
		\begin{tikzpicture}[scale=0.4,baseline=-0.3em]
			\coordinate (0) at(0,0);
			\node[draw,circle,scale=0.3,fill] at (0) {};
		\end{tikzpicture} 
		\q , \q
		\mathcal{K}_{3}  \, = \, 
		\begin{tikzpicture}[scale=0.4,baseline=-0.3em]
			\coordinate (0) at(-1,0);
			\coordinate (1) at (1,0);
			\draw[] (0) to (1);
			\node[draw,circle,scale=0.3,fill] at (0) {};
			\node[draw,circle,scale=0.3,fill] at (1) {};
		\end{tikzpicture} 
	\end{equation*}
	\begin{equation*}
		\mathcal{K}_4  \, = \,
		\begin{tikzpicture}[scale=0.4,baseline=-0.3em]
			\coordinate (0) at(0,1);
			\coordinate (1) at ({sin(72)} ,{cos(72)} );
			\coordinate (2) at ({sin(144)} ,{-cos(36)} );
			\coordinate (3) at ({-sin(144)} ,{-cos(36)} );
			\coordinate (4) at ({-sin(72)} ,{cos(72)} );
			\draw[] (0) to (1);
			\draw[] (1) to (2);
			\draw[] (2) to (3);
			\draw[] (3) to (4);
			\draw[] (4) to (0);
		\end{tikzpicture} 
		\q , \q
		\mathcal{K}_5  \, = \,
		\begin{tikzpicture}[scale=0.15,baseline=1.5em]
			\coordinate (0) at(0,0);
			\coordinate (1) at (-1,2);
			\coordinate (4) at (3,1);
			\coordinate (3) at (2,3);
			\coordinate (2) at (0,3);
			\coordinate (5) at (6,3);
			\coordinate (6) at (2,9);
			\coordinate (7) at (-1,10);
			\coordinate (8) at (3,4);
			\coordinate (9) at (-5,2);
			\coordinate (10) at (-6,4);
			\coordinate (11) at (-3,9);
			\coordinate (12) at (-1,8);
			\coordinate (13) at (1,8);
			\draw[] (0) to (1);
			\draw[] (1) to (2);
			\draw[] (2) to (3);
			\draw[] (3) to (4);
			\draw[] (4) to (0);
			\draw[] (4) to (5);
			\draw[] (5) to (6);
			\draw[] (6) to (7);
			\draw[opacity = 0.4] (7) to (8);
			\draw[opacity = 0.4] (8) to (5);
			\draw[] (0) to (9);
			\draw[] (9) to (10);
			\draw[] (10) to (1);
			\draw[] (10) to (11);
			\draw[] (6) to (13);
			\draw[] (7) to (11);	
			\draw[] (11) to (12);
			\draw[] (12) to (13);
			\draw[] (2) to (12);
			\draw[] (3) to (13);
			\draw[opacity = 0.4] (8) to (9);																	
		\end{tikzpicture} 
	\end{equation*}
	 Given a Stasheff polytope, a codimension-$p$ face is always isomorphic to a product of $p$+1 lower dimensional Stasheff polytopes. For instance, the pentagonal faces of $\mathcal{K}_5$ are obtained as $\mathcal{K}_2 \times \mathcal{K}_4$ while the quadrilateral faces are isomorhpic to $\mathcal{K}_3 \times \mathcal{K}_3$. This property has interesting consequences at the light of the correspondence between Stasheff polytopes and coherence relations of $n$-categories.
\end{leftbar}
\noindent
In the same way the polytope $\mathcal{K}_4$ naturally describes the combinatorics of the pentagon equation which ensures the coherence of  the $\mathcal{P}_{2 \mapsto 2}$ moves, we will emphasize later the fact that the polytope $\mathcal{K}_5$ is associated with the higher-dimensional coherence relation appearing in 3d string net models.

\subsection{Triangle relations}
\noindent
The triangular relations are coherence relations for the maps $\ell$ and $r$ and translate the fact that adding a line labeled by the vacuum sector as in \eqref{loc4} or \eqref{loc5} commutes with the $\mathcal{P}_{2 \mapsto 2}$ move. This guarantees the ``invisibility'' of the vacuum sector. The triangle relation can be represented via the following commutative diagram
\begin{equation}
	\label{triangle2D}
	\begin{tikzcd}[column sep=-1.5em, row sep=1.5em]
	{}
	& \triONE{0.8}{1}
	\ar[dl,"{{r}_{(01')} }"',leftarrow, pos =1, shorten >= -1em]
	\ar[dr,"{{l}_{(12)} }",leftarrow, pos=1, shorten >= -1em]
	& {}
	\\
	\triTWO{0.8}{1}
	\ar[rr, "{\alpha_{(01'12)}}"',rightarrow]
	& {}
	& \triTHREE{0.8}{1}
	\end{tikzcd}
\end{equation} 
such that the vertices are ordered as follows: $0<1'<1<2$.
Later we will refer to this triangle relation as the \emph{fundamental} triangle relation since it is possible to deduce corollary triangle coherence relations using the fundamental one together with the pentagon relation.

The local unitary transformations \eqref{loc1}--\eqref{loc6} associated with the input data $\{\mathcal{C},N,\alpha,\ell,r\}$ satisfying the pentagon equation as well as the fundamental triangle equation fully characterize the fixed point wave functions we are interested in.  It turns out that the defining properties for the fixed point wave functions that we have been spelling out are nothing but the defining axioms of a (weak) monoidal category. 
 
\subsection{Categorical aspects}
\noindent
In the following we will make explicit use of the category theoretical language in order to construct (3+1)d gauge models of topological orders. More precisely, we will emphasize how it is possible to recover higher gauge theory models by simply weakening some of the defining axioms of the category. As such, we will now review some basic definitions of category theory \cite{etingof2016tensor} and emphasize how it relates exactly to the previous string net construction.

\medskip \noindent
A (weak) monoidal category consists of:
\begin{enumerate}[itemsep=0.4em,parsep=1pt,leftmargin=*]
	\item[$\circ$] A category $\mathcal{C}$ whose collection of objects is denoted by Ob($\mathcal{C}$) and for each $x_1,x_2 \in \text{Ob}(\mathcal{C})$, the collection of morphisms between them is denoted by $\text{Hom}_{\mathcal{C}}(x_1,x_2)$. 
	\item[$\circ$] A functor $m : \mathcal{C} \times \mathcal{C} \rightarrow \mathcal{C}$ where we use the notation $m(x_1,x_2) = x_1 \otimes x_2$.
	\item[$\circ$] An identity object $\mathbbm{1} \in \mathcal{C}$.
	\item[$\circ$] Natural isomorphisms:
	\begin{align*}
		\alpha_{x_1,x_2,x_3}\, : \, (x_1 \otimes x_2) \otimes x_3 \, &\mapsto \, x_1 \otimes (x_2 \otimes x_3) \\
		\ell_{x} \, : \, \mathbbm{1} \otimes x \, &\mapsto \, x \\
		r_{x} \, : \, x \otimes \mathbbm{1} \, & \mapsto \, x
	\end{align*}
	referred to as the \emph{associator}, the \emph{left unitor} and the \emph{right unitor}. These structural morphisms must satisfy coherence relations which are encoded into commutative diagrams. For all objects $x_1,x_2,x_3,x_4 \in \mathcal{C}$, the associator is defined such that the following diagram commutes
\begin{equation*}
	\begin{tikzcd}[column sep=-5em, row sep=4em]
	{}
	& {}
	& {\scriptstyle(x_1 \otimes x_2) \otimes (x_3 \otimes x_4) }
	\ar[drr,"{\alpha_{x_{1},x_{2},x_{3} \cdot x_{4}}}",rightarrow]
	& {}
	& {}
	\\
	{\scriptstyle ((x_1 \otimes x_2) \otimes x_3) \otimes x_4 }
	\ar[urr,"{\alpha_{x_{1}\cdot x_{2}, x_{3}, x_{4}}}",rightarrow]
	& {}
	& {} 
	& {}
	& {\scriptstyle x_1 \otimes (x_2 \otimes (x_3 \otimes x_4)) 	}
	\ar[dl,"{1_{x_1} \otimes \alpha_{x_{2},x_{3},x_{4}}}",leftarrow]
	\\
	{}
	& {\scriptstyle (x_1 \otimes (x_2 \otimes x_3))\otimes x_4	}\q\q\q\q
	\ar[ul,"{\alpha_{x_1,x_2,x_3} \otimes 1_{x_4}}",leftarrow]
	& {}
	& \q\q\q\q {\scriptstyle x_1 \otimes ((x_2 \otimes x_3) \otimes x_4) }	\ar[ll,"{\alpha_{x_1,x_2\cdot x_3,x_4}}"',rightarrow, shorten <= -1em, shorten >= -1em]
	& {}
	\end{tikzcd}
\end{equation*}
while the coherence relation for the left and right unitors reads
\begin{equation*}
	\begin{tikzcd}[column sep=1em, row sep=4em]
	{}
	& {\scriptstyle x_1 \otimes x_2}
	\ar[dl,"{{r}_{x_1}\otimes 1_{x_2} }"',leftarrow]
	\ar[dr,"{1_{x_1} \otimes {\ell}_{x_2} }",leftarrow]
	& {}
	\\
	{\scriptstyle (x_1 \otimes \mathbbm{1}) \otimes x_2}
	\ar[rr, "{\alpha_{x_1,\mathbbm{1}, x_2}}"',rightarrow]
	& {}
	& {\scriptstyle x_1 \otimes ( \mathbbm{1} \otimes x_2) }
	\end{tikzcd}
\end{equation*}
\end{enumerate} 
A {\it strict monoidal category} is obtained from a weak monoidal category by requiring the isomorphisms $\alpha_{x_1,x_2,x_3}$, $\ell_{x}$ and $r_{gx}$ to be identitiy isomorphisms so that
\begin{align}
	\label{strictCat1}
	(x_1 \otimes x_2) \otimes x_3 &= x_1 \otimes (x_2 \otimes x_3) \; ,\\
	\label{strictCat2}
	\mathbbm{1} \otimes x = x \q & \;\; , \;\; \q x \otimes \mathbbm{1} = x \; .
\end{align}
From the two previous commutative diagrams, we can deduce the following ones
\begin{equation*}
	\begin{tikzcd}[column sep=1em, row sep=4em]
		{}
		& {\scriptstyle x_1 \otimes x_2}
		\ar[dl,"{{\ell}_{x_1}\otimes 1_{x_2} }"',leftarrow]
		\ar[dr,"{{\ell}_{x_1 \otimes x_2} }",leftarrow]
		& {}
		\\
		{\scriptstyle (\mathbbm{1} \otimes x_1) \otimes x_2}
		\ar[rr, "{\alpha_{\mathbbm{1}, x_1, x_2}}"',rightarrow]
		& {}
		& {\scriptstyle \mathbbm{1} \otimes ( x_1 \otimes x_2) }
	\end{tikzcd}
\end{equation*} 
\begin{equation*}
	\begin{tikzcd}[column sep=1em, row sep=4em]
		{}
		& {\scriptstyle x_1 \otimes x_2}
		\ar[dl,"{{r}_{x_1 \otimes x_2} }"',leftarrow]
		\ar[dr,"{1_{x_1} \otimes {r}_{x_2} }",leftarrow]
		& {}
		\\
		{\scriptstyle (x_1 \otimes x_2) \otimes \mathbbm{1}}
		\ar[rr, "{\alpha_{x_1,x_2,\mathbbm{1}}}"',rightarrow]
		& {}
		& {\scriptstyle x_1 \otimes ( x_2 \otimes \mathbbm{1}) }
	\end{tikzcd}
\end{equation*} 
which are referred to as the corollary triangle relations.

From now on and for the rest of this paper, we will focus on a specific class of models, namely \emph{group cohomological models}. In particular, we will now describe the category which yields the \emph{Dijkgraaf-Witten} topological theory. This category will be the starting point of the (3+1)d construction.
 
The string net model which corresponds to an Hamiltonian extension of Dijkgraaf-Witten theory is built upon the weak monoidal category $\mathbb{C}$--$\text{Vec}_G^\alpha$ of $G$-graded vector spaces over the field of complex numbers $\mathbb{C}$.\footnote{The category $\mathbb{C}$--$\text{Vec}_G^\alpha$ is actually an example of \emph{pointed fusion category}. However, for our purpose it is not strictly necessary to provide the corresponding additional structures.}\footnote{Given a finite group $G$, it is possible to define two types of string net models based on the $\text{Vec}_G$ or $\text{Rep}[G]$ fusion categories, respectively. This dichotomy is explained from a canonical quantization point of view in \cite{Delcamp:2018sef}.} A $G$-graded vector space is a vector space $V$ with a decomposition $V = \bigoplus_{g \in G}V_g$. The tensor product is then defined as 
\begin{equation}
(V \otimes W)_g = \bigoplus_{x,y \in G \atop xy=g} V_x \otimes W_y \; .
\end{equation}
This category has finitely many simple objects which are the 1-dimensional $G$-graded vector spaces, therefore they are in one-to-one correspondence with the elements of the finite group $G$. These simple objects are denoted by $\delta_g$, $g \in G$ and they are defined according to the formula $\text{Hom}(\delta_g, \delta_h) = \mathbb{C}$ if $g=h$, $0$ otherwise. The tensor product between simple objects reads
\begin{equation}
	\delta_{g_1} \otimes \delta_{g_2} \cong \delta_{g_1g_2}
\end{equation} 
which is nothing else than the group multiplication.
In order to define an associator $\alpha$ for the category $\mathbb{C}$--$\text{Vec}_G^\alpha$, it is only necessary to define it on the simple objects. We are therefore looking for a natural isomorphism which performs
\begin{equation}
		\alpha_{g_1,g_2,g_3}\, : \, (\delta_{g_1} \otimes \delta_{g_2}) \otimes \delta_{g_3} \, \mapsto \, \delta_{g_1} \otimes (\delta_{g_2} \otimes \delta_{g_3})
\end{equation}
but since $\delta_{g_1g_2g_3}$ is itself a simple object, it follows that the associator is valued in $\mathbb{C}$. We further require $\alpha$ to be invertible such that it lives in $\mathbb{C}^\times$. We denote this number by $\alpha(g_1,g_2,g_3) \in \mathbb{C}^\times$. It then follows from the commutativity of the pentagon diagram\footnote{More generally, we can think of the cocycles conditions as the result of applying a functor from the Stasheff polytopes to the category $\mathcal{C}$. We will show in the next section how this statement naturally applies to (3+1)d string net models.
} that
\begin{align}
	\label{coc3}
	\frac{\alpha(g_{1},g_{2},g_{3})\alpha(g_{1},g_{2} g_{3}, g_{4})
		\alpha(g_{2},g_{3},g_{4})}{\alpha(g_{1}g_{2},g_{3},g_{4})\alpha(g_{1},g_{2},g_{3}g_{4})} =1
\end{align}
where it is assumed that the group action on the $G$-module $\mathbb{C}^\times$ is chosen to be trivial. Equation \eqref{coc3} is nothing else than the defining relation for a group 3-cocyle $\alpha \in Z^3(G,\mathbb{C}^\times)$ (see app.~\ref{app:coho}). 

Interestingly, we can read off from the corollary triangle relations that 
\begin{align}
	\ell_g &= \alpha(\mathbbm{1}, \mathbbm{1}, g)^{-1}1_g  \\
	r_g &= \alpha(g, \mathbbm{1}, \mathbbm{1})1_g \; .
\end{align} 
The fundamental triangle relation then provides the equality
\begin{equation}
	\alpha(g, \mathbbm{1}, h) = \alpha(g, \mathbbm{1}, \mathbbm{1})
	\alpha(\mathbbm{1}, \mathbbm{1},h)
\end{equation}
which implies that the maps $\ell_g$ and $r_g$ are trivial if and only if 
\begin{equation}
	\alpha(g,\mathbbm{1},h) = 1, \forall \, g,h \in G
\end{equation}
{\it i.e.} $\alpha$ is a normalized 3-cocycle. Note that for any given equivalence class in ${H}^3(G, \mathbb{C}^\times)$, there is a normalized 3-cocycle. It follows from the fact that two cocycles $\alpha$, $\alpha'$ in the equivalence class $[\alpha]$ are related by $\alpha' = \alpha d^{(2)} \beta$ with $\beta$ a 2-cochain so that
\begin{equation}
	d^{(2)} \beta (g_1,g_2,g_3) = \frac{\beta(g_2,g_3)\beta(g_1,g_2g_3)}{\beta(g_1g_2,g_3)\beta(g_1,g_2)} \; .
\end{equation} 
is a 3-coboundary. By definition a 3-coboundary is a 3-cocycle which is in the equivalence class of the trivial one (see app.~\ref{app:coho} for details).  
By choosing $\beta(\mathbbm{1},g) = \alpha^{-1}(\mathbbm{1},\mathbbm{1},g)$ and $\beta(g,\mathbbm{1}) = \alpha(g,\mathbbm{1},\mathbbm{1})$, $\alpha'=\alpha d^{(2)} \beta$ defines a normalized cocycle.
Therefore, for a given equivalence class $[\alpha] \in {H}^3(G,\mathbb{C}^\times)$, there always exists a normalized cocycle $\alpha_{\rm norm.} \in [\alpha]$ such that the monoidal category whose associator is given by $\alpha_{\rm norm.}$ has trivial right and left unitors. Note that for a finite group $G$, there is no difference between the cohomology groups $H^d(G,\mathbb{C}^\times)$ and $H^d(G,{\rm U}(1))$. In the following, we will therefore interchange freely $\mathbb{C}^\times$-valued and ${\rm U}(1)$-valued group cocycles.

\bigskip \noindent
Let us now spell out the correspondence between (2+1)d string net models and monoidal categories in the case of $\mathbb{C}$--$\text{Vec}_G^\alpha$. The super-selection sectors labeling the 1-simplices of the triangulation correspond to the simple objects in the category $\mathbb{C}$--$\text{Vec}_G^\alpha$ which are in one-to-one correspondence with the group elements. The branching rules which dictate which colorings of the one-skeleton of the triangulation are admissible is directly provided by the functor $m$. For simple objects, the branching rules therefore boil down to the multiplication rule of the group $G$. Let us consider a 2-simplex labeled by $(abc)$ with $a<b<c$, we associate super-selection sectors $\delta_{g_{ab}}$ and $\delta_{g_{bc}}$ to the edges $(ab)$ and $(bc)$, respectively, such that the super-selection sector labeling the remaining edge $(ac)$ is obtained as $\delta_{g_{ab}} \otimes \delta_{g_{bc}} \cong \delta_{g_{ab}g_{bc}}$. 

Using the correspondence we just established, we also have
\begin{equation}
	\label{convCat}
	 \pent{02}{24}{0.4}{1} \longleftrightarrow (\delta_{g_1} \otimes \delta_{g_2}) \otimes (\delta_{g_3} \otimes \delta_{g_4}) \longleftrightarrow {\sss (\bul \bul)(\bul \bul)}
\end{equation}
which follows from the definition of the branching rules in terms of the functor, as well as the convention $\delta_{g_b} \equiv \delta_{g_{b-1b}}$, where $\delta_{g_{ab}}$ denotes the simple object in $\mathcal{C}$ labeling the edge $(ab)$. We also introduced in \eqref{convCat} a more abstract notation which keeps the objects as well as the tensor product implicit, we will use it when we only want to emphasize the bracketings and when no confusion is possible. Furthermore, we can identify each map $\alpha_{(abcd)}$ performing a $\mathcal{P}_{2 \mapsto 2}$ move with the corresponding associator, {\it e.g.}
\begin{align*}
 	\nn
	\alpha_{(0134)} \longleftrightarrow \alpha(g_{01}, g_{13}, g_{34}) &=  \alpha(g_{01},g_{12}g_{23},g_{34})\\ 
	& \equiv \alpha(g_1,g_2g_3, g_4) \; ,
\end{align*}
Applying the convention \eqref{convCat} to every pentagon appearing in \eqref{coher2D} together with the previous correspondence between $\alpha$ thought as an associator and $\alpha$ thought as a map performing a $\mathcal{P}_{2 \mapsto 2}$ move, we obtain the coherence relation for the associator. Similarly, we obtain from the triangle equation \eqref{triangle2D} the coherence relation for the left and right unitors $\ell_g$ and $r_g$. Using these two relations, we can also deduce the corollary triangle relations. 

To complete our review of (2+1)d cohomological string net models, it remains to describe the topological invariant underlying this model as well as the corresponding lattice Hamiltonian whose ground states are described by the fixed point wave functions.

\subsection{State-sum invariant and lattice Hamiltonian \label{sec:ham}}
\noindent
The topological invariant associated with the Hamiltonian model under consideration was introduced by Dijkgraaf and Witten in \onlinecite{dijkgraaf1990topological}. We will provide a more detailed definition of the model further, but for our current purpose, it is enough to use its explicit formulation in terms of triangulation and group cohomology. Let us consider a simplicial decomposition of a manifold $\mathcal{M}$ and introduce an ordering of the vertices so as to orientate the simplices. To every 1-simplex, we assign a group element $g_{ab} \in G$ such that for every 2-simplex $\triangle_2 = (abc)$ with $a<b<c$, we impose the flatness condition
\begin{equation*}
g_{ac} = g_{ab} \cdot 	g_{bc} \; .
\end{equation*}
To every oriented 3-simplex $\triangle_3 = (0123)$, we then assign the following topological action in terms of the group 3-cocycle $\alpha$
\begin{equation*}
	\alpha(\triangle_3) = \alpha(g_{01}, g_{12}, g_{23})
\end{equation*}
such that the state-sum reads
\begin{equation}
	\label{ZDW1}
	\mathcal{Z}_{\rm \alpha}^G(\mathcal{M}) = \frac{1}{|G|^{|\triangle_0|}} \sum_{g}\prod_{\triangle_3}\alpha^{\epsilon(\triangle_3)}(g) \; . 
\end{equation}
The topological invariance follows from the fact that the cocycle condition ensures that the quantity above remains unchanged under transformations of the triangulation.

Let us now define the lattice Hamiltonian whose ground states are given by the fixed point wave functions previously defined. To each 0-simplex of the graph we associate an operator $\mathbb{A}_{(\triangle_0)}$ which enforces the \emph{twisted gauge invariance}. To every 2-simplex, we associate a projector $\mathbb{B}_{(\triangle_2)}$ which enforces the \emph{zero-flux condition}. 

The zero-flux condition for a given 2-simplex is nothing else than the branching rules. Therefore, the action of the $\mathbb{B}_{(\triangle_2)}$ operator is particularly straightforward as it simply enforces the super-selection sectors labeling the boundary of every 2-simplex to be compatible. We have for instance
\begin{equation}
	\mathbb{B}_{(012)} \triangleright 
	\bigg| \triONE{0.7}{1} \bigg\rangle= \delta_{g_{01}g_{12}, g_{02}} \bigg| \triONE{0.7}{1} \bigg\rangle \; .
\end{equation}
The action of the operator $\mathbb{A}$ is more subtle, however there is particularly convenient way of writing it in terms of the partition function \eqref{ZDW1}. Let $\triangle_3=(0123)$ be a 3-simplex of the triangulation $\triangle$. The corresponding topological action is given by $\alpha(\triangle_3) \equiv \alpha(g_1,g_2,g_3)$ such that $g_{b} \equiv g_{b-1b}$. Let us now perform a gauge transformation at the vertex $(0)$ with gauge parameter $k$ such that $g _1\rightarrow kg_1 $. It follows from the cocycle condition that the change for the toplogical action under such gauge transformation is given by
\begin{equation}
	\label{gaugeDW}
	\alpha(\triangle_3) \rightarrow \alpha(\triangle_3) \frac{\alpha(k,g_1,g_2)\alpha(k,g_1g_2,g_3)}{\alpha(k,g_1,g_2g_3)} \; .
\end{equation} 
This suggests a way to construct the action of the operator $\mathbb{A}$. In the string net picture, the situation we are interested in consists in the action of the operator $\mathbb{A}$ on a vertex $v$ shared by several 2-simplices. In such situation, the action can be expressed in terms of the partition function as follows
\begin{equation}
	\label{actionA}
	\mathbb{A}_{v} = \frac{1}{|G|}\sum_{g_{v'v}}\mathcal{Z}_{\alpha}^G[v' \cup_{\rm j} \text{cl}(v)]
\end{equation}
where $\text{cl}(v)$ defines the minimal subcomplex of the triangulation containing all the simplices which contains $v$ as a subsimplex and $\cup_{\rm j}$ refers to the \emph{join} of the corresponding simplices \cite{Williamson:2016evv}. Furthermore, the vertex $v'$ appear just before $v$ in the ordered list of the vertices. For instance, let us consider a vertex $(0)$ shared by (and only by) three 2-simplices. In this case $\text{cl}(0)$ is just the subcomplex containing these three 2-simplices so that the join operation $(0') \cup_{\mathsf{j}}\text{cl}(0)$ reads
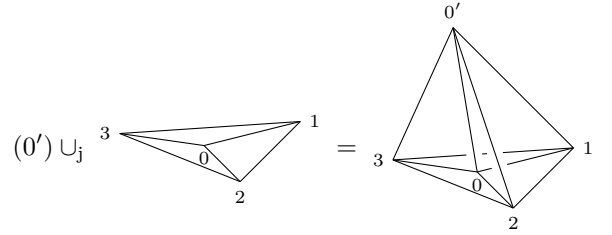
\begin{equation*}
	(0') \cup_{\rm j} \actionA{0.8}{1}{1} = \actionA{0.8}{2}{2}
\end{equation*}
with $0'<0<1<2<3$. In the following, we will refer to such operation as a \emph{tent move}. Using the convention regarding the orientation of the three-simplices, we apply \eqref{actionA} and we obtain the following action:
\begin{align}
	&\mathbb{A}_{(0)}\triangleright \bigg| \triPachner{1}{0} \bigg\rangle \\ \nn 
	& \q = \frac{1}{|G|}\sum_{g_{0'0}}
	\frac{\alpha(g_{0'0},g_1,g_2)\alpha(g_{0'0},g_1g_2,g_3)}{\alpha(g_{0'0},g_1,g_2g_3)}
	\bigg|\triPachner{1}{1} \bigg\rangle 
\end{align}
which matches the expression \eqref{gaugeDW} for $k=g_{0'0}$, as expected. We will use a similar argument later in the context of 2-group gauge theory models of (3+1)d topological phases. The operators $\mathbb{A}_{(\triangle_0)}$ and $\mathbb{B}_{(\triangle_2)}$ commute and the lattice Hamiltonian projector finally reads
\begin{equation}
	\label{latticeH}
	\mathbb{H} = -\sum_{\triangle_0} \mathbb{A}_{(\triangle_0)} - \sum_{\triangle_2} \mathbb{B}_{(\triangle_2)} \; .
\end{equation}
Note that the construction presented above, and in particular the equation \eqref{actionA}, is rather general. In fact we will see later how it can also be applied to higher gauge models. The starting point of our study of (3+1)d gauge models of topological phases is a straightforward three-dimensional generalization of the lattice Hamiltonian model \eqref{latticeH}, i.e. a Hamiltonian realization of the 4d Dijkgraaf-Witten model.

\newpage
\section{(3+1)d string net models}
\label{sec:3dstring}
\noindent
In the previous section, we briefly recalled the category-theoretical flavor of string net models. In particular, we emphasized how the Hamiltonian realization of the three-dimensional Dijkgraaf-Witten model is defined in terms of the category $\mathbb{C}$--$\text{Vec}_G^\alpha$ which is characterized by a discrete group $G$ and cohomology class $[\alpha] \in H^3(G,\mathbb{C}^\times)$. In this section, we consider a straightforward generalization of the previous construction where the underlying topological quantum field theory is the four-dimensional Dijkgraaf-Witten theory. As such, the corresponding Hamiltonian model will be characterized by a discrete group $G$ and a cohomology class  $[\pi] \in H^4(G,\mathbb{C}^\times)$. We will emphasize in particular the relation between (2+1)d and (3+1)d models from a geometrical and category theoretical point of view. This will serve as a motivation for the higher gauge theory models we consider in the following section.

\subsection{Fixed point wave functions}
\noindent
The string net model we are interested in is defined on the one-skeleton $\triangle$ of the triangulation of a three-manifold. The super-selection sectors still label the 1-simplices of $\triangle$ and the branching rules are enforced to hold at every 2-simplex. If we were to define the model on the dual complex $\triangle^*$ such that a 1-simplex of $\triangle$ is dual to a face in $\triangle^*$, we would obtain a  \emph{membrane net model} where super-selection sectors live on faces and branching rules are define at edges. Because $\triangle$ is a triangulation, the branching rules at every edge would only involve three super-selection sectors labeling the faces meeting at the edge.

The main difference between the (3+1)d model and its (2+1)d analogue is the replacement of the 2--2 Pachner move $\mathcal{P}_{2 \mapsto 2}$ (and its corollary 1--3 Pachner move) by the 2--3 Pachner move $\mathcal{P}_{2 \mapsto 3}$ (and its corollary 1--4 Pachner move). Using the same notation as previously, this local unitary transformation reads
\begin{equation}
	\nn
	\Bigg| \PPONE{1}{1} \Bigg\rangle = \sum_{x_{14}} \pi_{(01234)} \Bigg| \PPTWO{1}{1}{-0.3} \Bigg\rangle
\end{equation}
\noindent 
where $\pi_{(\triangle_4)}$ is a map which depends on the super-selection sectors labeling the edges of a 4-simplex $\triangle_4$. The 4-simplices $\{\triangle_4\}$ are identified by the vertices labeling the union of 3-simplices on which the $\mathcal{P}_{2 \mapsto 3}$ move acts. In the general case, the transformation of the corresponding states is performed by the map $\pi_{(\triangle_4)} {}^{\epsilon(\triangle_4)}$ where $\epsilon(\triangle_4) \pm 1 $ is a factor which is defined according to the following convention:\\~\\~\\

\noindent\textsc{Convention for the 2--3 Pachner move}\vspace{-0.7em}
\begin{leftbar}\noindent 
	Pick one of the 3-simplex $\triangle_3$ in the source polytope whose vertices define the 4-simplex $\triangle_4$. Let us assume this 3-simplex is labeled by $(abcd)$ such that $a < b < c < d$. The remaining vertex is labeled by $e$. Determine the orientation of the 3-simplex using the convention defined earlier. If it takes an odd number of permutations to bring the list $(e,a,b,c,d)$ to the ascending ordered one, then $\epsilon(\triangle_4)=\epsilon(\triangle_3)$, otherwise $\epsilon(\triangle_4)= -\epsilon(\triangle_3)$. For instance, in the example above, we can choose as 3-simplex $\triangle_3 = (0123)$. It takes four permutations to bring $(4,0,1,2,3)$ to $(0,1,2,3,4)$ and therefore $\pi_{(01234)}^{\epsilon(01234)} = \pi_{(01234)}^{-\epsilon(0123)} = \pi_{(01234)}$. 
\end{leftbar}
\noindent
The same $\mathcal{P}_{2 \mapsto 3}$ move appeared before in \eqref{pachner23} when describing the coherence relation \eqref{coher2D} of the $\mathcal{P}_{2 \mapsto 2}$ move. It turns out that the commutativity of the pentagon diagram can be interpreted as having a trivial map performing the corresponding $\mathcal{P}_{2 \mapsto 3}$ move. Conversely, we will see how the fact of having a non-trivial map $\pi$ performing the $\mathcal{P}_{2 \mapsto 3}$ can be understood as relaxing the commutativity of a given pentagon diagram.

Analogous to the pentagon relation, there is a coherence relation for the $\mathcal{P}_{2 \mapsto 3}$ move. Considering the union of three 3-simplices, there are two different ways to apply three $\mathcal{P}_{2 \mapsto 3}$ moves so as to obtain a complex which is the union of six 3-simplices. Applying the convention we just defined, this is illustrated by the following diagram
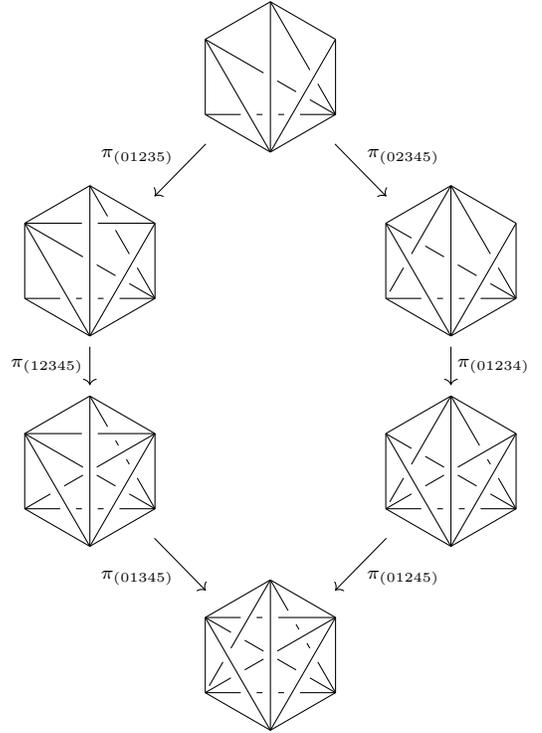
\begin{equation} \nn
	\begin{tikzcd}[column sep=1em, row sep=1.5em]
		{}
		& \PPPONE{0} \ar[dr,"{\pi_{(02345)}}",rightarrow, shorten <= -0.7em, shorten >= -0.7em]
		& {}
		\\[-1em]
		\PPPSIX{0} \ar[ur,"{\pi_{(01235)}}",leftarrow, shorten <= -0.7em, shorten >= -0.7em]
		& {}
		& \PPPTWO{0} 	\ar[d,"{ \pi_{(01234)}}",rightarrow]
		\\
		\PPPFIVE{0} 	\ar[u,"{\pi_{(12345)}}",leftarrow]
		& {}
		& \PPPTHREE{0}
		\\[-1em]
		{}
		& \PPPFOUR{0} 
		\ar[ur,"{\pi_{(01245)}}"',leftarrow, shorten <= -0.7em, shorten >= -0.7em]
		\ar[ul,"{\pi_{(01345)}}", leftarrow, shorten <= -0.7em, shorten >= -0.7em]
		& {}
	\end{tikzcd}
\end{equation}
from which we read off the following coherence relation
\begin{equation}
	\pi_{(02345)}\pi_{(01245)}\pi_{(01234)} = 
	\pi_{(01235)}\pi_{(01345)}\pi_{(12345)} \; . 
\end{equation}

\begin{figure*}[t]
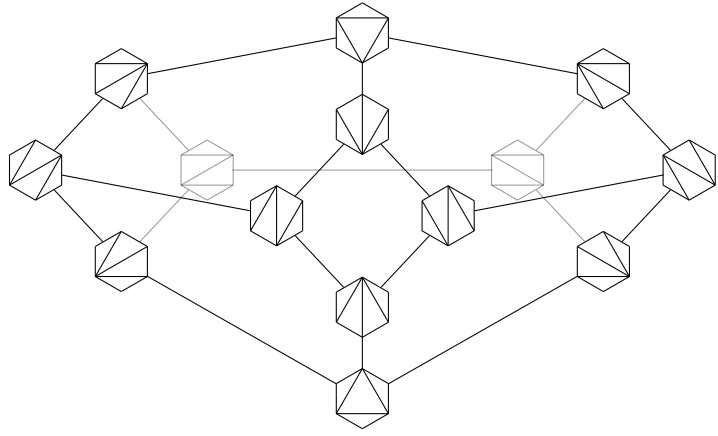

	\assocTWO{}	
	\caption{Graphical depiction of the fifth Stasheff polytope. Each vertex of this three-dimensional polytope is associated with a triangulation of a $6$-gon while every edge represents a 2--2 Pachner move. The combinatorics of this object is exactly the same as the one of the coherence relation of the pentagonator which is one of the structural 2-morphisms entering the definition of the monoidal 2-category we are interested in.}	
	\label{fifthS}	
\end{figure*}

\noindent
Before describing the categorical structure yielding such model, we want to make a geometrical remark. First of all, the six 4-simplices on which the map $\pi$ is defined bound a 5-simplex $(012345)$. The same 4-simplices appear in a higher-dimensional Pachner move, namely a 3--3 Pachner move. Furthermore, as mentioned above, the pentagon relation whose combinatorics corresponds to the one of the fourth Stasheff polytope is associated to a 2--3 Pachner move, and conversely, we can associate to each $\mathcal{P}_{2 \mapsto 3}$ move appearing in the coherence relation of the map $\pi_{(\triangle_4)}$ a Stasheff polytope $\mathcal{K}_4(\triangle_4)$. For instance, the top-right Pachner move performed by the map $\pi_{(02345)}$ is associated with a pentagon relation as follows:
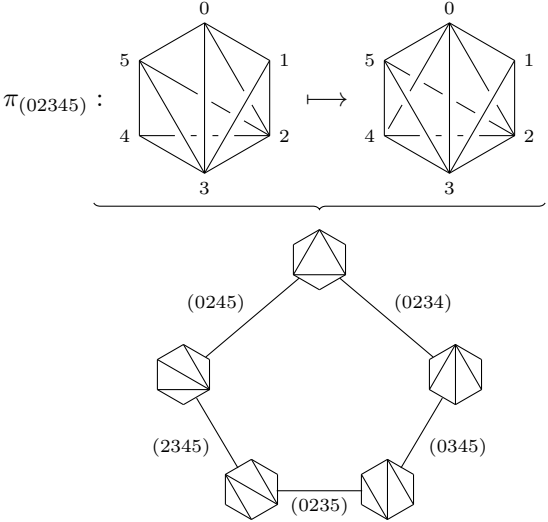
\begin{equation*} 
	\hspace{-3.5em}	
	\pi_{(02345)} : \PPPONE{1} \longmapsto \PPPTWO{1}
\end{equation*}\\[-3.9em]
\begin{equation*}
	\begin{tikzpicture}
	\draw[decoration={brace,mirror,raise=0em},decorate] (-3,0) -- (3,0);
	\end{tikzpicture}
\end{equation*}\\[-3.8em]
\begin{equation*}
	\begin{tikzcd}[column sep=-0.3em, row sep=1.3em]
	{}
	& {}
	& \hex{02}{24}{40}{0.4}{0}{1}
	\ar[drr,"{(0234)}",dash, shorten <= -0.9em, shorten >= -0.8em]
	& {}
	& {}
	\\
	\hex{02}{24}{25}{0.4}{0}{1} 
	\ar[urr,"{(0245)}",dash, shorten <= -0.8em, shorten >= -0.9em]
	& {}
	& {}
	& {}
	& \hex{02}{03}{04}{0.4}{0}{1}	
	\ar[dl,"{ (0345)}",dash, shorten <= -0.76em, shorten >= -0.85em]
	\\
	{}
	& \hex{02}{25}{53}{0.4}{0}{1}
	\ar[ul,"{(2345)}",dash, shorten >= -0.76em, shorten <= -0.85em]
	& {}
	& \hex{02}{35}{03}{0.4}{0}{1}
	\ar[ll,"{(0235)}", dash, shorten <= -0.5em, shorten >= -0.5em]
	& {}
	\end{tikzcd} 
\end{equation*}

\noindent
From these remarks, it is tempting to think that the coherence relation for the map $\pi$ should be associated with a fifth Stasheff polytope $\mathcal{K}_5(012345)$. It is indeed true and can be seen by drawing all the polytopes $\mathcal{K}_4$ associated with each $\mathcal{P}_{2 \mapsto 3}$ move as we did above and then build the 3d polytope by identifying identical vertices and edges. The result of this procedure is presented in fig.~\ref{fifthS}. This simple geometrical remark motivates the study of this model from a category theoretical point of view since the same combinatorial structure naturally appears in the definition of a \emph{monoidal 2-category}.

\subsection{2-categorical aspects}
\noindent
It has been conjectured for some time that the proper language to describe the input data of a (3+1)d state sum TQFT is the one of \emph{fusion 2-categories} (see for instance \onlinecite{Cui:2016bmd}). Over the years, there have been several attempts to define such structures, but there is no widely agreed upon definition yet. We will not attempt to fill this gap in this paper. We will merely focus on specific examples which fall within the proposals of Kapranov, Voevodsky and Mackaay \cite{kapranov19942, 1998math......5030M, 1999math......3003M}. More specifically, we will be interested in higher-gauge theory models which can be understood as special cases of Mackaay's construction of spherical 2-categories for finite groups.

We can think of \emph{2-categories} as a generalization of 1-categories where on top of the objects and 1-morphisms, we add 2-morphisms so that the 2-category consists of: $(i)$ Objects, $(ii)$ 1-morhpisms between objects and $(iii)$ 2-morphisms between 1-morphisms. Roughly speaking, these 2-morphisms can be used in order to weaken equalities between 1-morphisms in the same way as 1-morphisms are used to weaken the equalities \eqref{strictCat1} and $\eqref{strictCat2}$ of strict monoidal 1-categories. 

Recall that there are three natural 1-morphisms entering the definition of a monoidal 1-category, namely an associator, a right unitor and a left unitor. These 1-morphisms satisfy the pentagon equation and the three triangle equations. By adding specific 2-morphisms, it is possible to weaken such equalities. These 2-morphisms are structural 2-morphisms of the corresponding monoidal 2-category. 

The first structural 2-morphism is the \emph{pentagonator} $\pi$ which relaxes the pentagon equation. In the language of commutative diagrams, this is represented as follows:

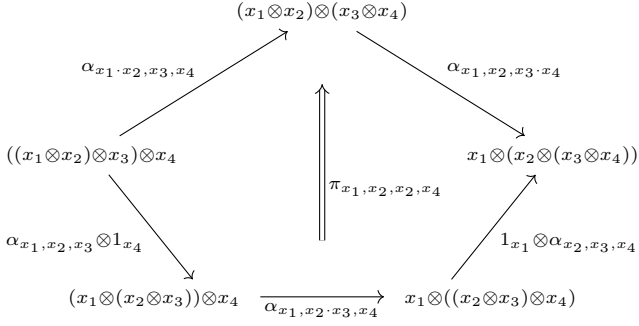
\begin{equation*}
	\begin{tikzcd}[column sep=-5em, row sep=4em]
	{}
	& {}
	& {\scriptstyle(x_1 \otimes x_2) \otimes (x_3 \otimes x_4) }
	\ar[drr,"{\alpha_{x_{1},x_{2},x_{3} \cdot x_{4}}}",rightarrow]
	\ar[dd,"\pi_{x_1,x_2,x_2,x_4}",Leftarrow, pos=0.6, shorten <= 2em, shorten >= 2em]	
	& {}
	& {}
	\\
	{\scriptstyle ((x_1 \otimes x_2) \otimes x_3) \otimes x_4 }
	\ar[urr,"{\alpha_{x_{1}\cdot x_{2}, x_{3}, x_{4}}}",rightarrow]
	& {}
	& {} 
	& {}
	& {\scriptstyle x_1 \otimes (x_2 \otimes (x_3 \otimes x_4)) 	}
	\ar[dl,"{1_{x_1} \otimes \alpha_{x_{2},x_{3},x_{4}}}",leftarrow]
	\\
	{}
	& {\scriptstyle (x_1 \otimes (x_2 \otimes x_3))\otimes x_4	}\q\q\q\q
	\ar[ul,"{\alpha_{x_1,x_2,x_3} \otimes 1_{x_4}}",leftarrow]
	& {}
	& \q\q\q\q {\scriptstyle x_1 \otimes ((x_2 \otimes x_3) \otimes x_4) }	\ar[ll,"{\alpha_{x_1,x_2\cdot x_3,x_4}}"',rightarrow, shorten <= -1em, shorten >= -1em]
	& {}
	\end{tikzcd}
\end{equation*}\\[-0.5em]

\noindent
where we define the 2-morphism $\pi$ as going from the composition of the three morphisms at the bottom to the composition of the two morphisms at the top. In general, simple arrows represent 1-morphisms while 2-morphisms are represented with a double arrow. Similarly we can introduce three structural 2-morphisms $\lambda$, $\mu$ and $\rho$ referred to as \emph{triangulators} which weaken the triangle relations:

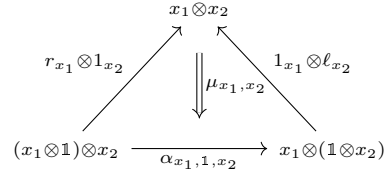
\begin{equation*}
	\begin{tikzcd}[column sep=1em, row sep=4em]
		{}
		& {\scriptstyle x_1 \otimes x_2}
		\ar[dl,"{{r}_{x_1}\otimes 1_{x_2} }"',leftarrow]
		\ar[dr,"{1_{x_1} \otimes {\ell}_{x_2} }",leftarrow]
		\ar[d,"\mu_{x_1,x_2}" ,Rightarrow, shorten <= 1em, shorten >= 1em]
		& {}
		\\
		{\scriptstyle (x_1 \otimes \mathbbm{1}) \otimes x_2}
		\ar[rr, "{\alpha_{x_1,\mathbbm{1}, x_2}}"',rightarrow]
		& {}
		& {\scriptstyle x_1 \otimes ( \mathbbm{1} \otimes x_2) }
	\end{tikzcd}
\end{equation*} 
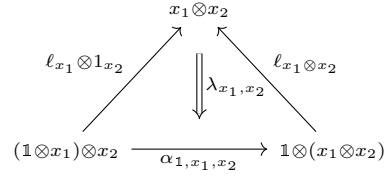
\vspace{-0.6em}
\begin{equation*}
	\begin{tikzcd}[column sep=1em, row sep=4em]
		{}
		& {\scriptstyle x_1 \otimes x_2}
		\ar[dl,"{{\ell}_{x_1}\otimes 1_{x_2} }"',leftarrow]
		\ar[dr,"{{\ell}_{x_1 \otimes x_2} }",leftarrow]
		\ar[d,"\lambda_{x_1,x_2}" ,Rightarrow, shorten <= 1em, shorten >= 1em]
		& {}
		\\
		{\scriptstyle (\mathbbm{1} \otimes x_1) \otimes x_2}
		\ar[rr, "{\alpha_{\mathbbm{1}, x_1, x_2}}"',rightarrow]
		& {}
		& {\scriptstyle \mathbbm{1} \otimes ( x_1 \otimes x_2) }
	\end{tikzcd}
\end{equation*} 
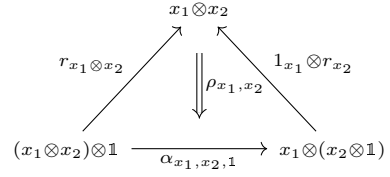
\vspace{-0.6em}
\begin{equation*}
	\begin{tikzcd}[column sep=1em, row sep=4em]
		{}
		& {\scriptstyle x_1 \otimes x_2}
		\ar[dl,"{{r}_{x_1 \otimes x_2} }"',leftarrow]
		\ar[dr,"{1_{x_1} \otimes {r}_{x_2} }",leftarrow]
		\ar[d,"\rho_{x_1,x_2}" ,Rightarrow, shorten <= 1em, shorten >= 1em]
		& {}
		\\
		{\scriptstyle (x_1 \otimes x_2) \otimes \mathbbm{1}}
		\ar[rr, "{\alpha_{x_1,x_2,\mathbbm{1}}}"',rightarrow]
		& {}
		& {\scriptstyle x_1 \otimes ( x_2 \otimes \mathbbm{1}) }
	\end{tikzcd}
\end{equation*} 
\noindent
Naturally theses 2-morphisms satisfy coherence relations of their own. We display below the commutative diagrams for the pentagonator $\pi$ together with one of the diagrams involving the triangulators:

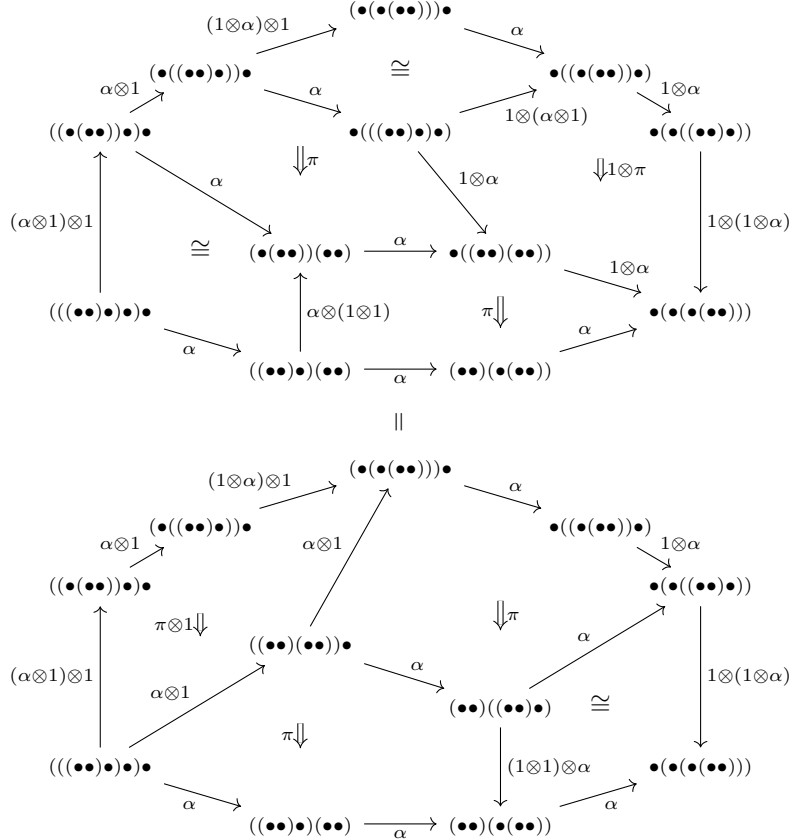
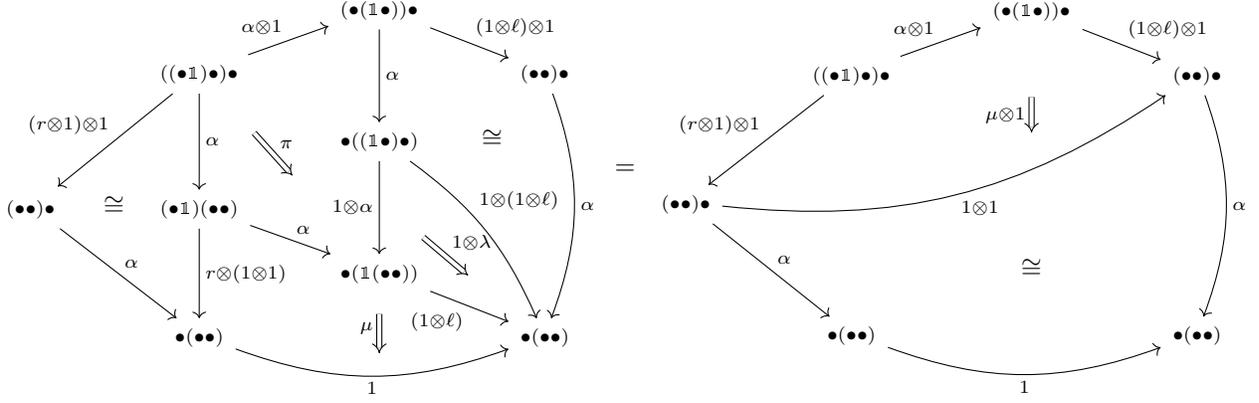
\begin{widetext}
	\begin{equation*}
		\begin{tikzcd}[column sep=-1em, row sep=0.8em]
			{} & {} & {}
			& {\sss(\bul(\bul (\bul \bul )))\bul}
			\ar[drr, "\alpha", rightarrow]
			& {} & {} & {}
			\\
			{} 
			& {\sss(\bul((\bul \bul) \bul ))\bul}
			\ar[urr, "(1 \otimes \alpha) \otimes 1", rightarrow]
			\ar[drr, "\alpha", rightarrow]
			& {}
			\ar[ddd, "\pi", Rightarrow, shorten <= 2.2em, shorten >= 2.2em] 
			& \cong & {}
			& {\sss \bul((\bul (\bul \bul ))\bul)}
			\ar[ddd, "1 \otimes \pi", Rightarrow, shorten <= 2.5em, shorten >= 2.5em]
			\ar[dr, "1 \otimes \alpha", rightarrow]
			& {}
			\\ 
			{\sss ((\bul(\bul \bul)) \bul )\bul}
			\ar[ur, "\alpha \otimes 1", rightarrow]
			\ar[ddrr, "\alpha", rightarrow]
			& {} & {} 
			& {\sss \bul(((\bul \bul) \bul )\bul)}
			\ar[urr, "1 \otimes (\alpha \otimes 1)"', rightarrow]
			\ar[ddr, "1 \otimes \alpha", rightarrow]
			& {} & {}
			& {\sss \bul(\bul((\bul \bul )\bul))}
			\ar[ddd, "1 \otimes ( 1 \otimes \alpha)", rightarrow]
			\\ \phantom{{\sss (\bul)}} \\ 
			{} & \cong
			& {\sss (\bul(\bul \bul)) (\bul \bul)}
			\ar[rr, "\alpha", rightarrow]
			& {} 
			& {\sss \bul((\bul \bul)( \bul \bul))}
			\ar[drr, "1 \otimes \alpha", rightarrow]
			& {} & {}
			\\
			{\sss (((\bul \bul)\bul )\bul) \bul}
			\ar[uuu, "(\alpha \otimes 1) \otimes 1", rightarrow]
			\ar[drr, "\alpha"', rightarrow]
			& {} & {} & {} & {} & {}
			& {\sss \bul(\bul (\bul( \bul \bul)))}
			\\
			{} & {}
			& {\sss ((\bul \bul) \bul)( \bul \bul)}
			\ar[uu, "\alpha \otimes (1 \otimes 1)"', rightarrow]
			\ar[rr, "\alpha"', rightarrow]
			& {}  
			& {\sss (\bul \bul)( \bul (\bul \bul))}
			\ar[uu, "\pi", Leftarrow, shorten <= .9em, shorten >= .9em]
			\ar[urr, "\alpha", rightarrow]
			& {} & {}
		\end{tikzcd} 
	\end{equation*}\vspace{-1.5em}
	\begin{equation}
		\label{CoherPent}
		\mathbin{\rotatebox[origin=c]{90}{=}}
	\end{equation}\vspace{-2em}
	\begin{equation*}		
		\begin{tikzcd}[column sep=-1em, row sep=0.8em]
			{} & {} & {}
			& {\sss (\bul(\bul (\bul \bul )))\bul}
			\ar[drr, "\alpha", rightarrow]
			& {} & {} & {}
			\\
			{} 
			& {\sss (\bul((\bul \bul) \bul ))\bul }
			\ar[ddd, "\pi \otimes 1"', Rightarrow, shorten <= 2.5em, shorten >= 2.5em]
			\ar[urr, "(1 \otimes \alpha) \otimes 1", rightarrow]
			& {} & {} & {}
			\ar[ddd, "\pi", Rightarrow, shorten <= 2.2em, shorten >= 2.2em]
			& {\sss \bul((\bul (\bul \bul ))\bul)}
			\ar[dr, "1 \otimes \alpha", rightarrow]	
			& {}
			\\ 
			{\sss ((\bul(\bul \bul)) \bul )\bul}
			\ar[ur, "\alpha \otimes 1", rightarrow]
			& {} & {} & {} & {} & {}
			& {\sss \bul(\bul((\bul \bul )\bul))}
			\ar[ddd, "1 \otimes ( 1 \otimes \alpha)", rightarrow]		
			\\
			{} & {}
			& {\sss ((\bul \bul)( \bul \bul)) \bul}
			\ar[uuur, "\alpha \otimes 1", rightarrow]
			\ar[drr, "\alpha", rightarrow]		
			& {} & {} & {} & {}
			\\
			{} & {} & {} & {} 
			& {\sss (\bul \bul)(( \bul \bul) \bul) }
			\ar[uurr, "\alpha", rightarrow]
			\ar[dd, "(1 \otimes 1)\otimes \alpha", rightarrow]		
			& \cong & {}
			\\
			{\sss (((\bul \bul) \bul) \bul) \bul}
			\ar[uuu, "(\alpha \otimes 1) \otimes 1", rightarrow]
			\ar[uurr, "\alpha \otimes 1", rightarrow]
			\ar[drr, "\alpha"', rightarrow]			
			& {} & {} & {} & {} & {}
			& {\sss \bul (\bul( \bul (\bul \bul)))}
			\\
			{} & {}
			& {\sss ((\bul \bul) \bul)( \bul \bul)}
			\ar[uuu, "\pi", Leftarrow, shorten <= 2.2em, shorten >= 2.2em]
			\ar[rr, "\alpha"', rightarrow]	
			& {} 
			& {\sss (\bul \bul)( \bul (\bul \bul))}
			\ar[urr, "\alpha"', rightarrow]	
			& {} & {}
		\end{tikzcd}
	\end{equation*}			
	\begin{equation}
		\begin{tikzcd}[column sep=3em, row sep=1em]
			{} & {}
			&{\sss (\bul(\uni \bul )) \bul}
			\ar[dr, "(1 \otimes \ell)\otimes 1", rightarrow]
			\ar[dd, "\alpha", rightarrow]
			& {}
			\\
			{}
			& {\sss ((\bul \uni) \bul ) \bul}
			\ar[dddr, "\pi", Rightarrow, shorten <= 2em, shorten >= 4em, pos=0.38] 
			\ar[ur, "\alpha \otimes 1", rightarrow]
			\ar[dd, "\alpha", rightarrow]
			& {}
			& {\sss (\bul \bul) \bul}
			\ar[dddd, "\alpha", rightarrow, bend left =20]
			\\
			{} & {}
			& {\sss \bul (( \uni \bul) \bul )}
			\ar[dd, "1 \otimes \alpha"'{name=R1}, rightarrow]
			\ar[dddr, "1\otimes (1 \otimes \ell)", rightarrow, pos=0.4, bend left =15] 
			& \hspace{-4em }\cong 
			\\
			{\sss (\bul \bul) \bul} \q\;\; \cong \hspace{-2.5em}
			\ar[uur, "(r \otimes 1) \otimes 1", leftarrow]
			\ar[ddr, "\alpha", rightarrow]
			& {\sss(  \bul \uni)(\bul \bul)}
			\ar[dr, "\alpha", rightarrow]
			& {} & {}
			\\
			{} & {}
			&{\sss \bul ( \uni (\bul \bul) )}
			\ar[d,"\mu"', pos=0.9, Rightarrow, shorten >= -0.8em, shorten <= 0.8em]
			\ar[dr, "(1 \otimes \ell)"', rightarrow]
			& {}
			\\
			{}
			& {\sss \bul (\bul \bul)}
			\ar[uu, "r \otimes (1 \otimes 1)"', leftarrow]
			\ar[rr, "1"', rightarrow, bend right =20]
			& {}
			& {\sss \bul ( \bul \bul )}
			\ar[Rightarrow, from=R1, to=6-4, shorten >= 2.5em, shorten <= 2.5em, "1 \otimes \lambda",pos=0.5, start anchor = {[yshift=1em]}]
		\end{tikzcd}
		\; = \;
		\begin{tikzcd}[column sep=3em, row sep=1em]
			{} & {}
			&{\sss (\bul(\uni \bul )) \bul}
			\ar[dr, "(1 \otimes \ell)\otimes 1", rightarrow]
			& {}
			\\
			{}
			& {\sss ((\bul \uni) \bul ) \bul} 
			\ar[ur, "\alpha \otimes 1", rightarrow]
			& {}
			& {\sss (\bul \bul) \bul}
			\ar[dddd, "\alpha", rightarrow, bend left =20]
			\\
			{} & {}
			& \phantom{{\sss \bul (( \uni \bul) \bul )}}
			& {}
			\\
			{\sss (\bul \bul) \bul} 
			\ar[uur, "(r \otimes 1) \otimes 1", leftarrow]
			\ar[ddr, "\alpha", rightarrow]
			\ar[uurrr, "1 \otimes 1"', rightarrow, bend right =20]
			& \phantom{{\sss ( \bul \uni)(\bul \bul)}}
			& {}
			\ar[uuu, "\mu \otimes 1", Leftarrow, shorten <= 2.5em, shorten >= 2.5em] 
			& {}
			\\
			{} 
			&\phantom{{\sss \bul ( \uni (\bul \bul) )}}
			& \cong & {}
			\\
			{}
			& {\sss \bul (\bul \bul)}
			\ar[rr, "1"', rightarrow, bend right =20]
			& {}
			& {\sss \bul ( \bul \bul )}
		\end{tikzcd}		
	\end{equation}
\end{widetext}
where we keep implicit the objects as well as the tensor product between them.
The first commutative diagram contains six pentagons. Each one of them corresponds to a pentagon relation weakened by a pentagonator. It turns out that the geometry and the combinatorics of this diagram is exactly the one the fifth Stasheff polytope drawn in fig.~\ref{fifthS}. This is the confirmation that the map $\pi$ performing a $\mathcal{P}_{2 \mapsto 3}$ move in our (3+1)d string net model correspond to the pentagonator of a given 2-category which we are about to describe. We further notice on the commutative diagram (or the corresponding Stasheff polytope $\mathcal{K}_5(012345)$) the presence of three squares decorated by $\cong$. In the categorical formalism, this indicates the fact that composition of the 1-morphisms bounding this square is isomorphic to the the trivial morphism. More specifically, it translates here the fact that two independent $\mathcal{P}_{2 \mapsto 2}$ moves commute.

\bigskip \noindent
We now have enough information to describe the 2-category yielding the lattice Hamiltonian whose ground state is described by the four-dimensional DW model, namely the category of $G$-graded 2-vector spaces denoted by $\mathbb{C}$--$2\text{Vec}_G^\pi$. First, let us define the category of \emph{2-vector spaces}. There exists several definitions for such structure which are more or less abstract. In order to keep the level of abstraction to a minimum, we will only present the so-called \emph{coordinatized} version. For more details about 2-vector spaces and monoidal 2-categories, refer to \onlinecite{kapranov19942}.  

Recall that the $\mathbb{C}$--$\text{Vec}$ category is the category whose objects are finite-dimensional complex vector spaces and 1-morphisms are linear maps. The coordinatized version of this monoidal category is a category whose objects are formal symbols $V[n]$ with $n=0,1,2,\dots$ and such that the collection ${\rm Hom}(V[m],V[n])$ of 1-morphisms from $V[m]$ to $V[n]$ is provided by the set of all $m \times n $ complex matrices $M$. 

We define a \emph{2-matrix} $\mathcal{M}$ as a matrix whose entries $\mathcal{M}_{ij}$ are finite-dimensional vector spaces valued in $\mathbb{C}$. The coordinatized version of $\mathbb{C}$--$2\text{Vec}$ is a 2-category whose objects are formal symbols $\mathcal{V}[n]$ with $n=0,1,2,\ldots$. The  collection ${\rm Hom}(\mathcal{V}[m],\mathcal{V}[n])$ of 1-morphisms from $\mathcal{V}[m]$ to $\mathcal{V}[n]$ is provided by the set of all $m \times n $ 2-matrices $\mathcal{M}$. The collection ${\rm Hom}(\mathcal{M},\mathcal{N})$ of 2-morphisms from $\mathcal{M}$ to $\mathcal{N}$ such that $\mathcal{M}, \mathcal{N}: \mathcal{V}[m] \rightarrow \mathcal{V}[n]$ is the set of all matrices $T$ of linear operators $T_{ij}: \mathcal{M}_{ij} \rightarrow \mathcal{N}_{ij}$. In the following, it will be enough to remember that $\mathbb{C}$--$2\text{Vec}$ \emph{is a 2-category whose objects are 2-vector spaces, 1-morphisms are matrices of vector spaces and 2-morphisms are matrices of linear maps}. 

Let us now describe the monoidal 2-category $\mathbb{C}$--$2\text{Vec}_G^\pi$. The objects of this category are $G$-graded 2-vector spaces such that the simple objects denoted by $\delta_{g}$ are the 1-dimensional $G$-graded 2-vector spaces and are in one-to-one correspondence with the group elements in $G$. As before, the tensor product between simple objects is given by the group multiplication. We suppose that the associator, the right unitor and the left unitor are all trivial 1-morphisms. However, we weaken the pentagon equation as well as the triangle equations by introducing a pentagonator $\pi$ and three triangulators $\lambda, \rho, \mu$. These can be represented by functions of the group valued in $\mathbb{C}^\times$ such that $\pi : G^4 \rightarrow \mathbb{C}^\times$ and $\lambda, \rho, \mu : G^2 \rightarrow \mathbb{C}^\times$. This means that despite the triviality of the structural 1-morphisms as functions on $G$, the coherence relations are satisfied only up to a complex phase.

The correspondence between the (3+1)d string net models and the category $\mathbb{C}$--$2\text{Vec}_G^\pi$ follows exactly the same rule as in the (2+1)d case. For instance, we have
\begin{equation*}
\hspace{-0.5em} \hex{04}{14}{24}{0.4}{1}{1} \longleftrightarrow (\delta_{g_1} \otimes ( \delta_{g_2} \otimes ( \delta_{g_3 }\otimes \delta_{g_4}))) \otimes \delta_{g_5} \longleftrightarrow {\sss (\bul (\bul(\bul \bul)))\bul}
\end{equation*}
as well as 
\begin{align*}
\pi_{(01245)} \longleftrightarrow \pi(g_{01}, g_{12}, g_{24}, g_{45}) &=  \pi(g_{01},g_{12}, g_{23}g_{34}, g_{45})\\ & \equiv \pi(g_1,g_2, g_3g_4, g_5) 
\end{align*}
where $g_b \equiv g_{b-1b}$. Applying this correspondence to the commutative diagram of the pentagonator, we obtain 
\begin{equation}
\nn
\frac{\pi(g_1g_2, g_3,g_4,g_5	)\pi(g_1,g_2,g_3 g_4,g_5)\pi(g_1,g_2,g_3,g_4)}{\pi(g_1,g_2,g_3,g_4g_5)\pi(g_1,g_2 g_3,g_4,g_5)\pi(g_2,g_3,g_4,g_5)} =1 
\end{equation}
which is the group 4-cocycle relation for a cohomology class $[\pi] \in H^4(G,\mathbb{C}^\times)$. Similarly, we can obtain from the coherence relation between the pentagonator and the triangulators different equations between functions on $G$. It turns out that these relations imply that the triangulators are trivial 2-morphisms if and only if the pentagonator is a normalized 4-cocycle. In the following, we will make the assumption that the triangulators are trivial so that the pentagonator is a normalized 4-cocycle. This is reminiscent of the fact the right and left unitors are trivial if and only if the associator is a normalized 3-cocycle. Therefore the Hamiltonian model under consideration is characterized by a discrete group $G$ and a cohomology class $[\pi] \in H^4(G,\mathbb{C}^\times)$.

The lattice Hamiltonian can finally be built following exactly the same procedure as in the (2+1)d case. In particular, the $\mathbb{B}_{(\triangle_2)}$ operator acts exactly in the same way since the branching rules are the same as before while the action of the $\mathbb{A}_{(\triangle_0)}$ is now defined in terms of the partition function of the four-dimensional DW model. Let $\triangle_4  = (01234)$ be a 4-simplex of the triangulation $\triangle$. The corresponding topological action is given by $\pi(\triangle_4) \equiv \pi(g_1,g_2,g_3,g_4)$. Let us now perform a gauge transformation at he vertex $(0)$ with gauge parameter $k$ such that $g_1 \rightarrow kg_1$. The 4-cocycle condition implies that the topological action changes under such gauge transformation as 
\begin{equation}
	\pi(\triangle_4) \rightarrow \pi(\triangle_4)
	\frac{\pi(k,g_1,g_2,g_3g_4)\pi(k,g_1g_2,g_3,g_4)}{\pi(k,g_1,g_2g_3,g_4)\pi(k,g_1,g_2,g_3)} \; .
\end{equation}
which suggests that we can use the same definition \eqref{actionA} for the operator $\mathbb{A}_{(\triangle_0)}$ as in the (2+1)d case but it is now associated to a four-dimensional tent move together with the four-dimensional DW model. The lattice Hamiltonian is finally provided by \eqref{latticeH}.

\subsection{Generalizations \label{sec:gen}}
\noindent
In the previous section, we presented the 2-category $\mathbb{C}$--$2\text{Vec}_G^\pi$ of $G$-graded 2-vector spaces. This 2-category is such that the simple objects are labeled by group elements in $G$, the underlying 1-category is strict, and there is a single non-trivial structural 2-morphism $\pi: G^4 \rightarrow \mathbb{C}^\times$ which is a normalized 4-cocycle of $H^4(G,\mathbb{C}^\times)$. This 2-morphism is a pentagonator weakening the pentagon equation. 

We emphasized, both in (2+1)d and (3+1)d, how the DW model has a lattice gauge theory interpretation. In particular, the state-sum model is invariant under gauge transformations acting at the vertices of the triangulation. We will now see how it is possible to define a new model with a higher gauge theory interpretation by weakening some axioms of the 2-category $\mathbb{C}$--$2\text{Vec}_G^\pi$. Effectively, it will amount to replace the gauge group $G$ by a gauge 2-group $\mathbb{G}$ which will correspond to the underlying 1-category of the new 2-category.

Roughly speaking, the strategy we follow consists in making our 2-category richer by weakening some of its defining axioms. Firstly, we would like to turn the underlying strict monoidal 1-category into a weak monoidal 1-category. Recall that the 1-morphisms of $\mathbb{C}$--$2\text{Vec}_G^\pi$ correspond to matrices of vector spaces. Let $H$ be an abelian group such that $G$ act trivially on it. We will now consider that the 1-morphisms correspond to matrices of $H$-graded vector spaces. We denote the corresponding category by $\mathbb{C}$--$2\text{Vec}_{G,H}^\pi$. Let us further introduce an associator $\alpha$ turning the underlying category of $\mathbb{C}$--$2\text{Vec}_{G,H}^\pi$ into a weak monoidal 1-category. As before, it is enough to define such associators on the simple objects which are labeled by elements in $G$. Furthermore, we make the assumption that all the 1-morphisms are simple. This implies that given a simple object $\delta_g$ we have $\text{Hom}(\delta_g, \delta_g) = H$. Together with the pentagon equation, this implies that the associator $\alpha$ is an element of the cohomology class $H^3(G,H)$ and we denote the new category by $\mathbb{C}$--$2\text{Vec}_{G,H}^{\pi, \alpha}$. We could complete the weak monoidal category by introducing right and left unitors, however we will pick them to be trivial which in turn implies that $\alpha$ is a normalized cocycle.

At this point we would like to emphasize the fact that the associator $\alpha$ is a group 3-cocycle valued in $H$ despite the fact that the 2-category $\mathbb{C}$--$2\text{Vec}_{G,H}^\pi$ possesses a pentagonator $\pi$ which weakens the corresponding pentagon equation. It is not a contradiction in the sense that the associator does satisfy the 3-cocycle condition as a 3-cochain valued in $H$ but the pentagon equation is only satisfied up to a morphism valued in $\mathbb{C}^\times$.

The 2-category under consideration is now a monoidal 2-category whose underlying 1-category is a weak monoidal 1-category and such that there is a pentagonator valued in $\mathbb{C}^\times$. We can enrich further the category by introducing additional strutural 2-morphisms. The most obvious possibility consists in weakening the associativity of the 1-morphisms by introducing a 1-associator. However, in order to keep the notation simpler, we will assume such 1-associator to be trivial. There is another possible set of 2-morphisms which can be introduced. We mentioned above that the commutative diagram of the pentagonator whose combinatorics is the one of the fifth Stasheff polytope involves three commutating squares which translates the fact that independent $\mathcal{P}_{2 \mapsto 2}$ should commute. We now would like to introduce three 2-morphisms $\tau^{\rm I}$, $\tau^{\rm II}$ and $\tau^{\rm III}$ which weaken these three square commutative diagrams as\footnote{In other words, the squarators define the pseudo-naturality of the associator $\alpha$.}
\begin{equation*}
	\begin{tikzcd}[column sep=1.5em, row sep=4em]
	{\sss ((\bul (\bul \bul))\bul)\bul}
	\ar[rr, "\alpha", rightarrow]
	&{}
	\ar[d,"\tau^{\rm I}", Leftarrow, shorten <= 1em, shorten >= 1em]
	&{\sss (\bul(\bul \bul))(\bul \bul)}
	\\
	{\sss (((\bul \bul) \bul ) \bul ) \bul}
	\ar[u, "(\alpha \otimes 1) \otimes 1", rightarrow]
	\ar[rr, "\alpha"', rightarrow]
	&{}
	&{\sss ((\bul \bul) \bul)(\bul \bul )}
	\ar[u,"\alpha \otimes (1 \otimes 1)"']
	\end{tikzcd}
\end{equation*}

\begin{equation*}
	\begin{tikzcd}[column sep=1.5em, row sep=4em]
	{\sss (\bul (\bul (\bul \bul)))\bul}
	\ar[rr, "\alpha", rightarrow]
	&{}
	\ar[d,"\tau^{\rm II}", Leftarrow, shorten <= 1em, shorten >= 1em]
	&{\sss \bul((\bul( \bul \bul )) \bul)}
	\\
	{\sss (\bul ((\bul \bul) \bul )) \bul}
	\ar[u, "(1 \otimes \alpha) \otimes 1", rightarrow]
	\ar[rr, "\alpha"', rightarrow]
	&{}
	&{\sss \bul((( \bul \bul)\bul) \bul )}
	\ar[u,"1 \otimes (\alpha \otimes 1)"']
	\end{tikzcd}
\end{equation*}

\begin{equation*}
	\begin{tikzcd}[column sep=1.5em, row sep=4em]
	{\sss (\bul \bul)(( \bul \bul)\bul)}
	\ar[rr, "\alpha", rightarrow]
	&{}
	\ar[d,"\tau^{\rm III}", Leftarrow, shorten <= 1em, shorten >= 1em]
	&{\sss \bul(\bul ((\bul \bul) \bul))}
	\\
	{\sss (\bul \bul)( \bul ( \bul  \bul))}
	\ar[u, "(1 \otimes 1) \otimes \alpha", rightarrow]
	\ar[rr, "\alpha"', rightarrow]
	&{}
	&{\sss \bul (\bul (\bul(\bul \bul )))}
	\ar[u,"1 \otimes (1 \otimes \alpha)"']
	\end{tikzcd}
\end{equation*}
We will refer to these 2-morphisms as \emph{squarators}. Remember that we work under the assumption that all the 1-morphisms are simple such that the set of invertible morphism between 1-morphisms is $\mathbb{C}^\times$. The first 2-morphism in the commutative diagrams above explicitly reads $\tau^{\rm I}_{\alpha(g_1g_2g_3, g_4,g_5),\alpha(g_1,g_2,g_3)}$. These squarators satisfy consistency conditions given by three-dimensional commutative diagrams which may involve additional morphisms. We will not reproduce them here but these can be found in \onlinecite{kapranov19942}.

After introduction of the squarators, the commutativity of the diagram  \eqref{CoherPent} now reads
\begin{align*}
	d^{(4)} \pi (g_1,g_2,g_3,g_4,g_5) = 
	&\tau^{\rm I}_{\alpha(g_1g_2g_3, g_4,g_5),\alpha(g_1,g_2,g_3)}\\
	\times & \tau^{\rm II}_{\alpha(g_1,g_2,g_3g_4g_5),\alpha(g_3,g_4,g_5)} \\
	\times & \tau^{\rm III}_{\alpha(g_1,g_2g_3g_4,g_5),\alpha(g_2,g_3,g_4)}					
\end{align*}
so that the pentagonator $\pi$ cannot be thought as a cocycle anymore but a 4-cochain. We can check that the right-hand-side of this equation is a 5-cocycle in $H^5(G,\mathbb{C}^\times)$ referred to as the \emph{obstruction class}. However, in order for this equation to be consistent, it is necessary for this obstruction class to be in the equivalence class of the trivial cocycle. In such a case, the model is qualified as ``obstruction free''.

Putting everything together, we obtain our final category denoted by $\mathbb{C}$--$2\text{Vec}_{G,H}^{\pi, \alpha, \tau}$. This is a monoidal 2-category which consists of:
\begin{enumerate}[itemsep=0.4em,parsep=1pt,leftmargin=*]
	\item[$\circ$] A collection of simple objects $\delta_{g}$ which correspond to the group elements of $G$. 
	\item[$\circ$] Associator $\alpha : G^3 \rightarrow H$ which is a 3-cocyle in $Z^3(G,H)$.
	\item[$\circ$] Pentagonator $\pi  :  G^4 \rightarrow \mathbb{C}^\times$
	which is a 4-cochain in $C^4(G,\mathbb{C}^\times)$.
	\item[$\circ$] Squarators $\tau^{\rm I}, \tau^{\rm II}, \tau^{\rm III} : H^2 \rightarrow \mathbb{C}^\times$.
\end{enumerate}
\noindent
We can make the restricting assumption that the squarators $\tau^{\rm I}$, $\tau^{\rm II}$ and $\tau^{\rm III}$ are chosen so as to evaluate to a function 
\begin{equation*}
	\tau: h_1,h_2 \mapsto e^{2\pi i B(h_1,h_2)}
\end{equation*} 
with $B({\sss \bul}, {\sss \bul})$ a bilinear form on $H$ valued in $\mathbb{Q}/\mathbb{Z}$ such that
\begin{align*} 
	B(h_1,h_3)+B(h_2,h_3) &= B(h_1+h_2,h_3) \\ 
	B(h_1,h_2)+B(h_1,h_3) &= B(h_1,h_2+h_3) \; . 
\end{align*}
We further require for simplicity that for all $h_1,h_2 \in H$, $\tau(h_1,h_2)\tau(h_2,h_1)=\unit$. The bilinearity of $B$ implies the following consistency conditions
\begin{align*}
	\tau(h_1,h_3)\tau(h_2,h_3) &= \tau(h_1+h_2,h_3) \\
	\tau(h_1,h_2)\tau(h_1,h_3) &= \tau(h_1,h_2+h_3) 	
\end{align*}
which correspond the horizontal and vertical compositions of the corresponding 2-morphisms, respectively. We can then check that these conditions are compatible with the coherence relations of the morphisms $\tau^{\rm I}$, $\tau^{\rm II}$ and $\tau^{\rm III}$ (still in the case where the 1-associator for the 1-morphisms is trivial). Under this assumption, we can make a connection with the state-sum model introduced in \onlinecite{Cheng:2017ftw} where the squarators are identified with the $R$-matrix of the abelian group $H$ seen as an abelian braided fusion category with trivial $F$-symbols. The consistency condition above is nothing else than the hexagon equation for such an $R$-matrix.

Let us assume for a moment that the abelian group $H$ is the permutation group of two elements $\mathbb{Z}_2$. In this case, we can choose the bilinear form $B$ so that $\tau_{h_1,h_2} = (-1)^{h_1+h_2}$. The coherence relation for the pentagonator then reads
\begin{align*}
	d^{(4)} \pi (g_1, \dots,g_5) = 
	&(-1)^{\alpha(g_1g_2g_3, g_4,g_5)+\alpha(g_1,g_2,g_3)}\\
	\times & (-1)^{\alpha(g_1,g_2,g_3g_4g_5)+\alpha(g_3,g_4,g_5)} \\
	\times & (-1)^{\alpha(g_1,g_2g_3g_4,g_5)+\alpha(g_2,g_3,g_4)}
\end{align*}
which can be more succinctly rewritten as 
\begin{equation}
	d^{(4)}\pi = (-1)^{{\rm Sq}^2(\alpha)}
\end{equation}
where ${\rm Sq}^2(\alpha)$ is the Steenrod square of the 3-cocycle $\alpha$. The Steenrod square which was first introduced in \onlinecite{steenrod1947products} can be expressed in terms of the $1$-cup product $\smile_1$ according to ${\rm Sq}^2(\alpha) := \alpha \smile_1 \alpha$ (see app.~\ref{app:coho} for more details). This obviously reminds of the super-cohomology appearing in fermionic topological orders \cite{2012arXiv1201.2648G, Bhardwaj:2016clt}.

With the introduction of the squarators 2-morphisms, the 4-cochain $\pi \in C^4(G, \mathbb{C}^\times)$ does not close anymore so that a DW state-sum model built out of $\pi$ would not describe a topological invariant anymore, i.e., invariance under the $\mathcal{P}_{3 \mapsto 3}$ would not be satisfied. Equivalently, for the corresponding Hamiltonian model, this means that the coherence relation of the map $\pi$ performing the $\mathcal{P}_{2 \mapsto 3}$ move would not be satisfied anymore. However, we can build a new state-sum model. 

As for the DW model, we consider a simplicial decomposition of the manifold $\mathcal{M}$ and introduce an ordering of the vertices. To every 1-simplex, we assign a group element $g_{ab} \in G$ such that for every 2-simplex $\triangle_2 = (abc)$ with $a<b<c$, we impose the flatness condition
\begin{equation*}
	g_{ac} = g_{ab} \cdot 	g_{bc} \; .
\end{equation*}
In order to incorporate the associator which is represented by a cohomology class $[\alpha] \in H^3(G,H)$ to the model, we assign to every 2-simplex a group element $h_{abc} \in H$ such that for every 3-simplex $\triangle_3 = (abcd)$, we impose the \emph{fake 2-flatness} condition
\begin{equation*}
	h_{bcd} - h_{acd} + h_{abd} - h_{abc} = \alpha(g_{ab}, g_{bc}, g_{cd} ) \equiv \alpha_{abcd} \; .
\end{equation*} 
Finally, to every oriented 4-simplex $\triangle_4 =(01234)$, we assign the topological action
\begin{align}
	\nn
	\omega(\triangle_4) = &\tau(h_{012},h_{234})\tau(h_{034}, \alpha_{0123})^{-1}
	\tau(h_{014}, \alpha_{1234})^{-1} \\ \times &\pi(g_{01},g_{12},g_{23},g_{34})
	\label{topo2Gr}
\end{align}
such that we can define the following state sum model
\begin{align}
	\label{SM2gr}
	\mathcal{Z}_{\omega}(\mathcal{M}) = \frac{1}{\#_{G,H,\triangle}} \sum_{g,h}\prod_{\triangle_4}\omega^{\epsilon(\triangle_4)}(g|h)
\end{align}
with $\#_{G,H, \triangle} = |G|^{|\triangle_0|}|H|^{|\triangle_1| - |\triangle_0|}$. It is a bit lengthy but straightforward to check, using the condition satisfied by $\pi$ which now can be rewritten
\begin{align*}
	d^{(4)} \pi (g_1,g_2,g_3,g_4,g_5) = 
	&\tau(\alpha_{0345},\alpha_{0123})\\
	\times & \tau(\alpha_{0125},\alpha_{2345}) \\
	\times & \tau(\alpha_{0145},\alpha_{1234})		\; ,			
\end{align*}
that the state-sum above is invariant under $\mathcal{P}_{3 \mapsto 3}$ moves \cite{Cheng:2017ftw}. In the following, we will explain how such model can be understood as a cohomological model built out of a 2-group $\mathbb{G}$. We will then sketch the construction of the corresponding Hamiltonian model.

\section{Higher gauge theory models}
\label{sec:higherGauge}
\noindent
Previously we introduced the 2-category $\mathbb{C}$--$2\text{Vec}_{G,H}^{\pi, \alpha, \tau}$ by progressively weakening some axioms of the category yielding the Hamiltonian realization of the four-dimensional DW model. We will now motivate this categorical structure from another point of view, namely \emph{higher gauge theory}. In particular, we are going to show to what extent the partition function \eqref{SM2gr} is a special case of topological gauge theories built from a 2-group. 

\subsection{2-connections and strict 2-groups}
\noindent
Ordinarily a gauge theory is built from a connection on a principle $G$-bundle where $G$ is a discrete or continuous group. For topologically trivial bundles, a connection is fully prescribed by holonomies of a 1-form field $A \in \Omega^{1}(\mathcal{M},\mathfrak g)$ where $\mathfrak g$ is the Lie algebra of $G$. In case $G$ is a discrete group, it is often more convenient to work with a local description wherein the analog of a 1-connection is a 1-cochain valued in $G$ denoted by $ {\bfg}${\footnote {Here ${\bfg}$ may be understood as a \v{C}ech 1-cochain for a general group $G$ (abelian or non-abelian) or as a simplicial 1-cochain for $G$ abelian. In either case, there is an equivalent description using a $G$-coloring of a triangulation of the manifold $\cM$ which we make explicit in the coming sections. We will often find it necessary to work with cochains or $G$-colorings satisfying some (possibly twisted) cocycle conditions. These structures can be generalized straightforwardly to higher-form groups $H_{[q]}$ and higher 2-groups $\mathbb G_{[q]}$.}}. A concise way to describe a $G$-connection is via the holonomy functor
\begin{align}
	\text{hol}_1: \mathcal P_1(\cM)\to G
\end{align}  
from the path groupoid of the manifold $\mathcal M$ to the group $G$. This definition requires some explanation.  A \emph{path groupoid} on a manifold $\mathcal M$ is a category where the objects are points on $\mathcal M$ and morphisms are paths between points. This is a groupoid as each morphism has an inverse that is provided by the reversed-path. In the path groupoid, composing morphisms is given by composition of paths. In the present discussion, we won't worry about smooth structure too much however it has been dealt with carefully in \cite{baez2011invitation, schreiber2007parallel, bartels2004higher}. 

 Similarly, in categorical terms, a group $G$ is a groupoid with a single object. Morphisms from the object to itself are labeled by elements in $G$, composition of objects is by group multiplication and the identity morphism is provided by the identity element of the group. Flat connections can have non-trvivial holonomies along non-contractible paths only and therefore a flat $G$-connection can be defined as a homorphism from the fundamental group $\pi_1(\cM)$ to $G$. Locally, this means that a flat $G$-connection is fully characterized by a 1-cocycle valued in $G$ satisfying\begin{equation*}
	d {\bfg} = \mathbbm{1} \; .
\end{equation*} 

\smallskip \noindent
In analogy, a 2-connection on a 2-bundle can be defined most succinctly as a 2-functor from a certain path 2-groupoid $\mathcal P_2(\cM)$ to a 2-group $\mathbb G$:
\begin{align}
	\text{hol}_2: \mathcal P_2(\cM)\to \mathbb G \; .
\end{align}
A \emph{path 2-groupoid} is a 2-groupoid, i.e, a 2-category in which every 1-morphism and every 2-morphism is invertible. More specifically a path 2-groupoid $\mathcal P_2(\cM)$ for a manifold $\mathcal M$ is a 2-category with points as objects, paths as 1-morphisms and surfaces between paths as 2-morphisms. A flat 2-connection may be defined in analogy to a flat connection as a homomorphism from the fundamental 2-group \cite{maclane19623} to a 2-group\footnote{Roughly speaking a fundamental 2-group of $\mathcal M$ may be thought of as a path 2-groupoid of $\mathcal M$ with paths and surfaces replaced homotopy classes of paths and surfaces, respectively.}.

 A \emph{2-group} $\mathbb G$ can be defined in several ways. In its weak version, it can be succinctly defined as a (weak) monoidal category whose 1-morphisms are all invertible and objects are all weakly invertible. We will make use of this definition eventually, however for now, we will consider \emph{strict 2-groups} which can be defined as a 2-groupoid with a single object such that all 1-morphisms and 2-morphisms are invertible. The 1-morphisms  are labeled by elements of a group $\Gamma_1$ which compose according to the group multiplication in $\Gamma_1$, i.e.
\begin{equation}
	\begin{tikzcd}[column sep=2em, row sep=1.5em]
		{\sss \bul}
		\ar[r, "g", leftarrow, bend left=60] 
		&
		{\sss \bul}
		\ar[r, "g'", leftarrow, bend left=60] 
		&
		{\sss \bul}
	\end{tikzcd}
	=
	\begin{tikzcd}[column sep=1.5em, row sep=1.5em]
		{\sss \bul}
		\ar[r, "gg'", leftarrow, bend left=60] 
		&
		{\sss \bul}
	\end{tikzcd}
	\nonumber
\end{equation}
where $g,g'\in \Gamma_1$ and ${\sss \bullet}$ refers to the single object in the category. The set of 2-morphisms from the identity 1-morphism $\mathbbm{1}\in \Gamma_1$ to $ g \in \Gamma_1$ form a group $\Gamma_2$, i.e
\begin{align}
	\Gamma_2:= \bigoplus_{g\in \Gamma_1}\text{Hom}(\mathbbm{1},g) \; .
\end{align}  
Given such a 2-morphism $h$, we denote the target 1-morphism by $t(h)$ such that
\begin{equation}
	\begin{tikzcd}[column sep=3em, row sep=1.5em]
		{\sss \bul}
		\ar[r, "\mathbbm{1}"{name=AR1}, leftarrow, bend left=60] 
		\ar[r, "t(h)"'{name=AR2}, leftarrow, bend right=60]
		& {\sss \bul}
		\ar[Rightarrow, from=AR1, to=AR2,"h",pos=0.5, shorten >= 0.5em, shorten <= 0.5em]
	\end{tikzcd}
	\equiv 
	\begin{tikzcd}[column sep=4em, row sep=1.5em]
		{\sss \bul}
		\ar[r, "\mathbbm{1}"{name=1}, leftarrow, bend left=65] 
		\ar[r, ""{name=2}, leftarrow] 
		\ar[r, "t(h)"'{name=3}, leftarrow, bend right=65]
		&
		{\sss \bul}
		\ar[Rightarrow, from=1, to=2, "h", pos=0.6, shorten >= -0.1em, shorten <= 0.5em]
		\ar[Rightarrow, from=2, to=3, "\mathbbm{1}", pos=0.4, shorten >= 0.5em, shorten <= 0.5em]   
	\end{tikzcd}
\end{equation}
where the identification follows from the vertical composition being given by the group multiplication in $\Gamma_2$. As a matter of fact, 2-morphisms can be composed both vertically and horizontally. The horizontal composition implies in particular that for two 2-morphisms $h,h'\in \Gamma_2$, it is simply the product $hh'\in \Gamma_2$, which in turn implies that the map $t:\Gamma_{2}\to \Gamma_1$ is a group homomorphism, i.e. 
\begin{equation}
	\begin{tikzcd}[column sep=3em, row sep=1.5em]
		{\sss \bul}
		\ar[r, "\mathbbm{1}"{name=AR1}, leftarrow, bend left=60] 
		\ar[r, "t(hh')"'{name=AR2}, leftarrow, bend right=60]
		& {\sss \bul}
		\ar[Rightarrow, from=AR1, to=AR2,"h h'",pos=0.5, shorten >= 0.5em, shorten <= 0.5em]
	\end{tikzcd}
		= 
	\begin{tikzcd}[column sep=3em, row sep=1.5em]
		{\sss \bul}
		\ar[r, "\mathbbm{1}"{name=1}, leftarrow, bend left=65] 
		\ar[r, "t(h)"'{name=2}, leftarrow, bend right=65]
		&
		{\sss \bul}
		\ar[Rightarrow, from=1, to=2,"h",pos=0.5, shorten >= 0.5em, shorten <= 0.5em]
		\ar[r, "\mathbbm{1}"{name=3}, leftarrow, bend left=65] 
		\ar[r, "t(h')"'{name=4}, leftarrow, bend right=65]
		&
		{\sss \bul}
		\ar[Rightarrow, from=3, to=4,"h'",pos=0.5, shorten >= 0.5em, shorten <= 0.5em]   
	\end{tikzcd} \; .
\end{equation}
The horizontal composition of 2-morphisms can also be used to define a $\Gamma_1$-group action $\tr$ on $\Gamma_2$ such that $\tr: \Gamma_{1}\times \Gamma_{2}\to \Gamma_2$ via
\begin{equation}
	\begin{tikzcd}[column sep=3em, row sep=1.5em]
		{\sss \bul}
		\ar[r, "\mathbbm{1}"{name=5}, leftarrow, bend left=60] 
		\ar[r, "t(g\tr h)"'{name=6}, leftarrow, bend right=60]
		&
		{\sss \bul}
		\ar[Rightarrow, from=5, to=6,"g\tr  h",pos=0.5, shorten >= 0.5em, shorten <= 0.5em]	
	\end{tikzcd}	
	:= 
	\begin{tikzcd}[column sep=3em, row sep=1.5em]	
		{\sss \bul}
		\ar[r, "g"{name=1}, leftarrow, bend left=65] 
		\ar[r, "g"'{name=2}, leftarrow, bend right=65]
		&
		{\sss \bul}
		\ar[Rightarrow, from=1, to=2,"1",pos=0.5, shorten >= 0.5em, shorten <= 0.5em]	
		\ar[r, "\mathbbm{1}"{name=3}, leftarrow, bend left=65] 
		\ar[r, "t(h)"'{name=4}, leftarrow, bend right=65]
		&
		{\sss \bul}
		\ar[Rightarrow, from=3, to=4,"h",pos=0.5, shorten >= 0.5em, shorten <= 0.5em]   
                \ar[r, "g^{-1}"{name=5}, leftarrow, bend left=65] 
		\ar[r, "g^{-1}"'{name=6}, leftarrow, bend right=65]
		&
		{\sss \bul}
		\ar[Rightarrow, from=5, to=6,"1",pos=0.5, shorten >= 0.5em, shorten <= 0.5em]   
	\end{tikzcd} \; .
	\label{hor_comp_id}
\end{equation}
It turns out that the quadruple $\mathbb G = (\Gamma_1,\Gamma_2,t,\tr)$ introduced above defines an algebraic structure known as a \emph{crossed module}\cite{whitehead1949combinatorial}. In particular, we can show that the set $(\Gamma_1,\Gamma_2,t,\tr)$ satisfies the properties 
\begin{align}
	t(g\tr h)=&\; gt(h)g^{-1} \nonumber \\
	t(h)\tr h'=&\; hh'h^{-1} 
\end{align}
which enter the definition of a crossed module. The first relation follows from the definition of $\tr$ while the second one is a bit more subtle and can be shown diagrammatically as follows
\begin{eqnarray}
	hh'h^{-1}&:=&
	\begin{tikzcd}[column sep=3em, row sep=1.5em]
	{\sss \bul}
	\ar[r, "\mathbbm{1}"{name=1}, leftarrow, bend left=65] 
	\ar[r, "t(h)"'{name=2}, leftarrow, bend right=65]
	&
	{\sss \bul}
	\ar[Rightarrow, from=1, to=2,"h", pos=0.5, shorten >= 0.5em, shorten <= 0.5em]
	\ar[r, "\mathbbm{1}"{name=3}, leftarrow, bend left=65] 
	\ar[r, "t(h')"'{name=4}, leftarrow, bend right=65]
	&
	{\sss \bul}
	\ar[Rightarrow, from=3, to=4,"h'",pos=0.5, shorten >= 0.5em, shorten <= 0.5em]   
	\ar[r, "\mathbbm{1}"{name=3}, leftarrow, bend left=65] 
	\ar[r, "t(h^{-1})"'{name=4}, leftarrow, bend right=65]
	&
	{\sss \bul}
	\ar[Rightarrow, from=3, to=4,"h^{-1}",pos=0.5, shorten >= 0.5em, shorten <= 0.5em]   
	\end{tikzcd} \nonumber \\
	&=&
	\begin{tikzcd}[column sep=4em, row sep=1.5em]
	{\sss \bul}
	\ar[r, "\mathbbm{1}"{name=1}, leftarrow, bend left=65] 
	\ar[r, ""{name=2}, leftarrow] 
	\ar[r, "t(h)"'{name=3}, leftarrow, bend right=65]
	&
	{\sss \bul}
	\ar[Rightarrow, from=1, to=2, "h", pos=0.6, shorten >= -0.1em, shorten <= 0.5em]
	\ar[Rightarrow, from=2, to=3, "\mathbbm{1}", pos=0.4, shorten >= 0.5em, shorten <= 0.5em]
	\ar[r, "\mathbbm{1}"{name=4}, leftarrow, bend left=65] 
	\ar[r, "t(h')"'{name=5}, leftarrow, bend right=65]
	\ar[r, ""{name=6}, leftarrow] 
	&
	{\sss \bul}
	\ar[Rightarrow, from=4, to=6, "\mathbbm{1}",pos=0.6, shorten >= -0.1em, shorten <= 0.5em]
	\ar[Rightarrow, from=6, to=5, "h'",pos=0.4, shorten >= 0.5em, shorten <= 0.5em]		
	\ar[r, "\mathbbm{1}"{name=7}, leftarrow, bend left=65] 
	\ar[r, "t(h^{-1})"'{name=8}, leftarrow, bend right=65]
	\ar[r, ""{name=9}, leftarrow] 
	&
	{\sss \bul}
	\ar[Rightarrow, from=7, to=9,"h^{-1}",pos=0.6, shorten >= -0.1em, shorten <= 0.5em]   
	\ar[Rightarrow, from=9, to=8,"\mathbbm{1}",pos=0.4, shorten >= 0.5em, shorten <= 0.5em]   
	\end{tikzcd}\nonumber  \\
	&=&
	\begin{tikzcd}[column sep=3em, row sep=1.5em]
	{\sss \bul}
	\ar[r, "t(h)"{name=1}, leftarrow, bend left=65] 
	\ar[r, "t(h)"'{name=2}, leftarrow, bend right=65]
	&
	{\sss \bul}
	\ar[Rightarrow, from=1, to=2, "\mathbbm{1}",pos=0.5, shorten >= 0.5em, shorten <= 0.5em]
	\ar[r, "\mathbbm{1}"{name=3}, leftarrow, bend left=65] 
	\ar[r, "t(h')"'{name=4}, leftarrow, bend right=65]
	&
	{\sss \bul}
	\ar[Rightarrow, from=3, to=4,"h'",pos=0.5, shorten >= 0.5em, shorten <= 0.5em]   
	\ar[Rightarrow, from=1, to=2,"\mathbbm{1}",pos=0.5, shorten >= 0.5em, shorten <= 0.5em]
	\ar[r, "t(h^{-1})"{name=3}, leftarrow, bend left=65] 
	\ar[r, "t(h^{-1})"'{name=4}, leftarrow, bend right=65]
	&
	{\sss \bul}
	\ar[Rightarrow, from=3, to=4,"\mathbbm{1}",pos=0.5, shorten >= 0.5em, shorten <= 0.5em]   
	\end{tikzcd}  \nonumber \\
	&=& \, t(h)\tr h' \; .
\end{eqnarray}
More generally, the domain of a 2-morphism need not to be the identity 1-morphism in which case we label it by a tuple $\lambda \sim (g,h)\in \Gamma_1\times \Gamma_2$ where $g$ refers to the source 1-morphism. When $\lambda:g\to g'$ we can introduce source and target maps $\underline{s},\underline{t}: \Gamma_1 \times \Gamma_2 \to \Gamma_{1}$ which map a 2-morphism to its source and target 1-morphisms, respectively. In terms of the data of the crossed module introduced above, these can be expressed as
\begin{align}
	\underline{s}(g,h)=&\;g  \nonumber \\
	\underline{t}(g,h)=&\; t(h)g =:g'\; . 
\end{align}
Note that a general 2-morphism is still valued in $\Gamma_2$, however the horizontal composition is not the group multiplication in $\Gamma_2$ anymore but instead it is a mutliplication in the semi-direct product $\Gamma_{2} \rtimes \Gamma_{1}$ which depends on the source 1-morphisms. Hence the labeling by an element of $\Gamma_1 \times \Gamma_2$. The horizontal composition is finally defined by the following diagram
\begin{equation}
	\begin{tikzcd}[column sep=4em, row sep=1.5em]
		{\sss \bul}
		\ar[r, "g_1g_2"{name=5}, leftarrow, bend left=60] 
		\ar[r, "g_1'g_2'"'{name=6}, leftarrow, bend right=60]
		&
		{\sss \bul}
		\ar[Rightarrow, from=5, to=6,"\lambda_1 \circ \lambda_2",pos=0.5, shorten >= 0.5em, shorten <= 0.5em]	
	\end{tikzcd}	
	= 
	\begin{tikzcd}[column sep=3em, row sep=1.5em]
		{\sss \bul}
		\ar[r, "g_1"{name=1}, leftarrow, bend left=65] 
		\ar[r, "g_1'"'{name=2}, leftarrow, bend right=65]
		&
		{\sss \bul}
		\ar[Rightarrow, from=1, to=2,"\lambda_1",pos=0.5, shorten >= 0.5em, shorten <= 0.5em]
		\ar[r, "g_2"{name=3}, leftarrow, bend left=65] 
		\ar[r, "g_2'"'{name=4}, leftarrow, bend right=65]
		&
		{\sss \bul}
		\ar[Rightarrow, from=3, to=4,"\lambda_2",pos=0.5, shorten >= 0.5em, shorten <= 0.5em]   
	\end{tikzcd} \; ,
	\label{hor_comp}
\end{equation}
where $\lambda_{i} =(g_i,h_{i})$, we have $\lambda_{1}\circ  \lambda_{2}= \big(g_1g_2, h_1 (g_1\tr h_2) \big)$. Furthermore, the vertical composition now reads
\begin{equation}
	\begin{tikzcd}[column sep=4em, row sep=1.5em]
		{\sss \bul}
		\ar[r, "g"{name=1}, leftarrow, bend left=65] 
		\ar[r, ""{name=2}, leftarrow] 
		\ar[r, "g''"'{name=3}, leftarrow, bend right=65]
		&
		{\sss \bul}
		\ar[Rightarrow, from=1, to=2,"\lambda",pos=0.5, shorten >= -0.1em, shorten <= 0.5em]
		\ar[Rightarrow, from=2, to=3,"\lambda'",pos=0.5, shorten >= 0.5em, shorten <= 0.5em]	
	\end{tikzcd}	
	=
	\begin{tikzcd}[column sep=4em, row sep=1.5em]
		{\sss \bul}
		\ar[r, "g"{name=1}, leftarrow, bend left=65] 
		\ar[r, "g''"'{name=3}, leftarrow, bend right=65]
		&
		{\sss \bul}
		\ar[Rightarrow, from=1, to=3,"\lambda'\cdot\lambda",pos=0.5, shorten >= 0.5em, shorten <= 0.5em]	
	\end{tikzcd}
	\label{vert_comp}	
\end{equation}
where $\lambda'\cdot\lambda=(g,h'h)$. Horizontal and vertical compositions must be compatible with one another. This is encoded in the fact that the diagram
\begin{equation}
	\begin{tikzcd}[column sep=4em, row sep=1.5em]
	{\sss \bul}
	\ar[r, "g_1", ""{name=1}, leftarrow, bend left=65] 
	\ar[r, "", ""{name=2}, leftarrow] 
	\ar[r, "g_1''"'{name=3}, leftarrow, bend right=65]
	&
	{\sss \bul}
	\ar[Rightarrow, from=1, to=2,"\lambda_1",pos=0.5, shorten >= -0.1em, shorten <= 0.5em]
	\ar[Rightarrow, from=2, to=3,"\lambda_1'",pos=0.5, shorten >= 0.5em, shorten <= 0.5em]	
	\ar[r, "g_2", ""{name=4}, leftarrow, bend left=65] 
	\ar[r, "", ""{name=5}, leftarrow] 
	\ar[r, "g_2''"'{name=6}, leftarrow, bend right=65]
	&
	{\sss \bul}
	\ar[Rightarrow, from=4, to=5,"\lambda_2",pos=0.5, shorten >= -0.1em, shorten <= 0.5em]
	\ar[Rightarrow, from=5, to=6,"\lambda_2'",pos=0.5, shorten >= 0.5em, shorten <= 0.5em]	
	\end{tikzcd}
\end{equation}
implies that the following equation must hold
\begin{align}
	(\lambda_1'\cdot\lambda_1)\circ (\lambda_2'\cdot\lambda_2)=(\lambda_1' \circ \lambda_2')\cdot (\lambda_1\circ \lambda_2) \; .
	\label{consistent_comp}
\end{align} 
It is straightforward to show that this is true for the composition rules define above.

\bigskip \noindent
Having defined the structure of a strict 2-group, we return to the original motivation of studying $\text{hol}_2$ which describes a 2-connection. The functor $\text{hol}_2$ would assign elements of $\Gamma_1$ to the paths on a manifold $\cM$ and elements of $\Gamma_{2}$ to paths between paths. Furthermore these assignments must compose in a coherent way adhering to the structure of the (strict) 2-group described above. Since our main purpose is to analyze topological models, we will only need to work with flat 2-connections. It turns out \cite{yetter1993tqft, martins2007yetter} that in analogy with flat connections, a flat 2-connection can be fully determined by a 1-cochain $g$ valued in $\Gamma_1$ satisfying 
\begin{equation*}
	d {\bfg} = t( {\bfh})
\end{equation*}
and a 2-cocycle ${\bfh}$ valued in $\Gamma_2$, but which compose as $\Gamma_1 \ltimes \Gamma_2$, satisfying
\begin{equation*}
	d_{g \triangleright}{h } =  \unit \; ,
\end{equation*}
where the differential $d_{g \triangleright}$ on 2-cochains evaluated on a 3-simplex $(abcd)$ reads
\begin{equation}
	\la d_{g \triangleright}h, (abcd)\ra = (g_{ab} \triangleright h_{bcd}) \cdot h_{acd}^{-1} \cdot h_{abd} \cdot h_{abc}^{-1} \; .
\end{equation}
D. Yetter constructed an interesting topological invariant for a manifold $\cM$ from a crossed module or 2-group $\mathbb G$ \cite{yetter1993tqft}. Let $\mathcal M$ be a compact oriented piecewise linear manifold. An admissible ``$\mathbb G$-configuration of $\mathcal M$'' is a configuration $(g,h)$ satisfying the above constraints. Furthermore, we denote by $b_{i}$ the $i$-th Betti number of $\mathcal M$. The Yetter's invariant $I_{\mathbb G}({\mathcal M})$ then reads
\begin{align}
	I_{\mathbb G}(M)=\frac{\#_{({\mathbb G}\text{-configurations on }{\mathcal M})}}{|\Gamma_1|^{b_0}|\Gamma_2|^{b_1-b_0}} \; .
\end{align}
Recall that while constructing the Hamiltonian extension of Dijkgraaf-Witten theory, we introduced a weak-version of a category whose group elements were labeled by a discrete group $G$.  By weakening the associativity constraint, we were able to construct the most general topological gauge theories from group-like categories. Similarly, it is natural to construct topological gauge theories from weak 2-group-like  2-categories.

\subsection{Weak 2-groups \label{sec:weak2groups}}
\noindent
Starting from the definition of strict 2-groups as 2-groupoids with simple objects, we can define a \emph{weak 2-group} as a 2-category where all 1-morphisms are weakly invertible and all 2-morphisms are invertible. It was shown by Baez and Lauda  \cite{baez2004higher}\footnote{Strictly speaking Baez and Lauda show this for coherent 2-groups but since there is an equivalence between the category of coherent and weak 2-groups, this statement holds.} that there is a one-to-one correspondence between equivalence classes of weak 2-groups and isomorphism classes of quadruples $(G,H,\tr, [\alpha])$ which consist of 
\begin{enumerate}[itemsep=0.4em,parsep=1pt,leftmargin=*]
	\item[$\circ$] A group $G=\text{coker}(t)$
	\item[$\circ$] An abelian group $H=\text{ker}(t)$
	\item[$\circ$] An action $\tr: G \to \text{Aut}(H)$
	\item[$\circ$] A cohomology class $[\alpha]\in H^{3}(G,H)$ which corresponds to the first Postnikov invariant.
\end{enumerate}
with $t$ is the group homomorphism introduced above.
Equivalently, if two 2-groups $\mathbb G$ and $\mathbb G'$ are isomorphic, their classifying spaces are homotopy equivalent and, according to a theorem by Maclane and Whitehead \cite{maclane19503}, captured by the above data.

\medskip \noindent Flat 2-connections corresponding to weak 2-groups are then fully determined by a 1-cochain ${\bfg}$ valued in $G$ satisfying
\begin{align*}
	d{\bfg}=1
\end{align*}
and a 2-cochain ${\bfh}$ valued in $H$ satisfying a cocycle condition twisted by an action $\tr$ of $G$ upto a cohomology class $\alpha({\bfg}) \in H^{3}(M,H)$ such that
\begin{align*}
	d_{\bfg \triangleright} {\bfh} = \alpha({\bfg})
\end{align*}
where the differential on 2-cochains evaluated on a 3-simplex $(abcd)$ now reads
\begin{align}
	\langle d_{g \triangleright}{\bfh}, (abcd)\rangle = {\bfg}_{ab}\tr {\bfh}_{bcd} - {\bfh}_{acd} + {\bfh}_{abd} - {\bfh }_{abc} \; ,
\end{align}
where we used an additive product rule since the group $H$ is abelian.
To summarize, ${\bfg}$ is a 1-cocycle and ${\bfh}$ is a 2-cochain satisfying a certain twisted cocycle condition controlled by $\alpha$.\footnote{Note that we denote the cochains by $({\bfg}, {\bfh})$ both in the case of a strict 2-group and an equivalence class of weak 2-groups. However in the former case they are valued in $(\Gamma_1, \Gamma_2)$ and in the latter one in $(G,H)$.} These cocycle conditions can be made natural by recalling that distinct flat 2-connections can be obtained via homotopy classes of maps from the manifold to a certain homotopy 2-type known as the classifying space of the 2-group. We defer a more detailed explanation to the next section.  

At this stage, we can use this result to redefine a weak 2-group as a weak monoidal category which consists of: Objects labeled by group elements in a group $G$ which are weakly invertible, 1-morphisms labeled by group elements in an abelian group $H$ which are invertible, and an associator $\alpha$ provided by a cohomology class in $H^3(G,H)$. Assuming a trivial action of $G$ on $H$,\footnote{When defining the category $\mathbb{C}$--$2\text{Vec}_{G,H}^{\pi, \alpha, \tau}$ it would have been possible to introduce a non-trivial action of $G$ on $H$ but we chose not to do so for notational convenience.}this is exactly the input data of the underlying  weak monoidal 1-category in $\mathbb{C}$--$2\text{Vec}_{G,H}^{\pi, \alpha, \tau}$ defined in sec.~\ref{sec:gen}. As we will see, it turns out that the state-sum introduced in \eqref{SM2gr} corresponds to a topological higher gauge model for the 2-group $(G,H,{\rm triv},[\alpha])$. In the following, we will always assume the the group $G$ acts trivially on $H$ for notational convenience.

\subsection{Topological gauge theories from groups and group-like generalizations}
\noindent
There is a natural way to define topological gauge theories as sigma models with the target space being the classifying space of a group, 2-group or some other group-like generalization. The general form of the partition function reads
\begin{align}
	\mathcal Z_{\omega}^{X}({\mathcal M})=\frac{1}{\mathcal N_{X,{\mathcal M}}}\sum_{[\gamma]:{\mathcal M}\to X}e^{2\pi i\langle \gamma^{\star}\omega,[{\mathcal M}] \rangle}
\end{align}
where $\omega \in C^{d+1}(X,\mathbb R/\mathbb Z)$, $\mathcal M$ is a compact oriented ($d$+1)-manifold, $[\mathcal{M}] \in H_{d+1}(\mathcal{M}, \mathbb{Z})$ its fundamental homology cycle and $\mathcal N_{X,{\mathcal M}}$ is a normalization constant that depends on the manifold and the choice of target space $X$. The sum in the partition function is over homotopy classes $[\gamma]$ of maps $\gamma$ from $\cM$ to $X$.

Given an oriented $(d+2)$-bordism $\mathcal{W}: \cM_{1}\sqcup \cM_2\to \cM_3$ we require that\cite{dijkgraaf1990topological} 
\begin{align}
	0=&\;\langle \gamma^{\star}\omega,[\cM_1] \rangle+\langle \gamma^{\star}\omega,[\cM_2] \rangle-\langle \gamma^{\star}\omega,[\cM_3] \rangle \nn \\
	=&\; \langle \gamma^{\star}\omega,[\partial \mathcal{W}] \rangle \nn \\
	=&\; \langle \gamma^{\star}d \omega ,[\mathcal{W}] \rangle
	\label{bordCond}
\end{align}
where $d: C^{d}(X,\mathbb R/\mathbb Z)\to C^{d+1}(X,\mathbb R/\mathbb Z)$ is the coboundary operator. Since we require \eqref{bordCond} to hold for all bordisms one has $\omega\in Z^{d+1}(X,\mathbb R/\mathbb Z)$. Similarly we may ask, what is the effect of modifying the cocycle $\omega$ by a coboundary $d \lambda$ where $\lambda \in C^{d}(X,\mathbb R/\mathbb Z)$. Clearly this has no effect when $\cM$ is closed. When $\cM$ is an open manifold it alters the action by a boundary term that can be absorbed into a ${\rm U}(1)$ phase when quantizing the theory. The redefined Hilbert space preserves amplitudes and hence describes the same theory. Therefore distinct topological sigma models are labeled by cohomology classes $[\omega]\in H^{d+1}(X,\mathbb R/\mathbb Z)$.\footnote{Here we have switched from labelling cohomological models by cocycles valued in {\rm U}(1) to an equivalent convention of labelling such models by cocycles valued in $\mathbb R/\mathbb Z$. Although these two formulations are clearly equivalent, we make such a switch because a lot of the discussion will be focussed on topological actions which are more naturally valued in $\mathbb R/\mathbb Z$.}

\bigskip \noindent In the following, we will be interested in situations corresponding to gauge theories built from: $(i)$ An ordinary group $G$, which could be abelian or non-abelian, $(ii)$ a \emph{higher-form} group which we denote $H_{[q]}$  where $q > 1$ such that it is necessarily abelian, $(iii)$ a 2-group $\mathbb G$ together with its higher form generalization $(iv)$ $\mathbb G_{[q]}$. In these examples, the corresponding classifying spaces will be $X=BG$, $B^{q+1}H$, $B\mathbb G$ and $B \mathbb G_{[q]}$,  respectively.
 
Furthermore, we will emphasize the lattice descriptions corresponding to these different kinds of topological gauge theories. As such we briefly recall here the notations we have been using so far: We are interested in a ($d$+1)-dimensional, compact, oriented manifold $\mathcal M$ with a triangulation $\triangle$. We denote the vertices by labels $a,b,c$ etc. and 1-simplices, 2-simplices, 3-simplices by $(ab)$, $(abc)$, $(abcd)$ etc. , respectively. Each $i$-simplex is assigned an orientation which we denote $\epsilon(abc \ldots)\in \pm 1$ depending on whether the orientation of the particular simplex coincides with the orientation of $ \mathcal M$. The number of $i$-simplices is denoted by $|\triangle_i|$.

\subsubsection{Dijkgraaf-Witten theory}
\noindent
The Dijkgraaf-Witten partition function on a compact, oriented, ($d$+1)-dimensional manifold $\cM$ takes the form 
\begin{align}
	\mathcal Z_{\omega}^{G}(\cM)=\frac{1}{|G|^{b_0}}\sum_{[\gamma]:\cM\to BG}e^{2\pi i\langle \gamma^{\star}\omega,[\cM]\rangle} \; .
	\label{DW_partn_fn}
\end{align} 
with $b_0$ the $0$-th Betti number.
A property of $BG$ that will be important for the present discussion is that its only non-vanishing homotopy group is in degree one, i.e $\pi_{i}(BG)$ is $G$ if $i=1$ and 0 otherwise. The topological action in \eqref{DW_partn_fn} can thus be realized as a lattice TQFT action. Let $\cM$ be equipped with an oriented triangulation $\triangle$. Since $BG$ is path-connected ($\pi_0(BG)=0$), one may smoothly deform a map $\gamma$ such that each 0-simplex (vertex) on $\cM$ is mapped to the same point in $BG$. Edges or 1-simplices of the triangulation are mapped to the space of paths in $BG$ which, up to homotopy, is $G$. Contractible paths are mapped to the identity element in $G$. This is obviously a rather familiar construction in lattice gauge theories as this is simply the lattice implementation of flat connection, i.e given a 2-simplex $\triangle_2 = (abc)$ we assign group elements to the edges such that   
\begin{align}
	g_{ac} = g_{ab} \cdot g_{bc} \; .
\end{align}
The Dijkgraaf-Witten partition function \eqref{DW_partn_fn} can now be recast as a lattice gauge theory. Consider $\cM$ with a $G$-coloring, i.e an assignment of group elements $g\in G$ to every 1-simplex of $\cM$ such that the above cocycle constraint is everywhere satisfied. Let the set of colorings be denoted by $\text{Col}(\cM,G)$. It is easy to check that the coloring of each ($d$+1)-simplex depends on $d$+1 group elements. The topological action assigns a ${\rm U}(1)$ phase to each ($d$+1)-simplex which depends on the coloring $\bfg \in \text{Col}(\cM,G)$ and a group cohomology class representative of $[\omega]\in H^{d+1}(G,\mathbb R/\mathbb Z)$. For example in (3+1)d, consider a simplex $\triangle_{4}=(abcde)$, the topological actions reads
\begin{align}
	e^{2 \pi iS_{\omega}(\bfg \, , \, \triangle_{4})} 
	\equiv e^{2\pi i \epsilon(\triangle_{4})\omega(g_{ab},g_{bc},g_{cd},g_{de})}
\end{align}
so that the Dijkgraaf-Witten partition function takes the form
\begin{align}
	\label{DWact1}
	\mathcal Z^{G}_{\omega}(\cM)=\frac{1}{|G|^{|\triangle_0|}}\sum_{\bfg\in \text{Col}(\cM,G)}\prod_{\triangle_{4}}e^{2\pi i S_{\omega}(\bfg \, , \, \triangle_{4})} \; .
\end{align} 
Note finally that if we would choose $\omega$ to be ${\rm U}(1)$-valued instead of $\mathbb R/\mathbb Z$-valued, we would recover exactly the expression \eqref{ZDW1} in ($2$+1)d.

\subsubsection{Higher-form topological gauge theory \label{sec:higherForm}}
\noindent Let us now consider an abelian group $H$ which we would get upon gauging a $q$-form global symmetry $H_{[q]}$. A topological gauge theory corresponding to this group is built from ($q$+1)-form $H$-valued cocycles. In ($d$+1)-dimensions these are classified by the cohomology group $H^{d+1}(B^{q+1}H,\mathbb R/\mathbb Z)$ where $B^{q+1}H$ is the classifying space such that $\pi_{q+1}(B^{q+1}H)=H$ and all other homotopy groups vanish. The partition function is defined analogously to the Dijkgraaf-Witten model:\\[-0.4em]
\begin{align}
	\mathcal Z^{H_{[q]}}_{\omega}(\cM)=\frac{1}{|H|^{b_{0 \rightarrow q}}}\sum_{[\gamma]:\cM\to B^{q+1}H}e^{2\pi i\langle \gamma^{\star}\omega , [\cM] \rangle }
\end{align}
with $b_{0 \rightarrow q} := \sum_{i=0}^{q}(-1)^i b_{q-i}$.
Similar to the case discussed above, homotopy classes of maps to $B^{q+1}H$ define distinct flat ($q$+1)-form fields which are labeled by elements in $\text{Hom}(\pi_{q+1}(\cM),H)$ which, since $H$ is abelian, is $H^{q+1}(\cM,H)$. Furthermore, since $H$ is abelian, we can always construct a simplicial expression for the topological action in terms of cup products and $H$-valued ($q$+1)-cochains $\bfh^{(q+1)}$ such that 
\begin{align}
	S_{\omega}(\bfh^{(q+1)} \, , \, \cM)=\langle \gamma^{\star}\omega,[\cM] \rangle \; .
\end{align}
We can construct an explicit lattice gauge theory realization by considering colorings ${\rm Col}(\mathcal{M}, H_{[q]})$ of $\mathcal{M}$, i.e. an assignment of group elements $h^{(q+1)} \in H$ to every ($q$+1)-simplex which satisfy the cocycle condition $dh^{(q+1)}=0$. The partition function for the topological gauge theory takes the form
\begin{align}
	\mathcal{Z}_{\omega}^{H_{[q]}}(\cM)
	=&\; \frac{1}{|H|^{|\triangle_{0 \rightarrow q}|}} 
	\sum_{\bfh^{(q+1)}}  \prod_{\triangle_{d+1}}e^{2\pi iS_{\omega}(\bfh^{(q+1)} \, , \, \triangle_{d+1})}
\end{align}
with $|\triangle_{0 \rightarrow q}| := \sum_{i=0}^q(-1)^i|\triangle_{q-i}|$. In particular, for a trivial cocycle $\omega \sim 0\in H^{d+1}(B^{q+1}H,\mathbb R/\mathbb Z) $, the partition function simplifies and reads
\begin{align}
	\mathcal Z_0^{H_{[q]}}(\cM)=|H|^{b_{0 \rightarrow q+1}} = |H|^{\sum_{i=0}^{q+1}(-1)^{i}b_{q+1-i}} \; .
\end{align}
For $q=1$, these topological gauge theories were studied in more detail in \cite{gaiotto2015generalized, kapustin2014coupling, putrov2016braiding}.

\subsubsection{2-group topological gauge theory}
\noindent
Similar to the group case, one can construct a topological gauge theory from a 2-group $\mathbb G=(G,H,\tr,\alpha)$. The partition function mimics the Dijkgraaf-Witten model:\\[-0.4em]
\begin{align}
	\label{2DW}
	\mathcal Z_{\omega}^{\mathbb G}(\cM)=\frac{1}{|G|^{b_0}|H|^{b_1-b_0}}\sum_{[\gamma]: \cM\to B\mathbb G}e^{2\pi i\langle \gamma^{\star} \omega, [\cM] \rangle}
\end{align} 
for some $\omega \in H^{d+1}(B\mathbb G,\mathbb R/\mathbb Z)$ where $B\mathbb G$ is the classifying space of the 2-group. The classifying space $B\mathbb G$ has non-vanishing homotopy in degree one and two such that
\begin{align*}
	\pi_1(B\mathbb G)=&\;G \; ,\\
	\pi_2(B\mathbb G)=&\;H \; .
\end{align*}
In fact $B\mathbb G$ can be understood conveniently as a $B^2H$ fibration on $BG$, where $B^{2}H$ is the classifying space for 2-form $H$-connections. 
Therefore, the 2-group classifying space can be understood through the sequence
\begin{align}
B^{2}H \to B\mathbb G \to BG
\end{align}
which is captured by the Postnikov class  $[\alpha]\in H^{3}(BG,H)$. In analogy to lattice Dijkgraaf-Witten theory, we may finally understand isomorphism classes of flat $\mathbb G$ 2-bundles via homotopy classes of maps $\gamma: \cM\to B\mathbb G$  \cite{baez2009classifying}. 

\bigskip \noindent
Let $\cM$ be a piecewise linear triangulated manifold (or more generally a \v{C}ech or simplicial complex) such that the simplices are oriented following the same convention as before. Since $B\mathbb G$ is path connected, we may deform $\gamma$ so that each 0-simplex is mapped to the same point in $B \mathbb G$, then each 1-simplex gets mapped to an element of $G$ such that contractible paths are mapped to the identity element which in turn imposes the cocycle condition corresponding to a flat bundle. In practice, it means that we assign to every 1-simplex a group element $g_{ab} \in G$, such that for every 2-simplex $\triangle_2 = (abc)$ we impose the flatness condition
\begin{equation}
	\label{str2Grcoc1}
	g_{ac} = g_{ab} \cdot 	g_{bc} \; .
\end{equation}
Similarly, 2-simplices are identified with group elements $h_{abc} \in H$, such that for every 3-simplex $\triangle_3 = (abcd)$ we impose the condition
\begin{equation}
	\label{str2Grcoc2}
	h_{bcd} - h_{acd} + h_{abd} - h_{abc} = \alpha(g_{ab}, g_{bc}, g_{cd}) \; .
\end{equation} 
These two conditions are only a simplicial translation of the cocycle conditions satisfied by the cochains ${\bfg}$ and ${\bfh}$\footnote{Note that we do not put a superscript ($q$+1) on $\bfh$, i.e, unless specified, $\bfh$ always denotes a 2-cochain corresponding to $H_{[1]}$.}. 
As expected, these conditions match the ones introduced in \ref{sec:gen}. The topological action then assigns a ${\rm U}(1)$-valued phase to every ($d$+1)-simplex, namely $e^{2\pi iS_\omega(\bfg|\bfh \, , \, \triangle_{d+1})}$ so that the partition function takes the form\\[-0.4em]
\begin{align}
	\mathcal Z^{\mathbb G}_{\omega}(\cM)=\frac{1}{\#_{G,H,\triangle}}\sum_{(\bfg,\bfh) }\prod_{\triangle_{d+1}}e^{2\pi i S_{\omega}(\bfg|\bfh \, , \, 
	\triangle_{d+1})} 
	\label{latticeZ2Gr}
\end{align} 
with  $\#_{G,H, \triangle} = |G|^{|\triangle_0|}|H|^{|\triangle_1| - |\triangle_0|}$ and $(\bfg,\bfh)\in \text{Col}(\cM,\mathbb G)$ the set of colorings for which the cocycle conditions written above are satisfied.
It turns out that the topological action $S_{\omega}$ defined in \eqref{topo2Gr} was nothing else than an explicit representation of a non-trivial cocycle in $H^4(B \mathbb{G}, {\rm U}(1))$. By linearizing this cocycle we would obtain a representative of $H^4(B \mathbb{G}, \mathbb R / \mathbb Z)$ and the partition function \eqref{SM2gr} would match \eqref{latticeZ2Gr} in (3+1)d. In \cite{kapustin2017higher} different classes of cocycles $[\omega]\in H^{4}(B\mathbb G,\mathbb R/\mathbb Z)$ were enumerated so that a general  2-group cocycle can be expressed as a sum of such classes.
 
\noindent
We distinguish three types:
\begin{enumerate}[itemsep=0.4em,parsep=1pt,leftmargin=*]
	 \item[$\circ$] Cocycles for the ordinary group $G$ which are naturally classified by $H^{4}(BG,\mathbb R/\mathbb Z)$. 
	 \item[$\circ$] Cocycles corresponding to the 2-form group $H_{[1]}$ and are classified by $H^{4}(B^2H,\mathbb R/\mathbb Z)$. 
	 \item[$\circ$] Non-trivial combination of cochains valued in $G,H$ which can be expressed in the form of a 4-cocycle $\omega\in Z^{4}(\cM,\mathbb R/\mathbb Z)$  as $\omega(\bfg,\bfh)=\bfh \smilo \lambda(\bfg) $ where $\lambda \in H^{2}(G,\hat{H})$ and $\hat{H}=H^{1}(H,\mathbb R/\mathbb Z)$.
\end{enumerate}
It turns out that the 4-cocycle obtained in \eqref{topo2Gr} can be understood as 2-group 4-cocycle which accounts for the first two classes presented above.\footnote{This can be made more explicit by realizing that the squarators terms in \eqref{topo2Gr} reproduces the Pontryagin square defined as $\mathfrak{P}f = f \smilo f - f \smilo_1 df$ out of which the cocycles corresponding to the 2-form group $H_{[1]}$ are constructed. }

\bigskip \noindent
Generally, in the same way as the topological action of the Dijkgraaf-Witten model \eqref{DW_partn_fn} depends on a class in $H^d(BG,{\rm U}(1))$ which can be represented by a group cocycle as a function of $d$ variables living in $G$, we can write the topological action of the 2-group state-sum model \eqref{latticeZ2Gr} as a single function of $d$ variables in $G$ and $d(d-1)/2$ variables in $H$ which explicitly represents an equivalence class in $H^4(B\mathbb{G}, {\rm U}(1))$. Note finally that starting from the general formula \eqref{latticeZ2Gr} for the topological theory constructed from the 2-group $\mathbb{G}$ and assuming that the 2-group cohomology class is trivial, we recover the lattice version of Yetter's invariant, but this time for a weak 2-group, still denoted by $I_{\mathbb G }(\cM)$ and defined as
\begin{align}
	I_{\mathbb G}(\cM) =\frac{\#_{(\mathbb G\text{--colorings of }\cM)}}{|G|^{|\triangle_0|}|H|^{|\triangle_1| - |\triangle_0|}} \; .
\end{align}
where $\#_{(\mathbb G\text{--colorings of }\cM)}$ is the number of colorings of the 1-simplices and 2-simplices satisfying the closure constraints \eqref{str2Grcoc1} and \eqref{str2Grcoc2}. Here we have made use of the fact that there is a bijection between equivalence classes of admissible $\mathbb G$--colorings of $\mathcal{M}$ and homotopy classes of maps $[\gamma]:\mathcal{M }\to B\mathbb G$ with $B \mathbb{G}$. Also, starting with the model constructed from the 2-group $\mathbb G$ and assuming that the group $G$ is trivial, we recover the 2-form gauge theory model presented in sec.~\ref{sec:higherForm}.

\subsubsection{Further generalizations \label{further_section}}

\noindent
Recently, interesting generalizations of the above 2-group structure were discussed by Tachikawa \cite{tachikawa2017gauging}. Let us call such a generalization $\mathbb G_{[q]}$. This structure can be defined via its classifying space which is a fibration
\begin{align}
	B^{q+1}H\to B\mathbb G_{[q]}\to BG \; .
\end{align} 
As a topological space, $B\mathbb G_{[q]}$ has non-vanishing homotopy in degree 1 and $q+1$. More precisely, $\pi_{1}(B\mathbb G_{[q]})=G$, $\pi_{q+1}(B\mathbb G_{[q]})=H$ and all other homotopy groups vanish. The bundle obtained upon pulling back from $B\mathbb G_{[q]}$ can be understood in terms of a local system of fields ${\bfg}$ and ${\bfh}
^{(q+1)}$ that satisfy 
\begin{align}
	d{\bfg}=&\; 1 \nonumber \\
	d{\bfh}^{(q+1)}=&\; \alpha^{(q+2)}({\bfg})
\end{align}
where ${\bfg}$ is a 1-cocycle valued in $G$, ${\bfh}^{(q+1)}$ is a ($q$+1)-cochain valued in $H$ and $[\alpha^{(q+2)}] \in H^{q+2}(BG,H)$. Similar to the above constructions, one can construct a topological gauge theory as a sigma model with target space $B\mathbb G_{[q]}$. Such topological gauge theories are also classified by cohomology classes $[\omega]\in H^{d+1}(B\mathbb G_{[q]},\mathbb R/\mathbb Z)$ and the partition function takes the form
\begin{align}
	\mathcal Z_{\omega}^{\mathbb G_{[q]}}(\cM)
	=\frac{1}{|G|^{b_0}|H|^{b_{0 \rightarrow q}}}
	\sum_{[\gamma]: \cM \to B\mathbb G_{[q]}}e^{2\pi i\langle \gamma^{\star}\omega,[\cM ]\rangle} \; .
\end{align}
This can be realized more explicitly as a topological lattice gauge theory by considering a simplicial triangulation for the compact oriented manifold $\cM$ with a coloring $\text{Col}(\cM,\mathbb G_{[q]})$ which involves assigning $G$ elements to 1-simplices and $H$-elements to ($q$+1)-simplices subject to the above cocycle conditions. The topological action assigns a ${\rm U}(1)$ phase which depends on a representative of the cohomology class $[\omega]$. 

\subsection{2-group gauge transformations}
\noindent
We presented topological models based on 2-groups which have a higher gauge theory interpretation. In this short section we emphasize what the corresponding gauge transformations are. 2-group gauge theory models have two sets of gauge symmetries. First we have the usual gauge transformations which will now be referred to as 0-form gauge transformations. These are parameterized by a gauge parameter $k \in C^{0}(\mathcal{M},G)$ and act at 0-simplices such that the $G$-group variables labeling the 1-simplices transform as
\begin{align}
	g_{ab} \, \rightarrow \,  g_{ab}^{k}:= k_a g_{ab}k_b^{-1} \; .
	\label{0formgt1}
\end{align}
The $H$-group variables labeling the 2-simplices are modified as well according to
\begin{align}
	h_{abc} \, \rightarrow \,  h_{abc}^{k}:= h_{abc}+\zeta(g,k)_{abc}
\label{0formgt2}
\end{align} 
such that $\zeta(g,k)_{abc} \equiv \la \zeta(g,k),(abc)\ra$ where $\zeta(g,k)$ is a descendent of $\alpha(g)$, i.e it satisfies
\begin{align}
	d\zeta(g,k)=\alpha(g^k)- \alpha(g) \; .
\end{align}
This condition ensures that the twisted cocycle condition is preserved under 0-gauge transformations:
\begin{align}
	dh^{k} &= d(h+\zeta(g,k))\nn = dh+ \alpha (g^k)- \alpha(g) =  \alpha(g^k) \; .
\end{align}
We also have 1-form gauge transformations. These are parameterized by a gauge parameter $\lambda \in C^{1}(\mathcal{M},H)$ and act at 1-simplices such that the $H$-group variables labeling the 2-simplices transform as
\begin{align}
	\nn
	h_{abc} \, \rightarrow \, h^\lambda_{abc} &:= h_{abc} + d\lambda_{abc} \\
	 &= h_{abc} + \lambda_{bc} - \lambda_{ac} + \lambda_{ab} \; .
\end{align}
with $d\lambda_{abc} = \la d\lambda, (abc) \ra$.
Together the gauge transformations form a crossed module so that their action on the fields satisfies the following multiplication rule
\begin{align}
	(k_1,\lambda_1)\circ (k_2,\lambda_2) = (k_1k_2, \lambda_1+ (k_1\tr \lambda_2)) 
\end{align}
where we have reintroduced the group action $\tr$, despite choosing it to be trivial, so as to make the semigroup structure of gauge transformations manifest.
\subsection{Lattice Hamiltonian realization \label{sec:expLat1}}
\noindent
We obtained earlier a generalization of DW model in (3+1)d by weakening some of the axioms of the corresponding category. This resulted in the definition of the 2-category  $\mathbb{C}$--$2\text{Vec}_{G,H}^{\pi, \alpha, \tau}$. We emphasized above how the underlying 1-category is actually the weak 2-group $(G,H,\text{triv},[\alpha])$, and in which sense the state sum \eqref{SM2gr} is actually a twisted 2-group model whose 4-cocycle $\omega$ defined in \eqref{topo2Gr} represents a class in $H^4(B \mathbb{G}, {\rm U}(1))$. In particular, this suggests a way to construct the lattice Hamiltonian corresponding to the string net model built upon $\mathbb{C}$--$2\text{Vec}_{G,H}^{\pi, \alpha, \tau}$ using the notions we just recalled about 2-groups, higher gauge theory and 2-connections. We will now sketch the general construction of such lattice Hamiltonian for a given 2-group cocycle which in (3+1)d would comprise the case of the $\mathbb{C}$--$2\text{Vec}_{G,H}^{\pi, \alpha, \tau}$ string net model. 

So we would like to construct a lattice Hamiltonian whose ground state is described by a 2-group TQFT using an approach analoguous to the one described in sec.~\ref{sec:ham}. To do so without burdening the reader with unwieldy notations, we will consider a representative of an equivalence class $[\omega_3] \in H^3(B \mathbb{G},{\rm U}(1))$. Note, however, that the model retains all the features of the four-dimensional case as far as the 2-group is concerned. The labeling convention for a 2-group 3-cocycle is the following:\\

\noindent\textsc{Labeling convention for 2-groups}\\[-1.7em]
\begin{leftbar}\noindent 
	Let $\triangle_3 = (abcd)$ with $a < b < c <d$ be a 3-simplex of the triangulation $\triangle$. We consider a flat $\mathbb G$-connection by assigning group variables $g_{ab} \in G$ to the 1-simplices and group variables $h_{abc} \in H$ to the 2-simplices. Using the differential on cochains, these group variables must satisfy $dg=1$ and $dh = \alpha(g)$ at every 2-simplex and 3-simplex, respectively. Because of the closure constraints, a 3-cocyle in $H^3(B \mathbb{G},{\rm U}(1))$ for the 2-group $\mathbb{G}$, depends on three group variables $g_1,g_2,g_3 \in G$ such that $g_{b}\equiv g_{b-1b}$ and three group variables $h_1,h_2,h_3 \in H$ such that $h_1 \equiv h_{abc}$, $h_2 \equiv h_{acd}$, $h_3 \equiv h_{abd}$. We denote such cocycle by $\omega_3(g_1,g_2,g_3 | h_1,h_2,h_3)$.
\end{leftbar}
\noindent
Using this labeling convention, we show in app.~\ref{app:cocycle2Gr} how to obtain the 2-group 3-cocycle condition. We reproduce the corresponding equation below in a slightly different form:
\begin{align*}
&\omega_3(g_2,g_3,g_4|h_1,h_2,h_3)
\\ \nn \times & \omega_3(g_1,g_2,g_3|h_4,h_1+h_5-h_4-\alpha(g_1,g_2,g_3),h_5)
\\ \nn \times &\omega_3(g_1,g_2g_3,g_4|h_5,h_2+h_6-h_5-\alpha(g_1,g_2g_3,g_4),h_6)
\\ \nn = \;  &\omega_3(g_1,g_2,g_3g_4|h_4,h_3+h_6-h_4-\alpha(g_1,g_2,g_3g_4),h_6)
\\ \nn \times & \omega_3(g_1g_2,g_3,g_4|h_1+h_5-h_4-\alpha(g_1,g_2,g_3),
\\ & \hspace{6.3em} h_2+h_6-h_5-\alpha(g_1,g_2g_3,g_4),
\\ & \hspace{6.3em} h_3+h_6-h_4-\alpha(g_1,g_2,g_3g_4))	
\; .
\end{align*}
In sec.~\ref{sec:ham}, we explained how the cocycle condition could be used to write down how the topological action is modified under a (0-form) gauge transformation. The result could then be used to define the corresponding projection operator in the lattice Hamiltonian picture. Let us now study how this strategy generalizes to the 2-group case using the cocycle condition written above. 

Let $\triangle_3 = (0123)$ be a 3-simplex of the triangulation $\triangle$. The corresponding topological action is given by $\omega_3(\triangle_3) \equiv \omega_3(g_1,g_2,g_3 | h_1,h_2,h_3)$. Let us now perform simultaneously: $(i)$ A 0-form  transformation at the vertex $(0)$ with gauge parameter $k \in G$ such that $g_1 \rightarrow kg_1$, $h_1 \rightarrow h_1 + \zeta(g,k)_{012}$, $h_2 \rightarrow h_2 + \zeta(g,k)_{023}$ and $h_3 \rightarrow h_3 + \zeta(g,k)_{013}$. $(ii)$ Three 1-form gauge transformations  at the 1-simplices $(01)$, $(02)$ and $(03)$ with gauge parameters $\lambda_{01}, \lambda_{02}, \lambda_{03} \in H$ such that $h_1 \rightarrow h_1 - \lambda_{02} + \lambda_{01}$, $h_2 \rightarrow h_2 - \lambda_{03}+ \lambda_{02}$ and $h_3 \rightarrow h_3 -\lambda_{03} + \lambda_{01}$. Putting everything together, we obtain the following gauge transformation

\begin{align*}
	&\omega_3(g_1,g_2,g_3|h_1,h_2,h_3) \\
	&\q  \longrightarrow 	\omega_3(kg_1,g_2,g_3|h_1-\lambda_{02}+\lambda_{01}+\zeta(g,k)_{012},
	\\ & \hspace{9.1em} h_2-\lambda_{03}+\lambda_{02}+\zeta(g,k)_{023},
	\\ & \hspace{9.1em} h_3 - \lambda_{03} + \lambda_{01} + \zeta(g,k)_{013})
\end{align*}
where $\zeta(g,k)$ must satisfy the equation $d \zeta(g,k)_{0123} \equiv \la d \zeta(g,k), (0123)\ra =  \alpha(kg_1,g_2,g_3)- \alpha(g_1,g_2,g_3)$. We can therefore choose $\zeta(g,k)$ such that 
\begin{align*}
	&\zeta_{123} = 0\\
	&\zeta_{012} = - \alpha(k,g_1,g_2) \\
	&\zeta_{023} = - \alpha(k,g_1g_2,g_3) \\
	&\zeta_{013} = -\alpha(k,g_1,g_2g_3).
\end{align*}
where $\zeta_{abc} \equiv \zeta(g,k)_{abc}$. Using a simple relabeling of the variables, it is now possible to use the 3-cocycle condition for $\omega_3$ in order to rewrite how the topological actions transforms under the gauge transformation considered above. One has
\begin{equation*}
\omega_3(\triangle_3) \longrightarrow \omega_3(\triangle_3)\mathfrak{G}[g,h,k,\lambda]
\end{equation*}
with
\begin{align*}
	&\mathfrak{G}[g,h,k,\lambda]\\
	& \q =
	\omega_3(k,g_1,g_2|-\lambda_{01}, h_1-\lambda_{02}+\lambda_{01}+\zeta_{012},-\lambda_{02})
	\\ &  \hspace{1.1em } \times \omega_3(k,g_1g_2,g_3|-\lambda_{02}, h_3-\lambda_{03}+\lambda_{02}+\zeta_{023}, -\lambda_{03}) 
	\\ & \hspace{1.1em } \times \omega_3(k,g_1,g_2g_3|-\lambda_{01},  h_2 -\lambda_{03}+\lambda_{01}+\zeta_{013}, -\lambda_{03})^{-1} 
\end{align*} 
which can be used to confirm the gauge invariance of the state sum model. Furthermore it provides an explicit expression for the operator $\mathbb{A}_v$ of the lattice Hamiltonian model acting on a vertex $v$ shared by three 1-simplices and performing the gauge transformations described above: 
\begin{align}
	&\mathbb{A}_{(0)}\triangleright \bigg| \triPachner{1}{0} \bigg\rangle \\[-0.5em]  
	& \q = \frac{1}{|G||H|^3}\sum_{g_{0'0}}\sum_{\{h_{0'0i}\}_{i=1}^3}
	\mathfrak{G}[g,h,k,\lambda]
	\bigg|\triPachner{1}{1} \bigg\rangle 
\end{align}
which matches the previous expression when setting $g_{0'0}=k$, $h_{0'01} = -\lambda_{01}$, $h_{0'02} = -\lambda_{02}$ and $h_{0'03} = -\lambda_{03}$. The same strategy applies when considering different 0-form and 1-form gauge transformations. More generally, the action of the operator $\mathbb{A}_v$ can be obtained using expression \eqref{actionA} where the state-sum is the 2-group one defined in terms of $\omega_3$. We postpone a thorough study of the complete lattice Hamiltonian to another article, however the same strategy applies for a 2-group  4-cocycle as for the one corresponding to the model built upon the 2-category $\mathbb{C}$--$2\text{Vec}_{G,H}^{\pi, \alpha, \tau}$.

\section{Bosonic SPTs protected by higher symmetries: State sum models and their gauging}
\label{sec:spts}
\noindent
\emph{Symmetry protected topological phases} (SPTs) of matter are short-range entangled, gapped phases of matter that are trivial in the absence of symmetry (say $G$) in the sense that if we allow ourselves to break symmetry $G$ then these phases can all be adiabatically connected to the trivial phase. A manifestation of this triviality is that the partition function for SPTs on a topologically trivial manifold takes the form 
\begin{align}
	\mathcal Z (\cM)=1+\mathcal O(e^{-\frac{L}{\xi}})+ \dots
\end{align}
where $\xi$ is some correlation length determined by the microscopics of the model and $L$ is the system size. In other words the partition function of an SPT is unity up to corrections that are exponentially suppressed in the system size and therefore vanish in the thermodynamic limit. This is not unexpected since it is just a statement about the theory being short-range entangled and having a unique ground state. 

In order to unravel the non-trivial aspects of these phases, one must probe their symmetry properties. This is naturally done by coupling the phase to a background $G$-connection $A$ on some manifold $\cM$. Since these phases of matter are inherently short-range entangled (also referred to as invertible), the partition function in the presence of any background structure $A$ and $\cM$ (note that we treat the background topology and connection on a somewhat equal footing) can at most be a ${\rm U}(1)$ phase. Hence, distinct phases of matter must furnish topologically distinct actions built from data of $G$-connection $A$. Long ago Dijkgraaf and Witten \cite{dijkgraaf1990topological} classified possible topological $G$ gauge theories for a general group in dimension 3 and for a finite group in any dimension. Let us restrict ourselves to a finite group for simplicity. Then the topological actions take the form 
\begin{align}
	S=\langle \gamma^{\star}\omega , [\cM]\rangle 
	\label{gc_action}
\end{align} 
where $\omega \in H^{d+1}(BG,\mathbb R/\mathbb Z)$ and as before $\gamma\in \text{Map}[\cM,BG]$, $[\cM]\in H_{d+1}(\cM,\mathbb Z)$ is the fundamental homology cycle and $BG$ is the classifying space for $G$. This simply implies that bosonic SPT phases in ($d$+1)-dimensions protected by group $G$ are classified by group cohomology classes $H^{d+1}(BG,{\rm U}(1))$ \cite{Chen:2011pg}.\footnote{In all of this discussion there is an assumption that there is a bijection between equivalence classes of gapped phases (with suitable adjectives such as bosonic, $d$-dimensional, $G$-symmetric etc.) of matter and relativistic topological quantum field theories (with analogous adjectives). This assumption is by no means obvious but let us proceed anyway.}

In the above equation \eqref{gc_action}, $\gamma^{\star}\omega$ should be thought of as a functional of local data which in our case is essentially a $G$-valued cocycle $g$. Hence we will often denote this interchangeably as $\omega(g) \simeq \gamma^{\star}\omega$. Furthermore, when $G$ is a discrete abelian group, we can always construct a simplicial expression for $\omega(g)$, i.e  an expression in terms of cup products and codifferentials of local objects defined on a simplicial triangulation. The distinct SPTs protected by symmetry $G$, labeled by cocycles $[\omega]\in H^{d+1}(BG,\mathbb R/\mathbb Z) $ furnish the following partition functions when coupled to a background connection $g$ on a manifold $\cM$
\begin{align}
	\mathcal Z_{\omega|{\text{SPT}}}(g \, , \, \cM)=e^{2\pi i\langle \omega(g),[\cM]\rangle }\left[1+\mathcal O(e^{-\frac{L}{\xi}})+ \dots \right] \; .
\end{align}
More precisely we should specify that here $G$ is a 0-form symmetry i.e it acts on point-like objects. Below we will consider several different situations, namely SPTs protected by $(i)$ 0-form symmetry, $(ii)$ $q$-form symmetry, $(iii)$ 2-group symmetry and $(iv)$ further group-like generalizations. We construct fixed point models for each of these cases. In order to gauge the symmetry we couple the model to background $(i)$ 1-form connection, $(ii)$ 2-form connection, $(iii)$ 2-group connection and $(iv)$ generalized higher group connection, respectively. These phases can be labeled as above by the topological response actions they furnish which are classes in $H^{d+1}(BG,\mathbb R/\mathbb Z)$, $H^{d+1}(B^{q+1}H,\mathbb R/\mathbb Z)$, $H^{d+1}(B\mathbb G,\mathbb R/\mathbb Z)$ and $H^{d+1}(B\mathbb G_{[q]},\mathbb R/\mathbb Z)$, respectively. 

\subsection{SPTs protected by 0-form symmetry \label{sec:SPTgr}}
\noindent
Bosonic SPTs protected by 0-form symmetry $G$ and labeled by a cohomology class $[\omega]\in H^{d+1}(BG,\mathbb R,\mathbb Z)$ can be modeled as a lattice quantum field theory. Consider a compact, oriented ($d$+1)-dimensional manifold $\cM$ with a given triangulation. Let $k$ denote local matter degrees of freedom that live on the vertices of the triangulation and are valued in $G$. The model takes the form
\begin{align*}
	\mathcal Z^{G}_{\omega|{ \text{SPT}}}(\cM)
	=&\;\frac{1}{|G|^{|\triangle_0|}}\sum_{k \in 	C^{0}(\cM,G)}\prod_{\triangle_{d+1}}e^{2\pi iS_{\omega}(\unit^k \, , \, \triangle_{d+1}) }
\end{align*}
where we have the identification $S_\omega(\unit^k \, , \, \triangle_{d+1}) \equiv \la \omega(\unit^k),\triangle_{d+1}\ra \equiv   \epsilon(\triangle_{d+1})\omega(\unit^k)$ which stipulates that the topological action evaluated via a pairing with a ($d$+1)-simplex is the same as the corresponding group cohomology cocycle evaluated on this simplex via a choice of coloring. This is essentially \eqref{DWact1} with $g_{ab}$ replaced by $k_{a}k_{b}^{-1}$. We note that $\omega(g)$ may have pieces of the form $d g$ so that $\omega (\unit^k) \equiv \omega(d k)$ may contain terms with $d^2k$. When $G$ is abelian, we have an isomorphism $G\simeq \prod_{i=1}^{P}\mathbb Z_{n_i}$ for some $P$, $n_i \in \mathbb Z$, so that we may lift $k $ to a cochain valued in $\Z^{P}$. Therefore $G$ fits in the exact sequence
\begin{align}
	1\rightarrow \widetilde{G}\to \mathcal G\to G\to 1
\end{align}
where $\widetilde{G}=\prod_{i=1}^{P}n_i \mathbb Z$ and $\mathcal G=\mathbb Z^{P}$. Furthermore, there is a natural homomorphism associated with the previous short exact sequence known as the \emph{Bockstein homomorphism}
\begin{align}
	\beta: H^{1}(\cM,G)\longrightarrow H^{2}(\cM,\widetilde{G})
\end{align}
where we should think of $d^2k$ as being an element of $H^{2}(\cM,\widetilde{G})$.

\medskip \noindent
This model is invariant under the global symmetry transformation $k_a \rightarrow  k_a K^{-1} $ where $K\in Z^{0}(\cM,G)\simeq \text{Hom}(\pi_0(\cM),G)$ is a constant. We can gauge this global $G$-symmetry by firstly promoting $K$ so that it represents a 0-cochain $l$, and secondly introducing a flat background $G$-connection $g$ so that they tranform as
\begin{align}
	k_a \rightarrow k_a l_a^{-1} \q , \q g_{ab} \rightarrow l_a g_{ab}l_b^{-1} \; .
\end{align}
As before the flat connection is obtained by coloring the lattice with group variables $g_{ab} \in G$ on every 1-simplex such that for every 2-simplex $\triangle_2 = (abc)$ one has $g_{ac} = g_{ab} \cdot g_{bc}$. The set of such colorings is still denoted by ${\rm Col}(\cM,G)$.
Furthermore, we replace all the differentials on cochains by their covariant extensions, namely
\begin{align}
	dk \rightarrow d_g k \; ,
\end{align} 
i.e. we replace $\unit^k = k_a k_b^{-1}$ with $g_{ab}^k = k_ag_{ab} k_b^{-1}$. The gauged SPT model therefore reads
\begin{align*}
	\mathcal Z^{G}_{\omega}(\cM)
	=\frac{1}{|G|^{2|\triangle_0|}}
	\sum_{g,k}
	\prod_{\triangle_{d+1}}
	e^{2\pi i \langle \omega(d_gk)\, ,  \,\triangle_{d+1}\rangle } \; .
\end{align*}
We may gauge away the bosonic degrees of freedom $k$ by choosing $l=k$ so that the gauged partition function finally takes the form
\begin{align*}
	\mathcal Z^{G}_{\omega}(\cM)
	=&\;\frac{1}{|G|^{|\triangle_0|}}\sum_{g}\prod_{\triangle_{d+1}}e^{2\pi i S_{\omega}(g \, , \, \triangle_{d+1})}  
\end{align*} 
where we made use of the fact that $|G|^{|\triangle_{0}|}=\sum_{k \in C^{0}(\cM,G)}$. As before, we have the identification $S_\omega(g \, , \, \triangle_{d+1}) \equiv \la \omega(g), \triangle_{d+1}\ra $. As expected we recover exactly the Dijkgraaf-Witten partition function \eqref{DWact1}.

\subsection{SPTs protected by $q$-form symmetry \label{sec:SPTform}}
\noindent
We now consider models where the objects on which the symmetry group acts have support on $q$-dimensional manifolds embedded in spacetime submanifolds. In order to probe these symmetries we follow a procedure completely analogous to the one for 0-form symmetries, namely we introduce a flat background ($q$+1)-form connection $h^{(q+1)}$. Such connection is captured by ``holonomies'' ${\rm hol}_{h}\in \text{Hom}(\pi_{q+1}(\cM),H)$. Since $H$ is always abelian for $q > 0$ \cite{gaiotto2015generalized},  isomorphism classes of flat connections are elements in $H^{q+1}(\cM,H)$.    

We expect that phases of matter protected by $q$-form symmetry are in correspondence with topological actions that can be built from ($q$+1)-form connections $h^{(q+1)}$ via the Dijkgraaf-Witten prescription. 

So given a class $[\omega]\in H^{d}(B^{q+1}H,\mathbb R/\mathbb Z)$, one can construct an invertible lattice topological field theory that describes an SPT phase. The matter degrees of freedom denoted by $\lambda^{(q)}$ correspond to so-called $q$-connections (rather than $H$-valued $q$-cochains) whose space is provided by 
\begin{align}
	\lambda^{(q)}\in \widetilde{C}^{q}(\cM,H)=C^{q}(\cM,H)/ d \widetilde{C}^{q-1}(\cM,H) 
\end{align} 
such that $\widetilde{C}^1(\cM,H) \cong C^1(\cM,H)$.
This may be thought of as a lattice analog of what is commonplace in gauge theory, i.e taking a quotient on the space of lie algebra-valued forms by gauge transformations. The matter degrees of freedom $\lambda^{(q)}$ are mathematical objects that are identical to the gauge transformations of a $H$-valued $(q+1)$-connection. Since when $q>1$ the gauge transformations themselves have ``gauge transformations'', $\lambda^{(q)} \in \widetilde{C}^q(\cM,H)$\footnote{The configuration space of the matter degrees of freedom can alternately be motivated from the physical requirement that the partition function on a manifold $\cM$ without any background $H_{[q]}$ field turned on must be unity.}. The partition function then reads
\begin{align*}
	\mathcal Z^{H_{[q]}}_{\omega|\text{SPT}}(\cM) =&\;\frac{1}{|H|^{|\triangle_{0 \rightarrow q}|}} 
	\sum_{\lambda^{(q)} } \prod_{\triangle_{d+1}}
	e^{2\pi i S_{\omega}(d \lambda^{(q)} \, , \, \triangle_{d+1})} 
\end{align*}
where $|\triangle_{0 \rightarrow q}| = \sum_{i=0}^q (-1)^i|\triangle_{q-i}|$.
This model has a $q$-form global symmetry under $\lambda^{(q)} \rightarrow \lambda^{(q)} + \Lambda^{(q)}$ where $\Lambda^{(q)} \in {Z}^{q}(\cM,H)$. In order to gauge this model we promote $\Lambda^{(q)}$ to a $q$-connection $\xi^{(q)}$ and introduce a ($q$+1)-form flat $H$-connection $h^{(q+1)}$ satisfying $dh^{(q+1)} =0 $ along with the gauge transformation
\begin{align*}
	h^{(q+1)} &\rightarrow h^{(q+1)} + d\xi^{(q)} \\
	\lambda^{(q)} &\rightarrow \lambda^{(q)} - \xi^{(q)} \; .
\end{align*}
The procedure to obtain the response action is identical to that for 0-form SPTs, we first gauge away $\lambda^{(q)}$ and then perform the partition sum to obtain
\begin{align}
	\mathcal Z^{H_{[q]}}_{\omega}(\cM)
	=\frac{1}{|H|^{|\triangle_{0 \rightarrow q}|}} 
	\sum_{h^{(q+1)}} \prod_{\triangle_{d+1}} 
	e^{2\pi i S_{\omega}(h^{(q+1)} \, , \, \triangle_{d+1})}
\end{align}
which is a topological gauge theory for the symmetry group $H_{[q]}$, or equivalently a topological sigma model with target space $B^{q+1}H$ and topological action provided by a class representative of $[\omega]\in H^{d+1}(B^{q+1}H,\mathbb R/\mathbb Z)$.

\subsection{SPTs protected by 2-group symmetry}
\noindent
Let us now construct SPT phases protected by a 2-group $\mathbb G=(G,H,\tr,\alpha)$. These SPTs must reduce to the usual group cohomology SPTs described in sec.~\ref{sec:SPTgr} when $H$ is trivial and to the 2-form SPTs described in sec.~\ref{sec:SPTform} when $G$ is trivial. As such, these can be modeled as a state-sum model with degrees of freedom $k \in C^{0}(\cM,G)$ and $\lambda \in \widetilde{C}^{1}(\cM,H)$\footnote{Similar to $h^{(q+1)}$, we drop the superscript on $\lambda^{(q)}$ when $q=1$.}. For an SPT labeled by a cocycle $[\omega] \in H^{d+1}(B\mathbb G,\mathbb R/\mathbb Z)$, the model takes the form \cite{kapustin2017higher}
\begin{align*}
	\mathcal Z_{\omega| \text{SPT}}^{\mathbb G}(\cM)=&\;
	\frac{1}{\#_{G,H,\triangle}}
	\sum_{k,\lambda}\prod_{\triangle_{d+1}}
	e^{2\pi i S_{\omega}(\unit^k|d\lambda + \zeta (\unit,k) \, , \, \triangle_{d+1})} 
\end{align*}
with $\#_{G,H,\triangle} = |G|^{|\triangle_{0}|}|H|^{|\triangle_1|-|\triangle_0|}$. Note that as before we have the identification $S_\omega(\unit^k|d\lambda + \zeta(\unit,k) \, , \, \triangle_{d+1}) \equiv \la \omega(\unit^k|d\lambda + \zeta(\unit,k)), \triangle_{d+1}\ra$. The gauged SPT is obtained by coupling the model to a flat background $\mathbb G$-connection so that the partition function reads
\begin{align*}
	\mathcal Z^{\mathbb G}_{\omega}(\cM)=&\; 
	\frac{1}{\#_{G,H,\triangle}^2} 
	\sum_{\substack{({g},{h}) \\ k,\lambda}}\prod_{\triangle_{d+1}}
	e^{2\pi i S_\omega(d_g k| d_h \lambda + \zeta(g,k)  \, , \, \triangle_{d+1})}
\end{align*}
where $(g,h) \in {\rm Col}(\cM, \mathbb G)$ are such that $dg = \unit$ and $dh=\alpha(g)$. This model has two sets of gauge transfromations, namely a 0-form gauge transformation with gauge parameter $l$ which is a 0-cochain valued in $G$, and a 1-form gauge transformation with gauge parameter $\xi$ which is a 1-cochain valued in $H$. Under the 0-form gauge transformation the flat connection $(g,h)$ transform as
\begin{equation*}
	(g,h) \rightarrow (g^l,h+\zeta(g,l))
\end{equation*}
whereas the matter degrees of freedom transform as
\begin{equation*}
	k_a \rightarrow k_al_a^{-1}
\end{equation*}
and $\zeta(g,k)$ transforms as
\begin{equation*}
	\zeta(g,k) \rightarrow \zeta(g,k) - \zeta(g,l) \; .
\end{equation*} 
On the other hand, the 1-form gauge transformation acts as 
\begin{align*}
	h &\rightarrow h + d\xi \\
	\lambda &\rightarrow \lambda - \xi \; .
\end{align*}
The model can be gauge fixed by setting $l=k$ and $\xi=\lambda$ to gauge away the matter degrees of freedom $k$ and $\lambda$. The partition function then takes the form
\begin{align}
	\mathcal Z^{\mathbb G}_{\omega}(\cM)
	=&\; \frac{1}{\#_{G,H,\triangle}}
	\sum_{(g,h)}\prod_{\triangle_{d+1}}
	e^{2\pi i S_{\omega}(g|h \, , \, \triangle_{d+1}) }
\end{align} 
which is a 2-group topological gauge theory labeled by cohomology class $[\omega] \in H^{d+1}(B\mathbb G,\mathbb R/\mathbb Z)$. 

\subsection{Further generalizations}
\noindent
State-sum models for SPT phases protected by $\mathbb G_{[q]}$ can be constructed in a similar manner as the models described above. The matter degrees of freedom live on $q$-simplices and vertices of a triangulation of the ($d$+1)-manifold $\cM$. We denote the degrees of freedom living on $q$-simplices as $\lambda^{(q)}$ where $\lambda^{(q)}\in \widetilde{C}^{q}(\cM,H)$. The degrees of freedom living on vertices are denoted by $k$ and are valued in $G$ which may be non-abelian. Such SPTs are labeled by a cocycle $[\omega]\in H^{d+1}(B\mathbb G_{[q]},\mathbb R/\mathbb Z)$ and the state-sum model takes the form 
\begin{align*}
	&\mathcal Z^{\mathbb G_{[q]}}_{\omega|\text{SPT}}(\cM) 
	\\ & \q = \frac{1}{\#_{(G,H_{[q]}, \triangle)}}  \sum_{k,\lambda^{(q)}}\prod_{\triangle_{d+1}}
	e^{2\pi i S_{\omega}(\unit^k|d\lambda^{(q)} +\zeta(\unit,k) \, , \, \triangle_{d+1})} 
\end{align*} 
with $\#_{(G,H_{[q]}, \triangle)} = |G|^{|\triangle_0|}|H|^{|\triangle_{0 \rightarrow q}|}$ and  $\zeta \in C^{q}(\cM,H)$. This model has a global $\mathbb G_{[q]}$-symmetry which comprises of a 0-form part and a $q$-form part which act as
\begin{align}
	k_a &\rightarrow k_aK^{-1} \\ 
	\lambda^{(q)} &\rightarrow \lambda^{(q)} + \Lambda^{(q)}
\end{align}
where $K\in \text{Map}(\pi_0(\cM),G)$ is a constant and $\Lambda^{(q)}\in \text{Map}[\pi_{q}(\cM),H]$ is an element of $Z^q(\cM,H)$. The model can be gauged by promoting $K$ and $\Lambda^{(q)}$ to cochains $l$ and $\xi^{(q)}$ such that
\begin{align}
	l\in &\; \text{Map}[\triangle_{0},G] \nonumber \\
	\xi^{(q)} \in &\; \text{Map}[\triangle_{q},H]
\end{align} 
and by introducing 1-form gauge field $g$ and ($q$+1)-form gauge field $h^{(q+1)}$. Coupling the model with this background connection amounts to the following replacements
\begin{align}
	\unit^k \equiv d k &\rightarrow d_{g}k  \nonumber \\
	d \lambda^{(q)} &\rightarrow d_h \lambda^{(q)} +\zeta(g,k) \; .
\end{align}
Under $0$-form gauge transformations, we have
\begin{align*}
	(g,h^{(q+1)}) &\rightarrow (g^l,h^{(q+1)} + \zeta(g,l)) \\
	\zeta(g,k) &\rightarrow \zeta(g,k) - \zeta(g,l) \\
	k_a &\rightarrow k_al_a^{-1}
\end{align*}
while $1$-form gauge transformations read
\begin{align*}
	h^{(q+1)} &\rightarrow h^{(q+1)} + d \xi^{(q)} \\
	\lambda^{(q)} &\rightarrow \lambda^{(q)} - \xi^{(q)} \; .
\end{align*}
We can finally gauge away the matter degrees of freedom by choosing $l=k$ and $\xi^{(q)}=\lambda^{(q)}$. The gauged model finally takes the form
\begin{align*}
	\mathcal Z_{\omega}^{\mathbb G_{[q]}}(\cM)
	= \frac{1}{\#_{(G,H_{[q]}, \triangle)}}
	\! \sum_{(g,h^{(q+1)})}\prod_{\triangle_{d+1}} \!
	e^{2\pi i S_{\omega}(g|h^{(q+1)} \, , \, \triangle_{d+1} )}
\end{align*}
where the sum is over configurations $(g,h)$ satisfying $dg=0$ and $dh^{(q+1)}=\alpha^{(q+2)}(g)$ with $[\alpha^{(q+2)}]\in H^{q+2}(BG,H)$. This is nothing but a $\mathbb G_{[q]}$-topological gauge theory.

\subsection{Explicit lattice realization of 2-group SPTs \label{sec:expSPT}}
\noindent
As explained above, SPTs are phases of matter satisfying three properties: $(i)$ They display a global symmetry, $(ii)$ they are gapped, $(iii)$ the low-enery limit is described by a TQFT which is trivial. As for the corresponding intrinsic topological orders, these symmetry protected phases can be formulated as lattice quantum field theories. In this section, we will present in detail an explicit lattice realization of SPTs protected by a 2-group symmetry together with the corresponding gauging procedure.

In order to be as explicit as possible, we will focus, as in sec.~\ref{sec:expLat1}, on the three-dimensional case. However, the same strategy applies to any dimension. Let us consider a closed three-dimensional manifold $\mathcal{M}$ realized as a gluing of oriented $3$-simplices $\triangle_3$ by picking a triangulation $\triangle$. We then consider a matter field configuration which is obtained by assigning group elements $\{k\}$ of $G$ to every vertex and group elements $\{\lambda\}$ of $H$ to every 1-simplex. We then consider a ${\rm U}(1)$-valued function $\nu_3$ which is evaluated on each $3$-simplex so that partition function finally reads 
\begin{align}
	\label{SPT1}
	\mathcal{Z}_{\nu_3|{\rm SPT}}^{\mathbb G}(\cM) &= \frac{1}{\#_{(G,H,\triangle)}} \sum_{k,\lambda}\prod_{\triangle_3}\nu^{\epsilon(\triangle_3)}_3(k|\lambda) 
\end{align}
with $\#_{(G,H, \triangle)} = |G|^{|\triangle_0|}|H|^{|\triangle_1| - |\triangle_0|}$. We could also work with a linearized version of $\nu_3$ valued in $\mathbb{R}/ \mathbb{Z}$ as in the previous sections but chose not to for notational convenience. For a given 3-simplex labeled as follows
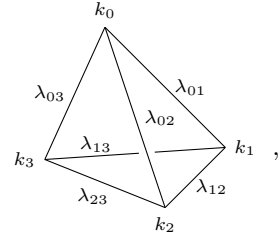
\begin{equation*}
	\SPTtetraONE{0.8}{2} \; ,
\end{equation*} 
the topological action explicitly reads $\nu_3^{\epsilon(0123)}(k|\lambda) \equiv \nu_3^{+1}(k_0, \dots, k_3 ; \lambda_{01}, \lambda_{02}, \lambda_{03}, \lambda_{12} ,\lambda_{13}, \lambda_{23})$.
Recall that the system must display a global 2-symmetry such that the 0-form symmetry group is $G$ and the 1-form symmetry group is $H$. This imposes the function $\nu_3$ to be \emph{homogeneous}, {\it i.e.}
\begin{align}
	&\nu_3(k_0, \dots, k_3 | \lambda_{01}, \dots, \lambda_{i<j}, \dots, \lambda_{23})
	\\ \nn & \q \equiv \nu_3(k_0K^{-1}, \dots, k_3K^{-1} | 
	\lambda_{01}+ \Lambda_a - \Lambda_b, 
	\\ \nn & \hspace{12.1em} \lambda_{02}+\Lambda_a, \lambda_{03}+\Lambda_c,  \lambda_{12}+\Lambda_b,
	\\ \nn & \hspace{12.1em} \lambda_{13}+\Lambda_c - \Lambda_a +\Lambda_b, \\ \nn & \hspace{12.1em} \lambda_{23}+\Lambda_c - \Lambda_a) 
\end{align}
such that $K \in Z^0(\cM,G)$ an $\Lambda \in Z^1(\cM,H)$.
Furthermore, the low energy limit is described by a TQFT which requires invariance of the partition function under Pachner moves which in turn imposes the function $\nu_3$ to satisfy the following condition
\begin{align} 
	\label{fakeCocycle}
	\prod_{j=0}^{4}\nu_3^{(-1)^j}(k_0, \dots, \widehat{k}_j, \dots, k_{4} | 
	\lambda_{01}, \dots, \widehat{\lambda}_{{\sss \bullet }j \atop j{\sss \bullet}}, \dots, \lambda_{34})= 1 \; ,
\end{align}
where the notation $\widehat{\sss \bul}$ means that the corresponding element is omitted from the list. For instance, we have 
\begin{align*}
	&(k_0, \dots, \widehat{k}_2, \dots, k_{4} |
	\lambda_{01}, \dots, \widehat{\lambda}_{{\sss \bullet }2 \atop 2{\sss \bullet }}, \dots, \lambda_{34}) \\
	& \q \equiv (k_0,k_1,k_3 | \lambda_{01},\lambda_{03}, \lambda_{04},\lambda_{13}, \lambda_{14}, \lambda_{34}) \; .
\end{align*}
This function $\nu_3$ turns out to represent a cocycle in $H^3(B\mathbb{G},{\rm U}(1))$.

\medskip\noindent

Let us now show how gauging the global $\mathbb{G}$-symmetry yields the topological model we discussed in the previous section. To do so we will generalize the procedure proposed in \onlinecite{2015PhRvB..91p5119W} to 2-groups. By fixing the global symmetry such that $K = k_3$, $\Lambda_a = - \lambda_{02}$, $\Lambda_b = - \lambda_{12}$ and $\Lambda_c = - \lambda_{03}$, we can rewrite the partition function \eqref{SPT1} in terms of the function 
\begin{align}
	\label{topoAct1}
	&\nu_3(k_0k_3^{-1}, \dots, k_2k_3^{-1},\unit |  \lambda_{01}-\lambda_{02}+\lambda_{12}, 0,0,0,
	\\ \nn &\hspace{10em} \lambda_{13}-\lambda_{03}+\lambda_{02}- \lambda_{12},
	\\ \nn &\hspace{10em} \lambda_{23}-\lambda_{03}+\lambda_{02}) \;.
\end{align} 
Let us now define another function $\omega_3$ such that
\begin{align}
	\label{defCocycle}
	&\omega_3(k_1,k_2,k_3|\lambda_{a}, \lambda_{b}, \lambda_{c}) 
	\\ \nn & \! \equiv \nu_3(k_1k_2k_3, k_2k_3, k_3,\unit | 
	\lambda_{a}-\zeta(\unit,k)_{012},0,0,0,
	\\ \nn& \hspace{10.6em} \lambda_{c}-\lambda_{a}-\zeta(\unit,k)_{013} + \zeta(\unit,k)_{012},
	\\ \nn & \hspace{10.6em} \lambda_b-\zeta(\unit,k)_{023}) \; .
\end{align}
It follows from the condition \eqref{fakeCocycle} that the function $\omega_3$ satisfies the usual 2-group $3$-cocycle condition (see app.~\ref{app:cocycle2Gr}).
We then deduce from the definition of $\omega_3$ that the topological action \eqref{topoAct1} can be rewritten  
\begin{align*}
	&\omega_3(k_0k_1^{-1},\dots, k_{2}k_3^{-1} | 
	\lambda_{12} - \lambda_{02}+ \lambda_{01} + \zeta(\unit,k)_{012}, 
	\\ & \hspace{9.1em} \lambda_{23} - \lambda_{03}+ \lambda_{02} + \zeta(\unit,k)_{023},
	\\& \hspace{9.1em} \lambda_{13} - \lambda_{03}+ \lambda_{01}+\zeta(\unit,k)_{013}) 
\end{align*}
which is manifestly invariant under global $\mathbb{G}$-symmetry (making in part use of the fact that we chose a normalization such that $\zeta(\unit^{k},K)=0$ if $K$ is constant). The previous expression can be rewritten in the synthetic form $\omega_3(\unit^{k}|d\lambda+\zeta(\unit,k))$.

Let us now couple our SPT phase cohmological model to a flat $\mathbb G$-connection. As explained in sec.~\ref{sec:weak2groups}, this can be performed by assigning a group variable $g_{ab} \in G$ to every 1-simplex of the lattice and a group element $h_{abc} \in H$ to every 2-simplex. Furthermore these variables must satisfy conditions which are the simplicial translations of cocycle conditions, namely $g_{ac} = g_{ab} \cdot g_{bc}$ and $h_{bcd}-h_{acd}+h_{abd}-h_{abc} = \alpha(g_{ab},g_{bc},g_{cd})$. Graphically, the coupling to the flat 1-connection can be represented as follows in the case of a 3-simplex $\triangle_3 = (0123)$:
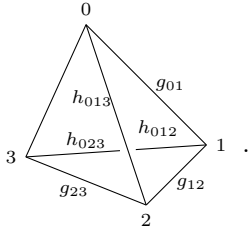
\begin{equation*}
	\SPTtetraONE{0.8}{1} \; .
\end{equation*}
More generally the convention regarding the labeling of the edges and faces is the same as the one presented in sec.~\ref{sec:expLat1}.
The insertion of these variables is such that the topological action now reads
\begin{align*}
	&\omega_3(k_0g_{01}k_1^{-1},\dots, k_{2}g_{23}k_3^{-1} | h_{012}+ d\lambda_{012} +\zeta(g,k)_{012}, 
	\\ & \hspace{11.8em} h_{023}+ d\lambda_{023} + \zeta(g,k)_{023},
	\\& \hspace{11.8em} h_{013} +d\lambda_{013} + \zeta(g,k)_{013}) \; ,
\end{align*}
where we made use of the notation $d \lambda_{abc} \equiv \la d\lambda, (abc) \ra =  \lambda_{bc}-\lambda_{ac}+\lambda_{ab}$. 
The new topological action is now invariant under the 1-form gauge transformation
\begin{equation*}
	h_{abc} \rightarrow h_{abc} + \xi_{bc} - \xi_{ac} + \xi_{ab} \q , \q \lambda_{ab} \rightarrow \lambda_{ab} - \xi_{ab}
\end{equation*}
and the 0-form gauge transformation
\begin{align*}
	&k_a \rightarrow k_al_a^{-1} \q , \q g_{ab} \rightarrow l_a g_{ab} l_b^{-1} \\
	& \q h_{abc} \rightarrow h_{abc} + \zeta(g,l)_{abc} \\
	& \zeta(g,k)_{abc} \rightarrow \zeta(g,k)_{abc} - \zeta(g,l)_{abc}
\end{align*}
with $k_a$ a $G$-valued 0-form gauge parameter and $\xi_{ab}$ is an $H$-valued 1-gauge parameter. These are exactly the 2-group gauge transformations we introduced above. 

By choosing $l_a=k_a$ and $\xi_{ab}=\lambda_{ab}$ we can set the matter field to the identity which shifts at the same time the flat $\mathbb G$-connection configuration. If we now decide to integrate over the space of flat $\mathbb G$-connections using the discrete measure $1/ \#_{G,H,\triangle}$, we can perform explicitly the sum over the matter field so as to recover the higher gauge model for the 2-group $\mathbb{G}$:
\begin{align*}
	\mathcal Z^{\mathbb G}_{\omega_3}(\cM)=\frac{1}{\#_{G,H,\triangle}}
	\sum_{(g,h)\in \text{Col}(\cM,\mathbb G) }\prod_{\triangle_{3}}\omega_3^{\epsilon(\triangle_3)}(g|h) \; .
\end{align*}

\section{'t Hooft anomalies for higher symmetry structures \label{sec:hooft}}
\noindent
\emph{'t Hooft anomalies} for global symmetries in $d$-dimensional quantum field theories and symmetry protected topological phases of matter in ($d$+1) dimensions are very intimately related. Consider a QFT on a $d$-dimensional manifold $ \mathcal N$ with a global symmetry. Global symmetries can be of different varieties. Restricting to unitary symmetries that do not act on spacetime indices, one can have 0-form symmetries, $q$-form symmetries, or symmetries corresponding categorically-higher versions of groups such as $\mathbb G$ or $\mathbb G_{[q]}$, as mentioned in the previous section. 

One might be interested in gauging these global symmetries, some subgroup or subcategory of them as the case may be. The usual method to carry out such a gauging procedure is to couple the theory to a background field and ask whether the theory is invariant under gauge transformations. More specifically, a background field is equivalent to a \emph{network of symmetry defects} so that invariance under gauge transformations is equivalent to invariance under rearrangements of the defect structure.

For a quantum field theory on a $d$-dimensional manifold $\cN$, defects corresponding to $0$-form symmetries are localized on ($d$$-$1)-dimensional submanifolds of spacetime while $q$-form symmetry defects are localized on ($d-$$q-$1)-dimensional submanifolds. Defects corresponding to other categorified generalizations can also be appropriately defined. The goal in such case is to build a fine enough mesh with these symmetry defects so as to be able to triangulate the manifold $\mathcal N$ using a network of defects. If the symmetry is healthy, i.e non-anomalous, the path integral of the quantum field theory assigns to the network of defects a ${\rm U}(1)$-valued number that is invariant under changes of triangulation, i.e. rearrangements of the network.

Let us denote the path integral on a manifold $\mathcal N$ in the presence of background defects $A_{X}$ by ${\mathcal Z}(\mathcal N,A_{X})$.\footnote{X denotes some general symmetry structure, i.e a group, a higher-form group or a categorical generalization thereof.} A reparamentrization corresponds to a ``gauge transformation'' of $A_{X}$ so that we require 
\begin{align}
	{\mathcal Z}(A_{X}, \mathcal N) = {\mathcal Z}(A_{X}+\delta A_{X}, \mathcal N) \; .
\end{align}
A violation of this requirement would then take the form 
\begin{align}
	{\mathcal Z}(A_X+\delta A_X, \mathcal N)= e^{i \langle f(A_X,\delta A_{X}), [\mathcal N] \rangle } {\mathcal Z}(A_X, \mathcal N)
\end{align}
which are most naturally encoded in cohomology classes of degree $d$+1, i.e $H^{d+1}(X,\mathbb R/\mathbb Z)$ where $X$ is the classifying space of the group-like global symmetry structure. In other words, in most cases of interest, it is possible to find a cohomology class $[\omega]\in H^{d+1}(X,\mathbb R/\mathbb Z)$ and a manifold $\mathcal M$ satisfying $\partial \mathcal M= \mathcal N$ such that the quantity 
\begin{align}
	{\mathcal Z}(A_{X}, \mathcal N) e^{2\pi i \langle \omega(A^{\#}_{X}),[\mathcal M] \rangle }
\end{align} 
depends unambiguously on the background field $A_{X}$. In the equation above, $\omega(A^{\#}_X)$ should be understood as a cocycle evaluated on a background structure on $\mathcal M$ such that it restricts to $A_{X}$ on $\mathcal N$ as discussed in the previous sections. Therefore, $d$-dimensional QFTs with 't Hooft anomalies can be realized as symmetric theories on the surface of appropriate SPTs.

Below we discuss several examples of quantum field theories with 't Hooft anomalies corresponding to higher-form symmetry and 2-group symmetries which can be realized on the surface of the corresponding higher-form and 2-group SPTs. Most of the examples we study correspond to gapped boundaries of SPTs.

\subsection{Scenario 1: 1-form symmetry anomalies in (2+1)d abelian Dijkgraaf-Witten theories}
\noindent 
In this subsection we study abelian Dijkgraaf-Witten theories in (2+1)d. We show that in general these theories have 2-group global symmetries. The 1-form symmetry subgroup of the 2-group is generated by the (abelian) line operators of the theory. Then gauging this 1-form symmetry implies condensing the corresponding lines. Such a condensation process may be anomalous if two or more of the lines are not-mutually local, i.e they have non-trivial braiding statistics. In this case, we say we have an abelian Dijkgraaf-Witten theory with an anomalous 1-form symmetry. We show that such a symmetry can be gauged on the surface of a 2-form SPT.  

\medskip \noindent Let us consider global symmetries of abelian Dijkgraaf-Witten theories in (2+1)d. Since any finite abelian group is a product of cyclic groups, it is convenient to work with $G=\mathbb Z_{n}^{k}$. This can be generalized to other finite abelian groups straightforwardly. Such Dijkgraaf-Witten theories admit a Lagrangian formulation of the form
\begin{align}
	S_{\omega|\text{DW}}(\cN)= \frac{2\pi i}{n}\int_{\mathcal N} \big(\delta_{IJ} \bfb^{I} \smilo d \bfa^{J} +\omega(a) \big)
	\label{DW_action}
\end{align}
where we use the Einstein summation convention. Here $\omega$ is a simplicial expression that represents a cohomology class in $H^{3}(G,\mathbb R/\mathbb Z)$ and the fields $a$, $b${\footnote{To avoid confusion we reserve using the notation $g,h$ etc. for fields appearing in the response theory of the bulk SPT.} 
are cochains valued in $G${\footnote{More precisely $b$ labels charge excitations in these twisted quantum doubles and should therefore be valued in $\text{Rep}(G)$ but for abelian groups $\text{Rep}(G) \simeq G$.}}. The above action \eqref{DW_action} is invariant under both a 0-form and a 1-form symmetry transformations which together have the structure of a 2-group. The 0-form global symmetry group is provided by $\text{Aut}_{\omega}(G)$, namely the subgroup of $\text{Aut}(G)$ which preserves the cohomology class $\omega$\footnote{More specifically, one defines ${\rm Aut}_\omega(G)$ as
	\begin{equation}
		\big\{ \varphi \in {\rm Aut}(G) \big| \omega(\varphi(g), \varphi(h), \varphi(k)) = \omega(g,h,k), \forall g,h,k \in G\big\} \; .
	\end{equation}
Note that here we only consider those global symmetries that are obtained from automorphisms of $G$. Notably we do
not consider electric-magnetic duality like symmetries that mix electric and magnetic sectors (a and b fields).}, such that a given $\varphi \in \text{Aut}_{\omega}(G)$ acts on the fields as
\begin{align}
	\varphi:\left[ \begin{array}{c}
	\bfa  \\
	\bfb  \end{array} \right] \mapsto 
	\left[ \begin{array}{c}
	\varphi(\bfa)  \\
	\varphi(\bfb)  \end{array} \right] \; .
\end{align}
The 1-form global symmetry group denoted by $(G\times G)_{[1]}$\footnote{More precisely this is the 1-form symmetry only for type-{\scriptsize (I)},{\scriptsize (II)} Dijkgraaf-Witten theories, not for type-{\scriptsize (III)} as we will see shortly.} such that $\bfla=(\lambda_a,\lambda_b) \in (G\times G)_{[1]}$ acts on the fields as
\begin{align}
	\bfla :\left[ \begin{array}{c}
	\bfa^I  \\
	\bfb^{I}  \end{array} \right] \mapsto \left[ \begin{array}{ccc}
	\bfa^{I}  \\
	\bfb^{I}  \end{array} \right]
	+
	\left[ \begin{array}{c}
	\bfla_\bfa^{I}  \\
	\bfla_\bfb^{I}  \end{array} \right] \; ,
	\label{global1}
\end{align}
where $\bfla_\bfa^{I},\bfla_\bfb^{I}\in Z^{1}(\mathcal M,\mathbb Z_n)$. Before attempting to gauge these global symmetries, let us explain how they form a 2-group. Earlier we saw that the data that prescribes a 2-group is: A group, an abelian (1-form symmetry) group, an action of the former on the latter, and a certain cohomology class. Let us call the 2-group of global symmetries of a particular Dijkgraaf-Witten labeled by $[\omega]$ by $\mathbb G^{(\omega,\alpha)}$. Clearly the first three pieces of data are the group $\text{Aut}_{\omega}(G)$, the 1-form symmetry group $(G\times G)_{[1]}$ and the obvious action of the former on the latter. In addition we must pick a 3-cocycle $[\alpha]\in H^{3}(\text{Aut}_{\omega}(G),G\times G)$. Then the 2-group $\mathbb G^{(\omega,\alpha)}$ fits in the exact sequence
\begin{align}
	1\rightarrow (G\times G)_{[1]} \rightarrow \mathbb G^{(\omega,\alpha)}\rightarrow \text{Aut}_{\omega}(G) \rightarrow 1 \; .
\end{align}
As expected gauging $\mathbb G^{(\omega,\alpha)}$ would involve turning on a background 1-cochain $g$ and 2-cochain $h$ valued in $\text{Aut}_{\omega}(G)$ and $G\times G$ respectively that satisfy the twisted cocycle conditions
\begin{align}
	dg=&\; 1 \; , \\
	d_{g \triangleright} h=&\; \alpha(g) \; .
\end{align} 
Let us now return to the more modest task of gauging the 1-form subgroup $(G\times G)_{[1]}$. As explained earlier, gauging a global symmetry is performed locally by relaxing the cocycle condition on $\bfla_\bfa^{I},\bfla_\bfb^{I}$ \eqref{global1}. 
We work explicitly by picking a cocycle in the group cohomology $H^{3}(G,{\rm U}(1))$. In general, discrete abelian groups have three families of 3-cocycles usually referred to as type-{\small (I)}, type-{\small (II)} and type-{\small (III)} \cite{propitius1995topological}:
\begin{align}
	\nn
	\omega_{\rm (I)}(g_\bfa,g_\bfb,g_\bfc) 
	&= {\rm exp}\Big( \frac{2 \pi i p_I}{n^2} g_\bfa^I \big (g_\bfb^I+g_\bfc^I-[g_\bfb^I+g_\bfc^I] \big)\Big)
	\\ \nn
	\omega_{\rm (II)}(g_\bfa,g_\bfb,g_\bfc) 
	&= {\rm exp}\Big( \frac{2 \pi i p_{IJ}}{n^2} g_\bfa^I \big (g_\bfb^J+g_\bfc^J-[g_\bfb^J+g_\bfc^J] \big)\Big)
	\\ \nn
	\omega_{\rm (III)}(g_\bfa,g_\bfb,g_\bfc) 
	&= {\rm exp}\Big( \frac{2 \pi i p_{IJK}}{n^2} g_\bfa^I g_\bfb^J g_\bfc^K \Big)
\end{align} 
where $g_\bfa,g_\bfb,g_\bfc\in G$, $p_I, p_{IJ}, p_{IJK} \in \mathbb{Z}/n\mathbb{Z}$ and $[\, {\sss \bullet} \, ] \equiv {\sss \bullet}\;  {\rm mod} \; n$. It is clear from the equations above that types-{\small (I)},{\small (II)} are quite similar so that we treat these cocycles separately from type-{\small (III)} cocycles. In the former case, the action \eqref{DW_action} takes the form
\begin{align}
	S_{{\rm (I,II)}|\text{DW}}(\cN)
	=\frac{2\pi i}{n}\int_{\cN}
	\big( \delta_{IJ}\bfb^{I}  \smilo d \bfa^{J} +p_{IJ} \bfa^{I} \smilo d \bfa^{J} \big) \; .
\end{align}
Under a 1-form gauge transformation (i.e treating $\lambda_{\bfa,\bfb}^{I}$ as a $\mathbb Z_{n}$-valued cochains instead of cocycles) the variation of the action reads\footnote{Note that we drop boundary terms as $\cN$ is a closed manifold.}
\begin{align}
	&\delta S_{{\rm (I,II)}|\text{DW}} (\cN) \\
	& \q = \frac{2\pi i}{n}
	\int_{\cN} \! \Big( \delta_{IJ} \big( \bfla_{\bfb}^{I} \smilo d \bfa^{J}+ \bfb^{I}\smilo d \bfla_{\bfa}^{J}+ \bfla_{\bfb}^{I}\smilo d \bfla_{\bfa}^{J}\big)
	\\ \nn 
	& \hspace{4.9em} + p_{IJ} \big(\bfla_{\bfa}^{I}\smilo d \bfa^{J} + \bfa^{I}\smilo d \bfla_{\bfa}^{J}+\bfla_{\bfa}^{I}\smilo d \bfla_{\bfa}^{J} \big) \Big) \; .
\end{align}
In order to compensate for the new terms which depends on the 1-form gauge parameters, we couple the model with background 2-form gauge fields $\bfh_{\bfa}^{I}$ and $\bfh_\bfb^{I}$ which transform as
\begin{align}
	\bfh_{\bfa}^{I}\; \rightarrow \; {\bfh}_{\bfa}^{I}-d \bfla_{\bfa}^{I} \q , \q
	\bfh_{\bfb}^{I} \; \rightarrow \; {\bfh}_{\bfb}^I -d \bfla_{\bfb}^{I}
\end{align} 
so that the gauged action reads 
\begin{align}
	&S_{\text{(I,II)}|\text{DW}}^{\text{gauged}}(\cN) \\ 
	& \q = \frac{2\pi i}{n}\int_{\cN}
	\Big(\delta_{IJ} \big( \bfb^{I}\smilo d \bfa^{J}   
	+ \bfh_\bfb^{I}\smilo \bfa^{J} + \bfb^I \smilo \bfh_\bfa^{J} \big) \\ 
	& \hspace{5em} + p_{IJ} \big( \bfa^{I} \smilo d \bfa^{J} + \bfh_\bfa^{I}\smilo \bfa^{J}+\bfa^{I}\smilo \bfh_\bfa^{J} \big) \Big) \; . 
\end{align} 
Under gauge transformations, the variation of its action is finally given by
\begin{align}
	&\delta S_{{\rm (I,II)}|\text{DW}}^{\rm gauged}(\cN)
	\\ \
	& \q = \frac{2\pi i}{n} \int_{\cN} \! \Big(\delta_{IJ}\big(-d \bfla_{\bfb}^I \smilo \lambda_{\bfa}^J +\bfla_{\bfb}^{I}\smilo \bfh_\bfa^{J} + \bfh_\bfb^{I}\smilo \bfla_{\bfa}^{J} \big)
	\\ \nn 
	& \hspace{4.85em} + p_{IJ} \big(\bfla_{\bfa}^{I} \smilo \bfh_\bfa^{J}+\bfh_\bfa^{I}\smilo \bfla_{\bfa}^J - \bfla_{\bfa}^{I} \smilo d \bfla_{\bfa}^{J} \big)\Big) \; .
\end{align}
It turns out that the variation of the gauged action exactly cancels the boundary term in the variation of 
\begin{align*}
	S_{{\rm (I,II)}|\text{SPT}}(\cM)=\frac{2\pi i}{n}\int_{\cM}
	\big( \delta_{IJ}{\bfh}^I_\bfb\smilo {\bfh}^J_{a}+p_{IJ}{\bfh}_\bfa^{I}\smilo \bfh_\bfa^{J} \big) \; .
\end{align*}
Together the 3+1d bulk and 2+1d boundary are gauge invariant under 1-form gauge transformations.
Hence we have shown that it is possible to gauge a 1-form symmetry corresponding to non-mutually local abelian lines at the cost of introducing a (3+1)d bulk with the topological response action proposed above. Gauging a 1-form symmetry corresponding to mutually non-local lines implies trying to proliferate these lines freely commonly known as anyon condensation. It is clear that such a process cannot be well defined (i.e is anomalous) for an inherently (2+1)d theory due to the phase ambiguity that the partition function would accrue every time two mutually non-local lines cross each other. Next we show that no such anomaly exists for type-{\small (III)} Dijkgraaf-Witten theory as all the abelian lines are mutually local for such theories.

\medskip \noindent Consider the following action that describes type-{\small(III)} Dijkgraaf-Witten theories
\begin{align}
	&S_{\text{(III)}|\text{DW}}(\cN)
	\\ & \q =\frac{2\pi i}{n}\int_{\cN}\big(\delta_{IJ}\bfb^{I}\smilo d \bfa^{J}+p_{IJK} \bfa^{I}\smilo \bfa^J \smilo \bfa^K \big)  \; .
\end{align}
Recall that a 1-form symmetry is generated by abelian lines in the theory. It is known that the type-{\small (III)} theory for $G=\mathbb Z_n^{k}$ has only $n^k$ abelian line operators instead of $|G^2|=n^{2k}$ \cite{propitius1995topological, he2017field}. Therefore in this case the 1-form symmetry is only $G_{[1]}$ instead of $(G\times G)_{[1]}$ and is generated by the following abelian Wilson line operators
\begin{align}
	\mathcal W_{\bfa^I}(\gamma)=\exp\Big\{i\oint_{\gamma}\bfa^{I}\Big\} \; .
\end{align}
Furthermore, the global 1-form symmetry acts as
\begin{align}
	\label{wilsonGauge}
	\bfla:\bfb^{I} \mapsto \bfb^{I} + \bfla_{\bfb}^{I} \q , \q \bfla_{\bfb}^{I}\in Z^{1}(\cN,G) \; .
\end{align}
Since all these lines are bosonic and mutually transparent, we are able to gauge the 1-form symmetry\footnote{Gauging a 1-form symmetry is synonymous to condensing abelian lines that generate the symmetry. If the lines are all bosonic and mutually transparent, they can be condensed and there is no 1-form 't Hooft anomaly \cite{Bhardwaj:2016clt,  tiwari2017bosonic}.} and the gauged action reads 
\begin{align}
	S_{\text{(III)}|\text{DW}}^{\text{gauged}}(\cN)
	& = \frac{2\pi i}{n}\int_{\cN}
	\Big( \delta_{IJ}\big(\bfb^{I}\smilo d \bfa^{J}+{\bfh}_\bfb^{I} \smilo \bfa^{J}\big) 
	\\ & \hspace{4em} + p_{IJK}\bfa^{I} \smilo \bfa^{J} \smilo \bfa^{K} \Big) \; .
\end{align}
It is then straightforward to show that this is invariant under the gauge transformation \eqref{wilsonGauge} and ${\bfh}^{I}_\bfb\rightarrow {\bfh}^{I}_{\bfb}- d \bfla_{\bfb}^{I}$ where $\bfla_{\bfb}^{I}\in C^{1}(\cN,G)$.

\subsection{Scenario 2: Anomalies from gauging 1-form subgroup of 2-group symmetry in (2+1)d QFTs}
\noindent We consider an $H$-topological gauge theory in (2+1)d where $H$ is an abelian group. As we saw in the previous section, such theories have a 1-form symmetry which always has a subgroup $H_{[1]}$. Let us consider the scenario where our QFT has a global 2-group symmetry with the 1-form part being $H_{[1]}$, 0-form part being $G$ and extension class being $\alpha\in H^{3}(BG,H)$. We will see that by gauging $H_{[1]}$, we realize a quantum field theory whose anomaly can be tuned by choosing $[\alpha]$ and can be cancelled by a 0-form SPT in (3+1)d protected by 0-form group $G\times H$.

\medskip \noindent We begin with a theory with a 2-group global symmetry $\mathbb G$, defined on a closed (2+1)d manifold $\cN$ whose partition function we denote by $\mathcal Z^{\text{th}}(\cN)${\footnote{The superscript ${\rm th}$ is meant to denote theory. Similarly the superscript ${\rm th}/{\sss \bullet}$ implies the partition function for the same theory after ${\sss \bullet}$ subgroup of the total global symmetry group has been gauged.}}. The global symmetry can be probed by coupling the theory to a background 2-group connection defined by the local data $(g,h)$ satisfying the usual twisted cocycle conditions $dg=1$ and $d_{g \triangleright} h=\alpha(g)$. We denote the partition function of the theory coupled to background 2-group connection by $\mathcal Z^{\text{th}}(g|h \, , \, \cN)$. The 1-form subgroup can be gauged by summing over equivalence classes of 2-chains $h$. The partition function after gauging $H_{[1]}\subset \mathbb G$ is
\begin{align}
	\mathcal Z^{\text{th}/H_{[1]}}(g \, , \, \cN)\propto \sum_{h}\mathcal Z^{\text{th}}(g|h \, , \, \cN) \; .
\end{align}
The gauged theory has a dual 0-form global symmetry $H$ in addition to the original 0-form symmetry $G$. Therefore we can probe $H$ by coupling to a background 1-cochain $\hat{h}$ as
\begin{align}
	\mathcal Z^{\text{th}/H_{[1]}}(g| \hat{h} \, , \, \cN)\propto \sum_{h}\mathcal Z^{\text{th}}(g|h \, , \, \cN)e^{i\int_{\cN} h \smile \hat{h}} \; .
\end{align}
Under a gauge transformation $\hat{h}\to \hat{h} -d \hat{\lambda}$, the partition function transforms as
\begin{align}
	\mathcal Z^{\text{th}/H_{[1]}}(g|\hat{h}-d \hat{\lambda} \, , \, \cN)= \mathcal Z^{\text{th}/H_{[1]}}(g|\hat{h} \, , \, \cN)e^{i\int_{\cN}\alpha(g) \smile \hat{\lambda}} \; .
\end{align}  
This anomaly can be absorbed by a bulk 0-form SPT in (3+1)d protected by the symmetry group $G\times H$ with the topological response 
\begin{align}
	S_{\text{SPT}}(\cM)=\int_{\cM} \alpha(g) \smilo \hat{h} \; .
	\label{anomalyDW_3+1}
\end{align}
More specifically, by choosing $H=\mathbb Z_n$ and $G=\mathbb Z_{n}^{k-1}$ we may obtain candidate surface theories for all (3+1)d bosonic SPTs protected by $G\times H=\mathbb Z_{n}^{k}$. Indeed, similar to (2+1)d abelian DW theories, (3+1)d DW are constructed from three kinds of cocycles which we call type-{\small (II)},{\small (III)},{\small (IV)}. The topological  action corresponding to type-{\small (II)},{\small (III)} takes the form\cite{Tiwari:2017wqf, Tiwari:2016zru}
\begin{align}
	S_{\text{(II,III)}|\text{SPT}}(\cM)\propto \int_{\cM} \hat{h}\smilo g \smilo dg
\end{align}
whereas the topological action corresponding to type-{\small (IV)} takes the form
\begin{align}
	S_{\text{(IV)}|\text{SPT}}(\cM)\propto \int_{\cM} \hat{h}\smilo g \smilo g \smilo g \; .
\end{align}
Surface theories corresponding to these anomalies can be constructed via the above recipe by choosing $\alpha(g)\propto g\smilo dg$ or $\alpha(g)\propto g\smilo g\smilo g$ with the appropriate coefficient in \eqref{anomalyDW_3+1}.

An alternate construction for gapless surfaces of (3+1)d 0-form bosonic SPTs was provided in \onlinecite{Tiwari:2017wqf}.

\subsection{Scenario 3: Mixed 0,1-form anomalies from gauging finite subgroups of (2+1)d QFTs}
\noindent 
We will now consider a situation which was recently studied in the interesting work by Tachikawa \onlinecite{tachikawa2017gauging} and which is closely related to the works in \onlinecite{kapustin2014anomalies, wang2017symmetric, wang2018tunneling}. Consider a (2+1)d QFT with a global symmetry $\Gamma$. Let $H$ be an abelian subgroup in the center of $\Gamma$. We consider the following short exact sequence
\begin{align}
	0\rightarrow H \rightarrow \Gamma \rightarrow G \rightarrow 0
\end{align}
which defines $\Gamma$ as a central extension of $G$ by $H$ such that $\Gamma / H$ is isomorphic to $G$.
Isomorphism classes of such central extensions are captured by the cohomology classes $[\beta]\in H^{2}(G,H)$. By gauging $H$, we obtain a gauge theory with gauge group $H$ together with a 0-form global symmetry $G$ and a 1-form global symmetry $\hat{H}=H^{1}(H,{\rm U}(1))$ which is generated by the line operators of the gauge theory. We can probe these global symmetries by coupling the model with background fields ${\bfg}\in \text{Hom}(\pi_1(\mathcal N),G)$ and $\hat{\bfh}\in H^{2}(\mathcal N,\hat{H})$. 

Such model has an anomaly which we can now express as
\begin{align}
	\int_{\mathcal M}\beta ({\bfg}) \smilo \hat{\bfh} \; .
\end{align}  
More precisely, let the partition function of the original theory coupled to a background $H$-connection ${\bfh}\in Z^{1}(\mathcal N,H)$ be denoted by $\mathcal Z^{{\rm th}}({\bfh} \, , \, \mathcal N)$. We can dynamically gauge this symmetry by summing over classes of flat bundles. The partition function for the gauged theory then reads
\begin{align}
	\mathcal Z^{{\rm th}/H}(\mathcal N)=\frac{1}{|H|^{b_{0}(\mathcal N)}}\sum_{[{\bfh}]\in H^{1}(\mathcal N,H)} \mathcal Z^{{\rm th}}( {\bfh} \, , \, \cN) \; .
\end{align}  
This theory has a mixed 0-form and 1-form global symmetry which we write $(G,\hat{H})$. This mixed symmetry can be probed by introducing a background system of fields $({\bfg},\hat{\bfh} )\in \big(Z^{1}(\mathcal N,G),Z^{2}(\mathcal N,\hat{H})\big)$. It is crucial to notice that in the presence of background ${\bfg}$, the gauge field ${\bfh}$ is no longer a cocycle but satisfies the relation $d{\bfh}=\beta({\bfg})$. The theory ${\rm th}/H$ coupled to background $({\bfg},\hat{{\bfh}})$ takes the form
\begin{align}
	\mathcal Z^{{\rm th}/H}({\bfg}|\hat{\bfh} \, , \, \mathcal N)\propto \sum_{{\bfh}}e^{i\int_{\mathcal N}{\bfh} \smile \hat{\bfh} }\; \mathcal Z^{{\rm th}}(g|h \, , \, \cN) \; .
\end{align}
This theory is not invariant under the gauge transformation $\hat{\bfh} \rightarrow \hat{\bfh} + \hat{\bfla} $ where $\hat{\bfla} \in C^{1}(\cM,\hat{H})$. It can be readily checked that it transforms as
\begin{align}
	\mathcal Z^{{\rm th}/H}({\bfg}| \hat{\bfh} + \hat{\bfla} \, , \, \cN)= \mathcal Z^{{\rm th}/H}({\bfg}|\hat{\bfh} \, , \, \cN) \, e^{i\int_{\mathcal N}\beta({\bfg})\smile \hat{\lambda}}
\end{align}
hence the anomaly valued in a 4d theory described by the topological response action $\int_{\mathcal M}\beta({\bfg})\smilo \hat{\bfh}$. These kind of response theories were studied previously in \onlinecite{kapustin2017higher}. The corresponding mixed 1,2-form topological topological gauge theory was analyzed in \onlinecite{Chan:2017eov} and shown to furnish interesting Borromean-ring like triple link invariants between two surfaces and a loop knotted in four-dimensional spacetime.

\subsection{Scenario 4: Anomalies from gauging 1-form subgroup of 2-group symmetry in (3+1)d QFTs}
\noindent 
Consider a 4d version of Scenario 2 above. Let there be a QFT with global symmetry structure described by a 2-group $\mathbb G$ captured by the sequence
\begin{align}
	0\rightarrow H_{[1]}\rightarrow \mathbb G \rightarrow G \rightarrow 0
\end{align}
where the extension class is $[\alpha]\in H^{3}(G,H)$. We can gauge the 1-form symmetry $H_{[1]}$ by coupling to a background ${\bfh}\in Z^{2}(\mathcal N,H)$\footnote{Note that in this subsection, $\cN$ is a closed 4-manifold and $\cM$ a 5-manifold such that $\partial \cM=\cN$.}. Let the partition function be labeled $\mathcal Z^{{\rm th}}({\bfh} \, , \, \cN)$. We can make the gauge field dynamical, i.e sum over isomorphism classes of flat 2-connections to obtain the gauged theory with partition function
\begin{align}
	\mathcal Z^{{\rm th}/H_{[1]}}(\cN)=\frac{1}{|H|^{b_{1}-b_0}}\sum_{[{\bfh}]\in H^{2}(\cN,H)} \mathcal Z^{\text{th}}({\bfh} \, , \, \mathcal N)
\end{align}
Then the gauged theory has a global symmetry group which as a set can be described as $(\hat{H}_{[1]},G)$. As described above and in \cite{tachikawa2017gauging}, the theory ${\rm th}/{H}_{[1]}$ has an anomaly given by 
\begin{align}
	\int_{\mathcal M}\alpha({\bfg}) \smilo \hat{\bfh}
	\label{5d_anomaly}
\end{align} 
where we have introduced background fields ${\bfg
} \in \text{Hom}(\pi_1(\mathcal M),G)$ and $\hat{\bfh} \in H^{2}(\cM,\hat{H})$ \footnote{Here we use $g,\hat{h}$ for both the fields on $\cN$ and their extension to the 5-manifold $\cM$.}. $\alpha(\bfg) \in Z^{3}(\mathcal M,H)$ refers to the cocycle $\alpha$ evaluated on ${\bfg}$. The background field $\hat{\bfh}$ enters the action via the coupling 
\begin{align}
	\int_{\mathcal N} {\bfh}\smilo \hat{\bfh} \; ,
\end{align}
i.e, the gauged theory ${\rm th}/H_{[1]}$ coupled to background fields $({\bfg},\hat{\bfh})$ takes the form
\begin{align*}
	\mathcal Z^{{\rm th}/H_{[1]}}({\bfg}|\hat{\bfh} \, , \, \cN)
	\propto \sum_{[{\bfh}]}  e^{i\int ({\bfh}\smile \hat{\bfh} + \omega({\bfg}|\hat{\bfh} ))}\mathcal Z({\bfg}|{\bfh} \, , \, \cN)
\end{align*} 
where we have also included $\omega \in H^{4}(G\times H_{[1]},\mathbb R/2\pi \mathbb Z)$ which is the analog of the discrete torsion phase that shows up in orbifold conformal field theories for four-dimensional spacetime with a 2-group instead of an ordinary group. The anomaly may be computed straightforwardly by performing a gauge transformation $\hat{\bfh}\rightarrow \hat{\bfh} +d \hat{\bfla}$ where $\hat{\bfla} \in C^{1}(\mathcal N, \hat{H})$. Since $d {\bfh} =\alpha({\bfg})$, it can be seen that a term of the form \eqref{5d_anomaly} is required to cancel this gauge variation. 

\bigskip \noindent  Alternately, we can consider a more interesting scenario where the theory `${\rm th}$' is itself anomalous and has an anomaly of the form 
\begin{align}
	\int_{\mathcal M}{\bfh}\smilo \widetilde{\alpha}({\bfg}) 
	\label{5d_anomaly2}
\end{align}
where $[\widetilde{\alpha}]\in H^{3}(G,\hat{H})$. Let us denote this theory by ``${\rm th}_{\text{anom}}$''. Then we may be interested in studying the anomaly structure of the theory one obtains by gauging the 1-form symmetry $H_{[1]}$ in theory ${\rm th}_{\text{anom}}$. In a recent study of SU($N$) gauge theories \cite{gaiotto2017theta} a similar phenomena was shown to occur for a mixed anomaly between 0-form $CP$ symmetry and 1-form center $\mathbb Z_{N}$ symmetry (at $\theta=\pi$). As before let us label the partition function of ${\rm th}_{\text{anom}}$ coupled to background fields $({\bfg},{\bfh})$ by $\mathcal Z^{{\rm th}_{\text{anom}}}({\bfg}|{\bfh} \, , \, \cN)$. Then we may gauge by summing over isomorphism classes of flat 2-form fields $[{\bfh}]$
\begin{align*}
	&\mathcal Z^{{\rm th}_{\text{anom}}/H_{[1]}}({\bfg} \, , \, \cN) \\
	& \q =\frac{1}{|H|^{b_1-b_0}}\sum_{[{\bfh}]\in H^{2}(\mathcal N,H)}\mathcal Z^{{\rm th}_{\text{anom}}}({\bfg}|{\bfh} \, , \, \cN) \; .
\end{align*} 
As before, the gauged theory ${\rm th}_{\text{anom}}/H_{[1]}$ has a dual 1-form symmetry which may be probed by coupling to a background field $\widehat{h}\in H^{2}(\cN,\hat{H}_{[1]})$
\begin{align}
	&\mathcal Z^{{\rm th}_{\text{anom}}/H_{[1]}}({\bfg}|\hat{\bfh} \, , \, \cN)
	\\ \nn & \q =\frac{1}{|H|^{b_1-b_0}}\sum_{[{\bfh}]\in H^{2}(\mathcal N,H)} \! \! e^{i\int_{\mathcal N}{\bfh}\smile \hat{\bfh} }\, \mathcal Z^{{\rm th}_{\text{anom}}}({\bfg}|{\bfh} \, , \, \cN) \; .
\end{align}
We require the theory ${\rm th}_{\text{anom}}/H_{[1]}$ to be invariant under gauge transformations, ${\bfh}\rightarrow {\bfh}+d \bfla$ where $\bfla \in C^{1}(\mathcal N,H)$. This requires imposing $d \hat{\bfh} =\widetilde{\alpha}({\bfg})$ in order to cancel the variation of the anomaly \eqref{5d_anomaly2} which reminds us that the symmetry of the gauged theory is not a direct product of a 0-form and 1-form symmetry but in fact forms a non-trivial 2-group whose extension class is $[\widetilde{\alpha}]\in H^{3}(G,\hat{H})$. Topologically distinct coupling to background $(\bfg, \hat{\bfh})$ are labeled by classes in $H^{4}(\widetilde{\mathbb G},\mathbb R/2\pi \mathbb Z)$ where $\widetilde{\mathbb G}$ is the 2-group which fits in the short exact sequence
\begin{align}
0\rightarrow \hat{H}_{[1]}\rightarrow \widetilde{\mathbb G} \rightarrow G \rightarrow 0
\end{align}
with extension class $[\tilde{\alpha}]$. 

\section{Discussion}
\noindent
Cohomological models of topological phases have been under intense investigation in the past years. So far they seem to encapsulate most of the known models of intrinsic topological orders in (3+1)d. For instance, it was argued in \onlinecite{lan2017classification} that bosonic topological orders with bosonic point-like excitations are classified by a pair $(G,[\omega])$ with $G$ a discrete group and $[\omega] \in H^4(BG,{\rm U}(1))$. Furthermore, these models typically have a topological lattice gauge theory interpretation which is made prominent in the design of lattice Hamiltonian realization. 

The goal of this manuscript was three-fold: Provide additional evidence that 2-category is the proper language to describe the input data of a (3+1)d string net model, emphasize how higher gauge theories naturally arise as a generalization of the group cohomological models, and finally study properties of these higher-gauge models built upon 2-groups.  

Starting with the study of the Hamiltonian realization of the four-dimensional Dijkgraaf-Witten model, we used simplicial arguments to explain how the consistency condition of the unitary map performing a three-dimensional Pachner move corresponds to the coherence relation of a structural 2-morphism of a given monoidal 2-category, namely the pentagonator. The combinatorics of the corresponding commutative diagram is the one of the fifth Stasheff polytope. We were then able to identify the 2-category which yields the four-dimensional Dijkgraaf-Witten, namely the category of $G$-graded 2-vector spaces, such that the cohomological twists plays the role of the pentagonator. Thinking of this 2-category as a base, we were then able to obtain richer topological phases by relaxing some of its defining axioms. Most interestingly, in the process we turned the underlying 1-category into a weak 2-group. The corresponding topological theory could therefore be understood as a cohomological model for a 2-group which has a higher gauge theory interpretation. Exploiting this relation with higher gauge theories is key in order to derive the corresponding lattice Hamiltonian. 

Following a strategy analogous to the one suitable to cohomological gauge models, we were able to describe the symmetry protected phases associated with the 2-group TQFTs, higher-form TQFTs and their generalizations. Further we described their lattice realization. Moreover, we explained in detail the gauging procedure which relate these symmetry protected phases to their respective topological orders. Finally, we provided a detailed discussion regarding the relation between these symmetry protected topological phases in $d$+1 dimensions and 't Hooft anomalies for global symmetries in $d$-dimensional quantum field theories. 

\medskip \noindent
An interesting question that we will address in the future is the characterization of the excitations for these higher gauge models of topological phases. In (2+1)d, it is well-known that the excitations of the Hamiltonian realization of the three-dimensional Dijkgraaf-Witten model are labeled by irreducible representations of the twisted Drinfel'd double \cite{Dijkgraaf1991,Kitaev1997,Hu:2012wx,DDR1}. Furthermore, it was shown in \onlinecite{Wan:2014woa, Wang:2014oya, Delcamp:2017pcw} using dimensional reduction techniques that the three-dimensional generalization of Kitaev's double model yields excitations labeled by irreducible representations of the so-called \emph{quantum triple}. We expect the excitations of topological models built upon 2-groups to be characterized by extensions of these algebraic structures. This will probably require us to make use of higher order representation theory \cite{2008arXiv0812.4969B}. A first step should be to study the excitations in the case of 2-form gauge theories. In three dimensions, these should yield point-like and membrane-like excitations.

Another prospect is the study of the Lagrangian formulation for such models in the continuum. We would then show how we can recover the lattice higher gauge models presented here upon canonical quantization. In particular, we would like to be able to provide explicit formulas in the spirit of \eqref{DW_action}. However, such explicit formula in terms of cochains are usually reserved for abelian groups. Therefore, we expect to study first the case of 2-groups built upon two abelian groups.

More generally, it would be very interesting to provide additional physical motivation for the study of such models. As we explained in this paper both from a category theoretical and gauge theory point of view, topological phases of matter built upon 2-groups appear very naturally. However, the role played by such models in physics needs further investigation.

\begin{acknowledgments}
\noindent
CD and AT contributed equally to this work.
CD is supported by an NSERC grant awarded to B.~Dittrich. 
AT is supported by NSF grant (under Grant No. NSF
PHY-1125915) awarded to S.~Ryu. We would like to thank Bianca Dittrich for discussions. AT would like to thank Davide Gaiotto, Shinsei Ryu, Lakshya Bhardwaj and Ken Shiozaki for numerous helpful conversations.
This research was supported in part by Perimeter Institute for Theoretical Physics.
Research at Perimeter Institute is supported by the Government of Canada through the Department of Innovation, Science and Economic Development Canada and by the Province of Ontario through the Ministry of Research, Innovation and Science.
\end{acknowledgments}

\appendix
\section{Group cohomology \label{app:coho}}
\noindent
Let us consider a finite group $G$ and a $G$-module $M$ which is an abelian group. There is an action $\triangleright$ of the group $G$ on $M$ which commutes with the abelian multiplication rule. Any function of the form $\omega_n: G^n \rightarrow M$ defines an $n$-{\it cochain}. The space of cochains $\{\omega_n\}$ is denoted by $C^n(G,M)$. We define the so-called {\it coboundary} operator $d^{(n)}: C^n(G,M) \rightarrow C^{n+1}(G,M)$ whose action reads
\begin{align} \nn
	&d^{(n)}\omega_n(g_1, \dots,g_{n+1}) \\ 
	\nn & \q = 
	g_1 \triangleright \omega_n(g_2, g_3, \dots, g_{n+1})\omega_n(g_1, \dots, g_n)^{(-1)^{n+1}} \\
	& \q \times \prod_{i=1}^n \omega(g_1, \dots, g_{i-1},g_{i} \cdot g_{i+1},g_{i+2}, \dots, 	g_{n+1})^{(-1)^i} \; .
\end{align}
We define {\it cocycles} as cochains satisfying the equation
\begin{equation}
	d^{(n)} \omega_n = 1
\end{equation}
which is commonly referred to as the {\it cocycle equation}. The subgroup of cocycles is denoted by $Z^n(G,M)$. If a cocycle is of the special form
\begin{equation}
	\omega_n = d^{(n-1)}\omega_{n-1} \; ,
\end{equation}
it defines a {\it coboundary}. The subgroup of coboundaries is denoted by $B^n(G,M)$. The equivalence classes of $n$-cocycles finally form the $n$-th {\it cohomology group}:
\begin{equation}
	H^n(G,M) := \frac{Z^n(G,M)}{B^n(G,M)} =
	\frac{\text{ker}(d^{(n)})}{\text{im}(d^{(n-1)})} \; .
\end{equation}
In order to visualize the cocycle conditions, it is convenient to introduce the corresponding dual homology theory. In particular, we can define \emph{n-chains} which are dual to $n$-cochains and form a group denoted by $C_n(G,M)$. Denoting the vertices of the $n$-simplex $(v_0, \dots,v_{d+1})$ such that $v_0 < \dots <v_{d+1}$, we can write $n$-chains as $(g_1, \dots, g_d)$ in terms of group elements labeling the 1-simplices $(v_0v_1), (v_1v_2), \dots, (v_{d}v_{d+1})$ of $\triangle_n$ such that $g_b \equiv g_{(v_av_b)}$.

By dualizing the coboundary operator, we obtain the \emph{boundary} operator $\partial^{(n)}: C_{n}(G,M) \rightarrow C_{n-1}(G,M)$ whose action reads
\begin{align} \nn
	&\partial^{(n)}(g_1, \dots,g_{n}) \\ 
	\nn & \q = 
	(g_2, g_3, \dots, g_{n})(g_1, \dots, g_{n-1})^{(-1)^{n}} \\
	& \q \times \prod_{i=1}^{n-1} (g_1, \dots, g_{i-1},g_{i} \cdot g_{i+1},g_{i+2}, \dots, 	g_{n})^{(-1)^i} \; .
\end{align}
We can now make the following statement: The coboundary $d^{(n)} \omega_n$ of $\omega_n$ evaluated on $\triangle_{n+1}$ is equal to the evaluation of the cocycle $\omega_n$ on the boundary $\partial^{(n+1)} \triangle_{n+1}$ of $\triangle_{n+1}$. This can be summarized as follows
\begin{equation}
	\la  d^{(n)} \omega_n, {\triangle_{n+1}} \ra = \la \omega_n , {\partial^{(n+1)} \triangle_{n+1}} \ra
\end{equation}
which is nothing else than the cohomological version of Stoke's theorem.

Given a $p$-cochain $\alpha \in C^p(G,M)$ and a $q$-cochain $\beta \in C^q(G,M)$, we define the \emph{cup product} $\alpha \smilo \beta : G^{p+q} \rightarrow M $ of $\alpha$ and $\beta$ as 
\begin{equation}
	[\alpha \smilo \beta](g_1,\dots,g_{p+q}) = \alpha(g_1, \dots, g_p)\beta(g_{p+1}, \dots, g_q) \; .
\end{equation}
Most importantly, we have the following property
\begin{equation}
	d^{(p+q)}(\alpha \smilo \beta) = d^{(p)}\alpha \smilo \beta + (-1)^p \alpha \smilo d^{(q)} \beta
\end{equation}
so that if $d^{(p)} \alpha = 1 = d^{(q)}\beta$ then $d^{(p+q)}(\alpha \smilo \beta) =1$. This cup product can then be used to turn a cohomology group into a cohomology ring where the cup product serves as ring multiplication. We can further define higher cup product. In the main text we only make use of the \emph{first higher cup product} $\smile_1$ such that
\begin{align} \nn
	&[\alpha \! \smile_1 \! \beta](g_1,\dots,g_{p+q-1}) \\ \nn
	& \q = \sum_{j=0}^{p-1}\alpha \Big(g_1,\dots,g_j, \prod_{i=1}^q g_{i+j},g_{j+p+1}, \dots, g_{p+q-1} \Big) \\ 
	& \hspace{2.65em} \times \beta(g_{j+1}, \dots, g_{j+q}) \; . 
\end{align}

\section{3-cocycle condition for 2-groups \label{app:cocycle2Gr}}
\noindent
Let us consider a 2-group 3-cocycle $\omega_3$ in $H^3(B\mathbb{G}, {\rm U}(1))$. In vertue of the fact that a flat 2-connection is realized locally by a 1-cocycle and a 2-cochain, the cocycle $\omega_3$ is a function of three group variables $g_1,g_2,g_3 \in G$ and three group variables $h_1,h_2,h_3 \in H$. Applying the convention defined in the main the text in sec.~\ref{sec:expLat1}, the tetrahedron associated with $\omega_3(g_1,g_2,g_3|h_1,h_2,h_3)$ is given by
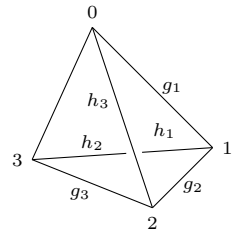
\begin{equation*}
\SPTtetraONE{0.8}{3}
\end{equation*} 
The coboundary of a 2-group 3-cochain reads
\begin{align}
	\label{cocycleTwoGr}
	& \hspace{-1em} d\omega_3(g_1,\dots,g_4|h_1, \dots h_6) \\ \nn
	= \, & \omega_3(g_2,g_3,g_4|h_1,h_2,h_3)
	\\ \nn \times & \, \omega_3(g_1,g_2,g_3|h_4,h_1+h_5-h_4-\alpha(g_1,g_2,g_3),h_5)
	\\ \nn \times & \, \omega_3(g_1,g_2g_3,g_4|h_5,h_2+h_6-h_5-\alpha(g_1,g_2g_3,g_4),h_6)
	\\ \nn \times & \, \omega_3(g_1,g_2,g_3g_4|h_4,h_3+h_6-h_4-\alpha(g_1,g_2,g_3g_4),h_6)^{-1}
	\\ \nn \times & \, \omega_3(g_1g_2,g_3,g_4|h_1+h_5-h_4-\alpha(g_1,g_2,g_3),
	\\ \nn & \, \hspace{6.3em} h_2+h_6-h_5-\alpha(g_1,g_2g_3,g_4),
	\\ \nn & \, \hspace{6.3em} h_3+h_6-h_4-\alpha(g_1,g_2,g_3g_4))^{-1}
	\; .
\end{align}
which is associated to the following $\mathcal{P}_{2 \mapsto 3}$ move:
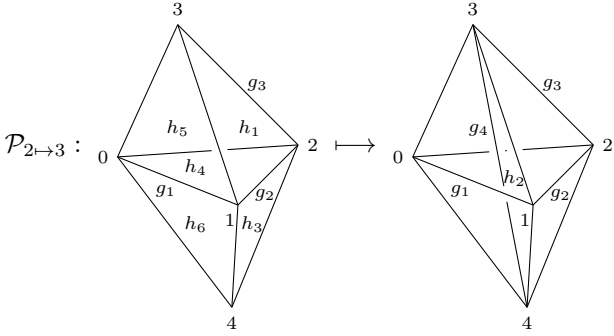
\begin{equation*}
	\mathcal{P}_{2 \mapsto 3}: \SPTtetraTWO{0.8}{1} \longmapsto
	\SPTtetraTWO{0.8}{2}
\end{equation*}
such that $h_1=h_{123}$, $h_2=h_{134}$, $h_3=h_{124}$, $h_4=h_{012}$, $h_5=h_{013}$ and $h_6=h_{014}$. 
The cocycle condition then simply reads $d^{(3)}\omega_3 =1 $. In the case where the abelian group $H$ is trivial, this reduces to the usual 3-cocycle condition for group cohomology. 

Recall that the SPT model studied in sec.~\ref{sec:expSPT} is defined in terms of a homogeneous function $\nu_3$ of four group variables in $G$ and six group variables in $H$ which satisfy the following condition ensuring the invariance under change of triangulation
\begin{align} 
	\prod_{j=0}^{4}\nu_3^{(-1)^i}(k_0, \dots, \widehat{k}_j, \dots, k_{4} | 
	\lambda_{01}, \dots, \widehat{\lambda}_{{\sss \bullet }j \atop j{\sss \bullet }}, \dots, \lambda_{34})= 1 \; .
\end{align}
As explained in the main text, the cocycle $\omega_3$ defined above can be expressed in terms of this function $\nu_3$ according to the formula:
\begin{align}
	&\omega_3(k_1,k_2,k_3|\lambda_{a}, \lambda_{b}, \lambda_{c}) 
	\\ \nn & \! \equiv \nu_3(k_1k_2k_3, k_2k_3, k_3,\unit | 
	\lambda_{a}-\zeta(\unit,k)_{012},0,0,0,
	\\ \nn& \hspace{10.6em} \lambda_{c}-\lambda_{a}-\zeta(\unit,k)_{013} + \zeta(\unit,k)_{012},
	\\ \nn & \hspace{10.6em} \lambda_b-\zeta(\unit,k)_{023}) \; .
\end{align}
and conversely, the functions $\nu_3$ can be written in terms of the 3-cocycle $\omega_3$ as
\begin{align*}
	&\nu_3(k_0, \dots, k_3 | \lambda_{01}, \dots, \lambda_{i<j}, \dots, \lambda_{23}) \\
	& \q = \omega_3(k_0k_1^{-1},\dots, k_{2}k_3^{-1} | d\lambda_{012} + \zeta(\unit,k)_{012}, 
	\\ & \hspace{11.5em}d\lambda_{023} + \zeta(\unit,k)_{023},
	\\& \hspace{11.5em} d\lambda_{013} + \zeta(\unit,k)_{013})
\end{align*}
using the notation $d \lambda_{abc} \equiv \lambda_{bc}-\lambda_{ac}+\lambda_{ab}$.
We can now show how the cocycle condition \eqref{cocycleTwoGr} follows from these two relations. We have the explicit formula: 
\begin{align*}
	&\prod_{j=0}^{4}\nu_3^{(-1)^j}(k_0, \dots, \widehat{k}_j, \dots, k_{4} | 
	\lambda_{01}, \dots, \widehat{\lambda}_{{\sss \bullet }j \atop j{\sss \bullet }}, \dots, \lambda_{34}) \\
	&\q = 
	\omega_3(k_1k_2^{-1},k_2k_3^{-1}, k_{3}k_4^{-1} | d\lambda_{123} + \zeta(\unit,k)_{123},	
	\\& \hspace{12.7em} d\lambda_{134} + \zeta (\unit,k)_{134},
	\\& \hspace{12.7em}d\lambda_{124}+\zeta(\unit,k)_{124}) 
	\\ & \q \times
	\omega_3(k_0k_2^{-1},k_2k_3^{-1}, k_{3}k_4^{-1} | d\lambda_{023} +  \zeta(\unit,k)_{023}, 
	\\& \hspace{12.7em} d\lambda_{034} +\zeta(\unit,k)_{034},
	\\& \hspace{12.7em} d\lambda_{024} +\zeta(\unit,k)_{024})^{-1}
	\\ & \q \times
	\omega_3(k_0k_1^{-1},k_1k_3^{-1}, k_3k_4^{-1} | d\lambda_{013} + \zeta(\unit,k)_{013}, 
	\\& \hspace{12.7em} d\lambda_{034} +\zeta(\unit,k)_{034},
	\\& \hspace{12.7em} d\lambda_{014} +\zeta(\unit,k)_{014})
	\\ & \q \times
	\omega_3(k_0k_1^{-1},k_1k_2^{-1}, k_2k_4^{-1} | d\lambda_{012} + \zeta(\unit,k)_{012}, 
	\\& \hspace{12.7em} d\lambda_{024} +\zeta(\unit,k)_{024},
	\\& \hspace{12.7em} d\lambda_{014} +\zeta(\unit,k)_{014})^{-1}
	\\ & \q \times
	\omega_3(k_0k_1^{-1},k_1k_2^{-1}, k_{2}k_3^{-1} | d\lambda_{012} + \zeta(\unit,k)_{012}, 
	\\& \hspace{12.7em} d\lambda_{023} +\zeta(\unit,k)_{023},
	\\& \hspace{12.7em} d\lambda_{013} +\zeta(\unit,k)_{013})  \; .
\end{align*}
Let us further define $g_i \equiv k_{i}k_{i+1}^{-1}$ for $i=1,\dots,4$,  $h_1=d\lambda_{123}+\zeta(\unit,k)_{123}$, $h_2=d\lambda_{134}+\zeta(\unit,k)_{134}$, $h_3=d\lambda_{124}+\zeta(\unit,k)_{124}$, $h_4=d\lambda_{012}+\zeta(\unit,k)_{012}$, $h_5=d\lambda_{013}+\zeta(\unit,k)_{013}$ and $h_6=d\lambda_{014}+\zeta(\unit,k)_{014}$. Using these conventions, we have the following relation
\begin{align*}
	&d\lambda_{023}+\zeta(\unit,k)_{023} \\
	&  \q = d\lambda_{123} +d\lambda_{013}  - d\lambda_{012} + \zeta(\unit,k)_{023} \\
	& \q = d\lambda_{123} +d\lambda_{013}  - d\lambda_{012} - d\zeta(\unit,k^{-1})_{0123}\\
	& \q \;\; +\zeta(\unit,k)_{123} + \zeta(\unit,k)_{013} - \zeta(\unit,k)_{012}
	\\   & \q =  h_1+h_5-h_4 - \alpha(g_1,g_2,g_3)
\end{align*}
where we used the fact $d\zeta(\unit,k)_{0123} = \alpha(\unit^{k})_{0123} = \alpha(g_1,g_2,g_3)$. Applying the same technique for every term in the previous expression, we finally recover the cocycle condition \eqref{cocycleTwoGr}.

\bibliography{ref_cat}	
\end{document}